\documentclass[10pt,journal]{IEEEtran}

\ifCLASSOPTIONcompsoc
  \usepackage[nocompress]{cite}
\else
  \usepackage{cite}
\fi

\ifCLASSINFOpdf
   \usepackage[pdftex]{graphicx}
\else
\fi
\usepackage{graphicx}

\usepackage{amsmath}

\usepackage{algorithmic}

\usepackage{array}

\usepackage{eqparbox}

\usepackage{stfloats}

\usepackage{url}

\usepackage{graphics} 
\usepackage{epsfig} 
\usepackage{mathptmx} 
\usepackage{times} 
\usepackage{amsmath} 
\usepackage{amssymb}  
\usepackage{graphicx}
\usepackage{caption}
\usepackage{url}
\usepackage{epstopdf}
\usepackage{balance}
\usepackage{subcaption}
\usepackage{booktabs}
\usepackage[table,xcdraw]{xcolor}
\usepackage{multirow}
\usepackage{todonotes}
\usepackage{slashbox} 
\usepackage{adjustbox}
\usepackage{fancyhdr} 
\usepackage{longtable}  
\usepackage{lipsum}
\usepackage{placeins}

\makeatletter
\def\endthebibliography{%
  \def\@noitemerr{\@latex@warning{Empty `thebibliography' environment}}%
  \endlist
}
\makeatother

\begin{document}
%
\title{A Survey of Controller Placement Problem in Software Defined Networks}
%
%

\author{Abha~Kumari,~\IEEEmembership{Student Member,~IEEE,}
                and~Ashok~Singh~Sairam,~\IEEEmembership{Senior~Member,~IEEE}
\IEEEcompsocitemizethanks{\IEEEcompsocthanksitem Abha~Kumari is pursuing her PhD from the Computer Science and Engineering Department, Indian Institute of Technology Patna.
\protect\\
E-mail: abha.pcs16@iitp.ac.in
\IEEEcompsocthanksitem Ashok Singh Sairam is an associate professor at Indian Institute of Technology Guwahati
\protect\\
E-mail: ashok@iitg.ac.in see http://www.iitg.ac.in/ashok}
\thanks{}}

%
%

\IEEEtitleabstractindextext{%
\begin{abstract}
\normalsize {Software Defined Network (SDN) is an emerging network paradigm which provides a centralized view of the network by decoupling the network control plane from the data plane. This strategy of maintaining a global view of the network optimizes resource management. However, implementation of SDN using a single physical controller lead to issues of scalability and robustness. A physically distributed but logically centralized SDN controller architecture promises to resolve both these issues. Distributed SDN along with its benefits brings along the problem of the number of controllers required and their placement in the network. This problem is referred to as the controller placement problem (CPP) and this paper is mainly concerned with the CPP solution techniques. The paper formally defines CPP, gives a comprehensive review of the various performance metrics and characteristics of the available CPP solutions. Finally, we point out the existing literature gap and discuss the future research direction in this domain.

}
\end{abstract} 

}

\maketitle

\IEEEdisplaynontitleabstractindextext

%
\IEEEpeerreviewmaketitle
\pagestyle{fancy}
\renewcommand{\headrulewidth}{0pt}
\fancyhead[L]{Abha et al.: A Survey of Controller Placement Problem in Software Defined Networks}
\fancyhead[R]{\thepage}
\fancyfoot[C]{}

\ifCLASSOPTIONcompsoc
\IEEEraisesectionheading{\section{Introduction}\label{sec:introduction}}
\else
\section{Introduction}
\label{sec:introduction}
\fi

%
%
%
%
\IEEEPARstart   With the growth of enterprise data centers, traffic pattern has changed significantly. Unlike the traditional client-server application where majority of the traffic flow is between a client and a server, today's application access a number of databases and servers which additionally generates substantial machine-to-machine traffic. At the same time, the explosion in the number of mobile devices and continuous generation of huge content, demand high-bandwidth, and a dynamic network environment. It is virtually impossible for the traditional network to cater to today's dynamic server environment due to its static nature and tight coupling between the control and data plane. 

SDN has grown as a networking paradigm that extends the adoption of virtualization technologies from servers to networks. Although SDN is still in its early days, a number of SDN based networking devices and software is being rolled out by the SDN industry, including major industry players such as B4\cite{jain2013b4} (Andromeda\cite{211243}) from Google,  SWAN\cite{hong2013achieving} from Microsoft (Windows Server 2016\cite{WindowsServer2016} ) and Application Centric Infrastructure\cite{ACI} from Cisco. 

One of the greatest advantages of SDN is that the network intelligence is centralized in software-based controllers. The network appears as a single control plane to applications which greatly simplifies network design and operation. Moreover, a centralized control layer facilitates network administrators management and optimization of network resources.     The logical centralization of SDN can be achieved using either a single controller or multiple controllers who share part of their local network state to form a global view. A single controller is an implementation of SDN in true spirit but it has serious issues with regard to scalability due to the limited resources of the SDN controller when handling a large number of requests. A single SDN controller also has robustness issues as it is a single point of failure.  To overcome these issues, a distributed SDN controller has been proposed. The general idea behind the deployment of multiple controllers is to distribute the load in the network equally. Further, one controller should take over when another crashes, thus solving both the issues of scalability and robustness. 

In a distributed SDN controller architecture, the network performance considerably depends on the placement of controllers, the so-called \textit{controller placement problem} (CPP). For that matter, controller placement is also a problem in a SDN with a single physical controller\cite{beheshti2012fast}, but the problem is less pronounced. Heller et al. \cite{heller2012controller} were the first to perform research on CPP. They formulated the problem as a facility location problem and it was shown to be NP-hard. Since then there have been numerous efforts to optimally place the controllers. In this survey, we comprehensively identify the performance metrics used in CPP and how they relate to the available CPP solutions. We group the various CPP solutions available in literature and identify the tradeoff between these approaches. Finally, we identify the research gaps to motivate further research in CPP.

\subsection{Why are we conducting the survey on CPP in SDN?}
In the recent past, there have been a number of surveys in the broad area of SDN. These surveys can be broadly classified into three categories - surveys dealing on a holistic view of the SDN paradigm, specific aspects of SDN and SDN for different network environments. Surveys that deal with the general aspect of SDN are those surveys that deal with SDN architecture, taxonomies, application domain and research issues. B. A. A. Nunes et. al. \cite{nunes2014survey}  emphasized the evaluation of SDN with previous variants of programmable networks. The authors of \cite{hu2014survey}-\cite{jarraya2014survey} discussed components of SDN like network virtualization, QoS, SDN languages abstraction etc. The other surveys on SDN in addition to the general discussion on SDN focused on the layers of SDN and overview of ongoing research in each layer \cite{xia2015survey}, controller design and its performance\cite{farhady2015software} and cross-layer issues of SDN \cite{kreutz2015software}.

 The surveys that focus on a single aspect of SDN are those dealing with distributed aspects of SDN design \cite{xie2015control}-\cite{oktian2017distributed} \cite{murat2017distributed} \cite{bannour2017distributed}. These surveys discuss the design aspects of a distributed control plane such as topology, controller synchronization and datastore. Troist et al.\cite{trois2016survey} presented an extensive investigation of \textit{OpenFlow based SDN programming languages}, its features and ongoing research efforts. The survey by Fonseca et al.\cite{fonseca2017survey}  addressed \textit{fault management} in SDN while the \textit{security} aspects of SDN has been investigated in \cite{scott2013sdn}-\cite{yan2016software}, \cite{ahmad2015security}.  The survey by Karakus et al.\cite{karakus2017quality} focused on the \textit{QoS} aspects in SDN. The other SDN survey focus areas include \textit{load balance}\cite{badirzadeh2018survey}-\cite{li2017load}, \textit{traffic engineering} \cite{mendiola2016survey}, \textit{resource allocation}\cite{zehra2017survey}, test-bed for large-scale SDN\cite{huang2016survey} and energy\cite{tuysuz2017survey}\cite{rawat2017software}. 
 
 There are a few surveys that focus on the application of SDN in networks other than wired. For instance, Liang et al.\cite{liang2015wireless}, Haque et al.\cite{haque2016wireless} have investigated SDN for wireless networks. SDN based optical network has been discussed in\cite{thyagaturu2016software}.

 To the best of our knowledge, there exist a very limited number of surveys that have addressed the CPP of SDN~\cite{wang2017controller}\cite{yoon2017controller}\cite{singh2018survey}. These few available surveys are not comprehensive, the numbers of papers reviewed are limited and the classification is based on a single performance parameter. The state of the art on CPP is large and the solution techniques vary widely. In this survey, we have tried to present a comprehensive survey covering all possible aspects of CPP. In order to present a consistent view of the control placement problems formulated in literature, we formulated the problems again using the same set of symbols and variables. 

\subsection{Survey Strategy}
In this survey, we identify the limitations of a single controller, advantages and additional concerns of a distributed control plane and then identify the issues concerning controller placement in SDN. We examine the current efforts to address these issues, the contributions and research gaps in the ongoing researches. As there is a large body of work on CPP, we compare these works using two broad comparison criteria - (i) objectives/evaluation metric and (ii) solution approach.
 
Performance metrics or evaluation criteria are important to judge the quality, performance, and efficiency of a proposed scheme. In traditional networks, latency is generally considered as an evaluation metric while designing network algorithms. However, in SDN, there are a number of additional performance metrics which are equally important and specific to SDN such as controller capacity, flow setup time etc. In this survey, we examine in detail the various metrics used in CPP research and try to understand their significance. 

The performance metrics are further categorized into \textit{independent} and \textit{dependent}. Independent metrics are those metrics which can be optimized either in isolation or in combination. These metrics may be constructively dependent in the sense that if we improve one it may positively affect the other, for example, flow setup time and latency. The independent metrics may also be conflicting in nature as in the case of latency and capacity. Dependent metrics, on the other hand, have a direct relation and they cannot be considered in isolation. These metrics are conflicting in nature because if we improve one it will negatively affect the other. For example, to improve energy, some controllers and links need to be switched off which will adversely affect the latency.

The CPP is a combinatorial problem which is known to be NP-hard. To tackle the problem, researchers bring about simplifying assumptions and have experimented with different solution approaches. In this survey, we compare the complexity and accuracy of the different CPP solution approaches. The placement of controllers depends on the type of network. Solutions to the CPP are thus structured according to those that target \textit{wired} and \textit{wireless} networks. There are two end results of a solution to the CPP, placement of the controllers and the mapping (or binding) of switches to a controller. Most of the early works on CPP assume that the mapping between the remains unchanged. The Internet traffic being bursty for small time scales, researchers have realized that a static controller to switch mapping cannot guarantee optimal performance for all time instants. To account for this temporal variation of network traffic, there is a large body of work which proposes to dynamically change the binding between the switch and the controller. In this survey, the CPP solutions for wired networks is further classified into \textit{static} and \textit{dynamic or adaptive}.

\begin{figure}[!htb]

\centering
\includegraphics[scale=0.65]{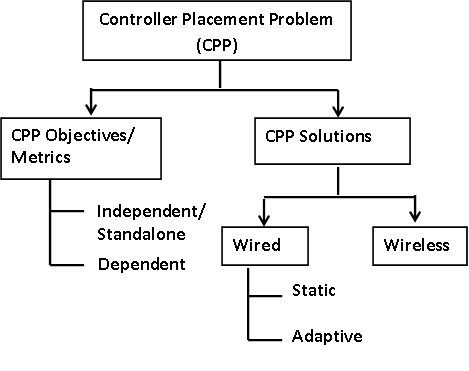}
\caption{Survey strategy}
\label{fig:survey}
\end{figure}

In the case of wireless networks, the link characteristic is highly dynamic and so a static controller to switch mapping will not work for such an environment. Thus most of the CPP solutions for wireless networks consider a dynamic mapping. Although we consider wireless SDN from a CPP perspective, the deployment issues of SDN in a wireless environment is not in the scope of this survey. The survey technique used in the paper is summarized in Figure \ref{fig:survey}.

\subsection{Roadmap}
The rest of the paper is organized as follows. Section \ref{background} briefly discuss the components and working principle of software defined networking. The limitations of a single control plane and the design issues of a distributed control plane architecture are discussed in Section \ref{distributed}. Section \ref{CPP} introduces the controller placement problem (CPP) and outlines the numerous objectives used in CPP. The solutions proposed to static solve CPP is presented in Section \ref{CPP_Solutions}. Literature related to dynamic controller placement is provided in this section~\ref{DCPP_Solutions}. The next section~\ref{sec:CPP_sol_Wireless} focus on CPP for wireless network. Section~\ref{sec:Ch_CPP_Sol} presents the overview of solution approaches and techniques that are used in literature.
 Section~\ref{sec:Disc} highlights open research problems of controller placement problem in SDN. Lastly, Section~\ref{Con}  concludes the conclusions.

\section{Background} \label{background}
In traditional networks, the tight coupling of the control plane which decides how to handle network traffic and the forwarding plane which forwards traffic as per the decisions of the control plane, restrict flexibility and evolution of the network. SDN is defined as a network architecture where the control plane is decoupled from the data plane and moved to an external entity called the SDN controller or NOS (Network Operating System). In SDN forwarding decisions are flow-based and the network is programmable through software applications running on top of the NOS \cite{farhady2015software}.
In the SDN architecture,  the network intelligence in SDN is logically centralized in software-based SDN controllers, who are authorized with the task of maintaining a global view of the network. As a result, the network is abstracted to the applications above as a single logical switch. Enterprises, carriers and end-users gain control over the entire network as a single logical point which simplifies network design and operation. At the same time, the design of the network devices underlying the SDN controllers is simplified as they do not need to understand protocols and policies but merely accept instructions from the controllers. A logical view of the SDN architecture, therefore, broadly consists of three layers or planes -\textit{management}, \textit{control} and \textit{infrastructure}. \textit{Interfaces} are used for communication among the layers.

\subsection{Layers in SDN}
The infrastructure layer consists of network devices such as switches, routers or middle-boxes. Each of this device has one or more \textit{flow} tables which consist of rule, action, and statistics. The rules in the flow table are populated by the controller either in a $proactive$ or $reactive$ manner.  The forwarding devices are responsible to receive packets coming from a host, match the incoming flow to a rule in its flow table and take action accordingly. In case of a miss, the rule is sent to the controller for making a decision.

The SDN controller or the control plane is considered as brain of the SDN and it is realized in software.  The controller provides the programmatic interface which facilitates network administrators to incorporate new features without vendor dependence. The centralized network state allows automation in configuring, management and optimization of network resources. Moreover, the SDN architecture supports a set of APIs to implement common network services such as security, access control, traffic engineering including routing. The centralized SDN architecture also makes it possible to provision resources, on-demand allocation of resources thereby truly virtualizing the network. 
 
SDN applications use the controller's API to configure flows to route packets through the best possible path, load balance traffic, handle link failures and redirect traffic for auditing, authentication, inspection and other such security-related tasks. The SDN applications are event-driven and their response may range from a simple default action to complex dynamic reaction.  

\subsection{Interfaces in SDN} Communication between a controller and the other two layers of SDN or between SDN controllers  in case of a distributed control plane take place through APIs provided by the controller. Communication between forwarding devices and controller take place through \textit{southbound API},  a well-known example of which  is OpenFlow\cite{mckeown2008openflow}.  This interface provides an abstract view of the infrastructure layer to the programmer. OpenFlow is not the only southbound API but there are others, such as \textit{ForCES}\cite{doria2010forwarding} and \textit{OpFlex}\cite{smith2014opflex}. The extensible messaging and presence protocol (XMPP)\cite{saint2009xmpp}, a communication protocol that allows messaging exchange among clients through a centralized server, has found a new application as a southbound API\cite{bakhshi2017state}.  The network devices run XMPP clients and announce their presence by setting up connections with a central server. The clients on receiving instructions from server update their configuration. 

The \textit{northbound API} is a low-level interface that provides SDN applications access to the network devices. This interface provides application developers an abstract view of the network topology. As a result, developers are able to deal with the whole network topology instead of a particular node.  The abstraction facilitates network virtualization and decoupling the network services from underlying network devices. Although there is no standard northbound API,  $REST API$ \cite{zhou2014rest} and OSGi\cite{cummins2013enterprise} are two popular northbound API. Most major SDN controllers include REST and OSGi support. 

A SDN domain \cite{yin2012sdni} is that portion of a network consisting of a SDN controller and SDN capable switches under its control. The controller controls the entire intradomain information including routes and bandwidth under its domain.  In large SDN environment, multiple controllers are deployed to reduce latency, minimize controller workload or minimize multiple network performance metrics in relation to controller placement. Thus a SDN controller can cover multiple NOS. Such distributed control plane architecture, as well as  independent SDN domains, require periodic information exchange between the controllers to allow synchronization, exchange of notification and messages. In the SDN architecture, the \textit{ East-Westbound} API has been proposed as an inter-SDN domain protocol.  To provide interoperability between different controllers, it is necessary to have standard  East-Westbound API. 

SDNi\cite{yin2012sdni} is a message exchange protocol to exchange information between domain SDN controllers. SDNi primarily allows two types of message to be exchanged - propagation of flow setup requirements and exchange of reachability information. SDNi can be implemented as an extension of Border Gateway Protocol (BGP) and Session Initiation Protocol(SIP) over stream control transmission protocols(SCTP). P. Lin et al.\cite{lin2015west} proposed an inter-domain mechanism West-East Bridge (WE-Bridge) for inter-domain SDN. WE-Bridge allows multiple SDN domains to coexist by sharing the network static aspects such as reachability, topology, network services as well as the dynamic aspects such as flow table entries in the switches, real-time bandwidth utilization and flow paths in the network.  F. Benamrane\cite{benamrane2017east} proposed an  East-Westbound API called communication interface for distributed control plane (CIDC). The proposed interface provides the exchange of notification service or exchange message between controllers and customizes the behavior of controllers. It also allows sharing of policies to support distributed services such as firewall, load balancing, and security.   

  \subsection{Packet forwarding in SDN}
  The \textit{flow table} is at the core of a SDN switch. Each entry of the flow table consists of three entries - header fields, counters, and actions. When a packet arrives at a forwarding device, the device extracts the header from the packet and matches it to its flow table's header fields. In case of a match, it will take action corresponding to that entry; it may forward or drop the packet. The counter field is used to track statistics relative to the flow. The header fields that are matched consists of twelve tuple - switch input port, VLAN ID, VLAN priority, source MAC (medium access control) address, destination MAC address, frame type, IP source address, IP destination address, IP protocol, IP ToS (type of service) bits, TCP/UDP source port and TCP/UDP destination port. The matching process stops when the first matching flow entry is found. The number of fields that are matched will depend upon the conformance level. In case of \textit{full conformance} all the twelve fields are matched, in \textit{layer two conformance} only the layer two fields are matched and in \textit{layer three conformance} the layer three header fields are supported. 
  
\begin{figure}
    \centering
    \begin{subfigure}[b]{0.4\textwidth}
        \centering
  \includegraphics[width=\textwidth]{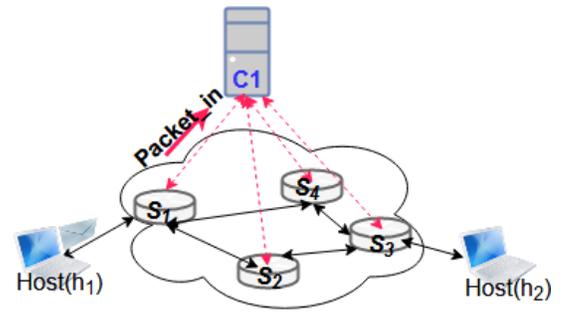}
   \caption{Switch Forwarding Packet to Controller}
       \label{fig:req}
\end{subfigure}
    \begin{subfigure}[b]{0.4\textwidth}
        \centering
         \includegraphics[width=\textwidth]{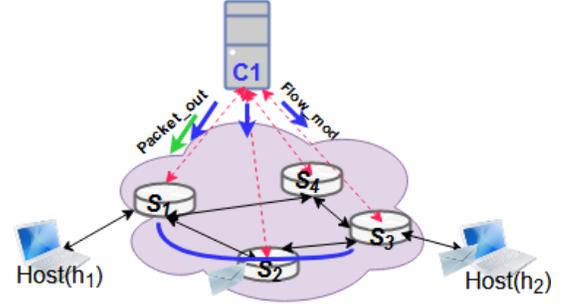}
            \caption{Command From Controller}
    \label{sdn:resp} 
\end{subfigure}
  \caption{Communication between Controller and Switch}
    \end{figure}

In case an arriving packet does not match any flow entry, the forwarding device considers it as a new packet, encapsulates the packet header in a $ PACKET\_IN$ message and forwards it to the controller for exception handling. However, for certain matching flow entries, the switch may always forward the packet to the controller such as a routing protocol control packet.   On receiving the $ PACKET\_IN$ message,  the controller runs an instance of the routing protocol to find a route and this would potentially require changing forwarding tables. The controller modifies the forwarding tables of the affected switches by sending the $FLOW\_MOD$ command and sends a $ PACKET\_OUT$ message to the source device. These forwarding devices are also responsible to forward the new packet after receiving the rule from the controller.
 Figure \ref{fig:req} reflects the working principle of SDN where the switch and controller both are assumed to be based on the $OpenFlow$ protocol. The $host(h_1)$ sends a packet to $switch(S_1)$ and results in a flow table miss. The exception results in a control packet $PACKET\_IN$ being forwarded to the $controller(C_1)$ containing the packet header. The controller computes the path using its interior gateway protocol. Assume the  path $S_1$--$S_2$--$S_3$ is computed. Figure \ref{sdn:resp} shows the response from the controller. The controller updates the routing table of the switches $S_1$, $S_2$, and $S_3$  by sending  $FLOW\_MOD$ control packets and simultaneously sends a $PACKET\_OUT$ message to the source $switch(S_1)$ defining how the packet should be processed by the switch.

 \subsection{SDN Signaling}
\begin{figure}[thb]
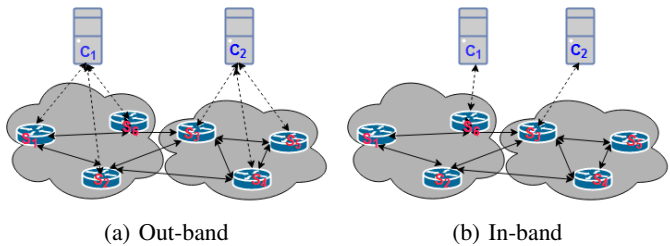

  \centering
\begin{subfigure}{.25\textwidth}
  \includegraphics[width=.99\linewidth]{Out_band_1_.png}
  \caption{Out-band}
  \label{fig:sfig1}
\end{subfigure}%
  \centering
\begin{subfigure}{.25\textwidth}
  \includegraphics[width=.99\linewidth]{In_band.png}
  \caption{In-band}
  \label{fig:sfig2}
\end{subfigure}
\caption{Types of Control Signaling}
\label{fig:control_type}
\end{figure}

The communication between the control plane and the forwarding devices happen using secure and reliable channel called control channel.  Figure \ref{fig:control_type} depicts the different types of control channel - out-band control and in-band control. The choice between these two channels is a trade-off between capital expenditure (CAPEX) and performance. In out-band signaling, there are separate channels for data and control traffic. Out-band signaling is generally used in a data center kind of environment, where nodes are in close proximity and performance is a primary objective. For large networks like carrier and ISP networks, which span over a country or a continent, the cost for out-band signaling will be prohibitive. In in-band signaling, the control and data traffic share the same network infrastructure. In-band signaling has lower CAPEX and is generally preferred. However, timely and reliable transmissions of control traffic are a challenge in in-band signaling. S.-C. Lin et al. \cite{lin2016control} proposed to solve this performance issue by load balancing the control traffic. The other issue with in-band signaling is that failure of the data plane will also affect the control plane.  

\section{SDN Architecture Issues}\label{distributed}
The flow-based routing in SDN has enabled networks to support flexible policies. However, the architecture of SDN which relies on directing the first packet of each flow to the controller places considerable load on the controller. Installing all the flow rules proactively is not an attractive solution since the rules can change over time and a huge initialization cost will have to be incurred. To address the limitations of a single controller, researchers have suggested a two-pronged strategy. The first approach is to \textit{offload} some of the tasks of the control plane to the data plane. The second approach is to revamp the control plane by adding more controllers.  

\subsection{Limitations of a single controller}
A single controller often suffices the latency requirement of a small to moderate size network. Some researchers have advocated the presence of one physical controller and they have concentrated on improving its performance. For instance, NOX\cite{gude2008nox} is a centralized, single-threaded OpenFlow controller system. Although NOX has simple programming model, it cannot exploit the advantages offered by current multi-core processors. To enhance the capacity of the single controller, some proposals such as Beacon \cite{erickson2013beacon}, McNettle \cite{McNettle} and Maestro \cite{ng2010maestro} have been proposed which utilizes the optimization techniques of parallelism provided by multi-core processors. 

  The choice of a controller can be based on different requirements of the user such as the programming language, learning curve, areas of focus, etc.  For example, controllers built using the programming language like Python degrades the controller performance but has an easy learning curve. Some controllers are focused to a specific domain, such as RouteFlow \cite{RouteFlow}, which is optimized for inter-domain routing. POX\cite{POX} is widely used in research and education because it is easy to understand and can be rapidly prototyped. However, POX performance in terms of throughput and response time is not as good as the others. OpenDaylight controller supports high modularity and is based on Java language which supports multi-threading and also enables to run cross-platform. Thus it is preferred as a production level controller, though it is difficult to learn. A comparison of the mainstream production level controllers and those used in academic research is given in Table \ref{tab:cont_list}.


\begin{table}[!htb]
\centering
\caption{Comparison of SDN controllers}
\label{tab:cont_list}
\begin{tabular}{@{}llll@{}}
\toprule
\textbf{Controller} & \textbf{Instance} & \textbf{Language} & \textbf{Organization} \\ \midrule
NOX \cite{gude2008nox} & Single & C++ & Nicira Networks \\ \midrule
POX\cite{POX} & Single & Python & Nicira Networks \\ \midrule
Ryu\cite{Ryu} & Single & Python & NTT \\ \midrule
Beacon \cite{Beacon} & Single  & Java & Standford University \\ \midrule
Floodlight\cite{Floodlight} & Single  & Java & Big Switch Networks \\ \midrule
McNettle \cite{McNettle} & Single & Haskell & Yale University \\ \midrule
ONOS \cite{berde2014onos} & Multiple  & Java & ON.Lab \\ \midrule
OpenDaylight\cite{OpenDaylight} & Multiple & Java & Linux Foundation \\ \midrule
McNettle\cite{McNettle} & Single  & Haskell & Yale University \\ \midrule
Terma\cite{Trema} &  & Ruby, c & NEC \\ \midrule
MUL\cite{MUL}& & C & Kulcloud \\ \midrule
RouteFlow\cite{RouteFlow}& &  &  \\ \midrule
OpenContrail\cite{opencontrail}  & Multiple & Python, c, Java & Juniper project \\ \bottomrule
\end{tabular}
\end{table}
However, as the size of the network grows, a single controller will not scale well due to the following reasons. 
\begin{itemize}
\item \textbf{Control traffic overhead:} In order to setup up a flow on a $N$-switch path, the one-way cost in terms of network load is about 94+144$N$ bytes\cite{curtis2011devoflow}. The control traffic overhead destined towards a central controller will grow as the number of switches increase. 
\item \textbf{Flow setup latency:} The flow setup time for switches farther from the controller will be higher. A network with a large diameter or a network with several node clusters will encounter long flow setup latencies irrespective of where the controller is placed. It has been shown that increasing the number of controllers from 1 to 3 reduces the average latency by half \cite{heller2012controller}.
\item \textbf{Intrinsic controller capacity:} A controller can only support a limited number of flow setups per second. For example, one NOX controller can handle 30$K$ new flows per second \cite{tavakoli2009applying}. As the flow setup times are bound by the processing power of the controller, the flow setup times will increase with the increase in network size.
\item \textbf{Delay in flow statistics gathering}: SDN schedulers \cite{al2010hedera}\cite{pereini2014espres} \cite{qin2015bandwidth}
can be used to improve static load balancing, reordering of the rule to fully utilize the switches and improve distributed processing framework such as Hadoop\cite{shvachko2010hadoop}. The schedulers need timely access to flow statistics in order to perform optimally. A centralized controller will frequently need to query the switches which will considerably increase the flow of control traffic in the network.    
\item \textbf{Availability:} The availability of a network can be affected either due to reliability or security \cite{blial2016overview}
The single point of failure problem is an issue in SDN with a centralized controller, although this can be overcome by deploying hot standby slave controllers. A single controller is also more vulnerable to security attacks. If the single controller gets compromised or comes under a denial of service attack, there will be  network outage. 
\end{itemize}
\begin{figure*}[thb]
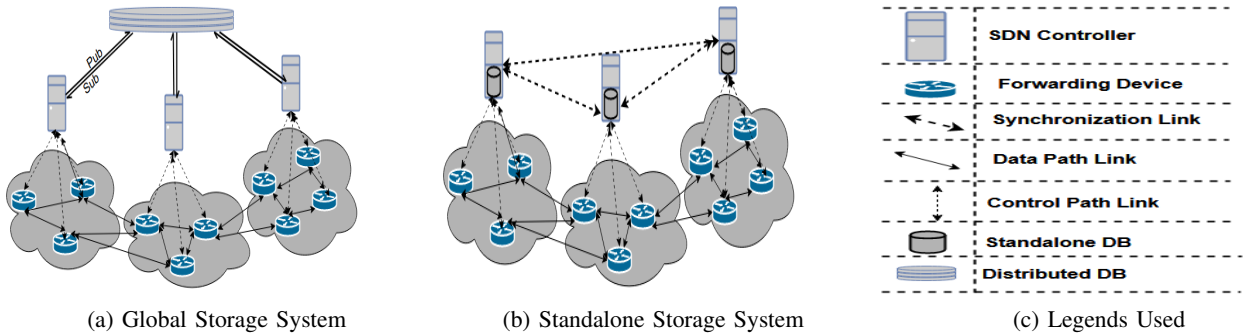

  \centering
\begin{subfigure}{.32\textwidth}
  \includegraphics[width=1.9in, height=1.5in]{distributedcontrolplane_1__2_.png}
  \caption{Global Storage System}
  \label{fig:sfig1}
\end{subfigure}%
  \centering
\begin{subfigure}{.32\textwidth}
 \includegraphics[width=1.9in, height=1.5in]{distributedcontrolplane2_2_.png}
  \caption{Standalone Storage System}
  \label{fig:sfig2}
\end{subfigure}
\begin{subfigure}{.32\textwidth}
 \includegraphics[width=1.9in, height=1.5in]{Symb.png}
  \caption{Legends Used}
  \label{fig:sfig3}
\end{subfigure}
\caption{Flat Distributed Control Plane}
\label{fig:fig2}
\end{figure*}

\subsection{ Modification in Data Plane}
SDN imposes high control traffic overhead, particularly for events that require frequent per packet processing (eg. traffic classification). Researchers have thus proposed an alternative perspective for handling events which require frequent control plane intervention. Both \textit{DIFANE}\cite{yu2010scalable} and \textit{DevoFlow}\cite{curtis2011devoflow} reduce the reliance of SDN on the control plane by attempting to keep flows in the data plane. In DIFANE, the entire traffic is kept in the data plane by partitioning the rules for forwarding and storing it in the appropriate switches. The task of the controller is reduced to partitioning the rules over the switches. In the DevoFlow  architecture, the authors have modified data plane, so that small flows (called mice) are handled by the data plane itself. 

These architectures, however, breach the very principle of SDN which is centralizing the intelligence and is not in the scope of our survey. 


\subsection{Distributed control plane}
To address the issues of a single controller, researchers have proposed a distributed control plane, where controllers  are physically distributed but logically centralized. Most of the current controllers listed in Table \ref{tab:cont_list}  support the logically centralized and physically distributed control plane.  An important and related issue often ignored by SDN researchers while handling CPP is  synchronization of the multiple controllers. Even though the controllers are physically distributed they should be logically centralized, keeping in view the central SDN principle that a controller should have a global view of the network. To achieve the consistent network-wide view, whenever the local view of one controller changes, it synchronizes the updated information with the other controllers. One of the main sources of overhead in multi-controller deployment is this synchronization traffic among the controllers. It has been reported that in a network with 20 nodes and two OpenDaylight controllers, the aggregated synchronization traffic is about 1 Mbps\cite{ros2016reliable}. Although the East-Westbound API has been proposed with the sole objective of communication among distributed SDN controllers, no such universally accepted interface exists.

Researchers have explored different design choices of the distributed control plane. Techniques to synchronize the distributed control plane in SDN differ primarily in the topology of the controllers, the way network state  information is stored and exchange of routing information.  

\subsubsection{Topology}
Distributed control plane (DCP) are categorized based on the arrangement of the controllers - flat, hierarchical and hybrid hierarchical. In the flat architecture (Figure \ref{fig:fig2}), all controllers are  at one level and communication takes place among the controllers to get a global view. In the hierarchical architecture, controllers are arranged according to a tree-like structure, thus all the controllers are not at the same level. The controllers at the lowest level manage the network devices directly connected to it and contain only their local view. The upper-level controllers manage the view of all the lower level controllers directly connected to it. The root controller will have the global view and will be responsible for coordination among the local controllers. The hybrid architecture is similar to hierarchical, except that there is more than one controller at the highest level which can communicate with one another (like flat architecture).  

\subsubsection{Data Storage}
Controllers need a datastore to store the status of network devices under its domain. The controllers can use either a relational or non-relational database and it may be global or a standalone database system. In the global datastore framework shown in Figure \ref{fig:sfig1}, a global network store is built  by aggregating the network state of the individual controllers. The commonly used global datastore are WheelFS\cite{stribling2009flexible}, MongoDB\cite{hows2014mongodb}, Cassandra\cite{wang2012nosql}, and  Hazelcast\cite{Hazelcast}.  In the case of a standalone datastore as shown in Figure \ref{fig:sfig2} the routing information of each controller domain is stored locally in the controllers. 

\subsubsection{Synchronization} 
The controllers need to frequently exchange information about the devices it manages, so as to build a  global view of the physical layer. In the global datastore framework, the publish/subscribe messaging paradigm is  widely used to accomplish this task. Publish/subscribe model provides an asynchronous means of communication in distributed systems. In this paradigm, the queue is an intervening entity that represents a topic or channel which enables the  publisher(sender) to interact with the subscriber(receiver). In this model, the sender publishes the message to the queue while the receiver subscribes the message from the queue.  

The other approach commonly used is custom messaging protocol. In the standalone datastore framework, the controllers exchange information among themselves using East/West API.  

\subsection{Overview of distributed control plane} \label{Over_DCP}
In this section, we review the mainstream distributed control plane architecture available in the literature. Onix \cite{koponen2010onix} provides a framework for a distributed control plane to scale networks. The scalability is achieved using three strategies - \textit{partitioning}, \textit{aggregation} and \textit{consistency and durability}. Onix allows the network to be partitioned, where each partition is controlled by an instance of an Onix controller. Details of the network entities are stored in a data structure called the network information base (NIB). Aggregation allows a cluster of Onix nodes to appear as a single node in a different cluster's NIB. The consistency and durability requirements of the network state can be tuned as per an application's requirement. For applications with the requirement of stronger consistency and durability, a replicated transactional database is used. For applications that are more tolerant to consistency and durability, a one-hop DHT is used.

HyperFlow \cite{tootoonchian2010hyperflow} uses a publish/subscribe messaging paradigm  to propagate the controller events using the distributed file system \textit{WheelFS} \cite{stribling2009flexible}. Each controller subscribes to three channels - data channel, control channel and its own channel. Application events of general interest are published on the data channel. Events and commands targeted to a specific controller are published in the controller channel. A controller also periodically advertises itself in its own channel to facilitate controller discovery and failure detection. HyperFlow has been implemented as an application for NOX. 

Phemius et. al. \cite{phemius2014disco} proposed DISCO, a distributed SDN control plane where the DISCO controller also communicates using the publish/subscribe messaging paradigm. The controller consists of two parts - intradomain part and an interdomain part. The intradomain part handles the network related service like link discovery and path computation while the interdomain part is used to exchange aggregated network-wide information with other controllers. DISCO was implemented on top of Floodlight, an open source OpenFlow controller.

In Kandoo \cite{hassas2012kandoo} all communications among the controllers are event based. It creates a two-level hierarchy of the controllers - multiple local controllers and  a root controller. The local controllers at the lower level are as close as possible to the switches in order to process local applications. Local controllers can be deployed on the same host where a software switch is housed or by changing the software of a physical switch. In case both the options are not possible, then Kandoo local controllers are placed as close as possible to the switches. The root controller subscribes to events in the local controller and after the local controller locally processes the event; it relays the event to the root controller for further processing. The root controller can be based on HyperFlow or Onix.

OpenDaylight \cite{khattak2014performance} is an open source controller platform designed for scalability and high availability. The network state information is stored in network state database (NSDB). To provide high availability the NSDB is partitioned, each partition managed by a controller. For high availability, the replica of the partitions is distributed across other SDN controllers. 

Open Network Operating System (ONOS) \cite{berde2014onos} is another open source project to provide a global network view of the application even if the controllers are physically distributed as well as provide fault tolerance. The network was implemented using the single-instance SDN controller Floodlight. The network view discovered by each ONOS instance was implemented using the graph database Blueprint. To construct the global network, RAMCloud \cite{RamCloud}  a distributed key-value store was implemented. Event notification between ONOS instances was implemented using the broadcast messaging system of Hazelcast \cite{Hazelcast} which is based on the publish-subscribe system. 

The logical xBar \cite{mccauley2013extending} concept envisions that xBars are the building blocks which can be recursively merged to form large xBars. The proposed control plane design consists of two building blocks - (i) the logical xBar, a programmable entity which is responsible for switching packets and (ii) the logical server responsible for forwarding table management and control plane computations. Although the network topology is not assumed to be explicitly hierarchical, the recursive aggregation of xBars imposes a natural hierarchy on the network.

While logical xBar envisions the basic building blocks to be recursive, IRIS \cite{lee2014iris} envisions a recursive network abstraction. The network is split into multiple \textit{unit SDN}, each managed by a separate controller. The interaction between these unit SDNs is managed by super-controllers which are hierarchically placed. 

The logically centralized SDN controllers that we have seen so far assume that the network is under a single or centralize administrative domain. That is the same organization administers the network. Decentralize-SDN or D-SDN proposes a SDN framework allowing  distribution of the control plane among administratively decentralized and heterogeneous networks. The control distribution is achieved by defining a hierarchy of main controllers and secondary controllers. Each controller has its own database and the network state among the main controllers is exchanged using  \textit{update}  messages.

Y. Fu et. al. \cite{fu2014orion} show that when the network size increases by $X$, the complexity of the control plane increases by $X^2$. In such networks, they argue that the flat distributed control plane architecture, like Onix and HyperFlow, cannot solve the super-linear complexity of the control plane. Although the hierarchal control plane such as KANDOO and Logical xBar solve the problem, they argue that the path between nodes get \textit{stretched}. To solve these problems, they propose Orion, a hybrid hierarchal  control plane. In this architecture,  a large-scale network is managed by multiple instances of \textit{area} controllers which in turn are managed by \textit{domain} controllers. The data store used to store node information is NoSQL database and the distribution of the routing rules is realized using the publish/subscribe mechanism. 

The distributed SDN controllers that we have encountered so far primarily focused on how to scale SDN for large networks. However, they assume that the traffic is static and hence the mapping between a switch and controller is static. ElastiCon \cite{dixit2014elasticon} focus on an \textit{elastic} distributed controller architecture. It allows the controller pool to expand or shrink dynamically as well as the mapping between a switch and controller to be dynamic depending on the traffic characteristics. 


A comparison of the mainstream distributed control plane architecture is given in Table \ref{tab:distrib_ctrls}. 

\begin{table*}[!htb]
\centering
\small
\caption{Comparison of Distributed Control Plane Architecture}
\label{tab:distrib_ctrls}
\begin{tabular}{@{}lllll@{}}
\toprule
\textit{\textbf{\begin{tabular}[c]{@{}l@{}} Distributed \\Architecture \end{tabular}}} & \textit{\textbf{\begin{tabular}[c]{@{}l@{}}Controller \\ Implemented \end{tabular}}} & \textit{\textbf{\begin{tabular}[c]{@{}l@{}}Data Store \\\end{tabular}}} & \textit{\textbf{\begin{tabular}[c]{@{}l@{}}Synchronization \\ Technique\end{tabular}}} & \textit{\textbf{Controller Topology}} \\ \midrule
Onix\cite{koponen2010onix}& OpenFlow based & \begin{tabular}[c]{@{}l@{}}Replicated \\ Database, \\ DHT \end{tabular} & \begin{tabular}[c]{@{}l@{}}Custom API\\ Pub/Sub (Zookeeper) \end{tabular} & Flat  \\ \hline
HyperFlow\cite{tootoonchian2010hyperflow} & NOX & WheelFS & Pub/Sub & Flat  \\ \hline
DISCO\cite{phemius2014disco} & Floodlight & \begin{tabular}[c]{@{}l@{}}Extended Database\end{tabular} & \begin{tabular}[c]{@{}l@{}}Pub/Sub\\ (AMQP)\end{tabular} & Flat \\ \hline
ONOS\cite{berde2014onos} & Floodlight & \begin{tabular}[c]{@{}l@{}} Blueprint \\ RamCloud \end{tabular} & Pub/Sub & Flat  \\ \hline
Opendaylight\cite{khattak2014performance} & Opendaylight & \begin{tabular}[c]{@{}l@{}}Network State\\ Database (NSDB)\end{tabular} & Raft [Replication] & Flat \\ \bottomrule \hline
KANDOO\cite{hassas2012kandoo} & \begin{tabular}[c]{@{}l@{}}Any OpenFlow \\ Controller\end{tabular} & \begin{tabular}[c]{@{}l@{}} Centralized \\ Repository \end{tabular} & Pub/Sub & Hierarchical \\ \hline
Logical xBar\cite{mccauley2013extending} &  -- & Logical Server & -- & Hierarchical \\ \hline
IRIS\cite{lee2014iris} & Floodlight/Beacon & MongoDB & Custom &  Hierarchical \\ \hline
D-SDN\cite{santos2014decentralizing} & Floodlight & \begin{tabular}[c]{@{}l@{}}Individual \\ Database \end{tabular}& Custom &  Hierarchical \\ \hline
ORION\cite{fu2014orion} & Floodlight & NoSQL database & Pub/Sub & Hybrid Hierarchical \\ \hline
LazyCtrl\cite{wang2015lazy} & Floodlight &Bloom Filter& &Hybrid \\ \hline
ElastiCon\cite{dixit2014elasticon} & Floodlight & Hazelcast & Pub/Sub & Flat  \\ \bottomrule \hline
\end{tabular}
\end{table*}


\section{Controller Placement Problem (CPP)} \label{CPP}
As the task of generating forwarding rules and populating it onto the switches is delegated to the controllers, the layout of the controllers will greatly influence the performance of the network. Multiple controllers are required not only to improve network performance but also for high availability of the network. There are two aspects to the controller placement problem, the number of controllers to be deployed and their placement. The number of controllers required is a trade-off between performance and deployment cost. 

Heller et al. \cite{heller2012controller} were among the first few researchers to address the controller placement problem. To formulate the problem mathematically, let us consider a network with $n$ number of switches represented by the set $S$=\{$s_1$, $s_2$, \ldots, $s_n$\} and $k$ number of controllers represented as $C$=\{$c_1$, $c_2$, \ldots, $c_k$\}, where $k\le n$. The network is modeled as an undirected graph $G=(V,E)$, where $V$ refers to the set of switches and $E$ refers to the links between the switches. The CPP problem is to place the controllers such that the network gives optimal performance with respect to one or more performance metrics. The input to the controller placement problem is thus the tuple ($C$, $S$, $G$). The output of the controller placement is the tuple ($<p_1,A_1>$, $<p_2,A_2>$, $\ldots$, $<p_k,A_k>$) where $p_i$ is the potential location of  $c_i$, the $i^{th}$ controller. $A_i$ is the set of switches controlled by the controller $c_i$ and they form the control domain of the controller. Let 
\begin{equation}
P_i=\{p_1, p_2, \ldots, p_k\} 
\label{eq:placement}
\end{equation} 
denote one possible placement of the $k$ controllers.  Given $k$  is the number of controllers, the total  possible number of placements is ${n}\choose{k}$. The set of all placements ($P$) is given as $P=\{P_i \in 2^V| |P_i|=k\}$.

 In general, the controllers are usually assumed to be co-located  with the switches\cite{killi2017capacitated}. Huque et al. \cite{hu2017energy} referred to such co-location of controllers with the switch as \textit{restricted search} and advocated the advantages of \textit{open search}, which allows the controller to be placed in any location within the geographical area. However, they also mention that open search may not be always feasible due to geographical restrictions. The symbols used are summarized in Table \ref{tab:symbols}.

\begin{table}[]
\centering
\scalebox{0.8}{
{\renewcommand{\arraystretch}{1.5}%
\begin{tabular}{@{}l|l@{}}
\rowcolor[HTML]{C0C0C0} 
\textbf{ Symbol} & \textbf {Description} \\
 $G(V,E)$&\begin{tabular}[c]{@{}l@{}}Graph $G$, where $V$ is a set of switches \\ \& $E$ is a set of links between the switches\end{tabular}  \\
\rowcolor[HTML]{C0C0C0} 
$S= \{ s_1, s_2, \ldots, s_n \}$ & Set of switches  \\
$C= \{ c_1, c_2, \ldots, c_k \}$ & Set of controllers, where  $k \le n$\\
\rowcolor[HTML]{C0C0C0} 
$P=\{ P_1, P_2, \ldots, P_{{n}\choose{k}} \}$ & Set of placements \\
$P_i= \{ p_1, p_2, \ldots, p_k \}$ & One possible placement of the $k$ controllers  \\
\rowcolor[HTML]{C0C0C0} 
 $p_i$ & Potential location of  $c_i$\\
$A_i$ & Set of switches controlled by the controller $c_i$\\
\rowcolor[HTML]{C0C0C0} 
 $d(u,v)$ & Delay between the nodes $u$ and $v$, where $u,v \in V$\\
$a_c$ & Binary Variable (Active controller) \\
\rowcolor[HTML]{C0C0C0} 
$a_e$ & Binary Variable (Active link)\\
$x_f$ &  Probability of failure of a component $x$, where $x \in \{ V \cup E \}$\\
\rowcolor[HTML]{C0C0C0} 
 ${E'}_c(x_f)$ & \begin{tabular}[c]{@{}l@{}} Total number of control paths directly or indirectly \\ affected path due to failure probability $x_f$\end{tabular}\\
$\mathbb{F}$ & Set of all failure scenarios\\
$f \in \mathbb{F}$ & A failure scenario\\
\rowcolor[HTML]{C0C0C0} 
$F$& Set of all controller failure scenarios\\
$f_{i} \in F$ & A Controller failure scenario\\
 \rowcolor[HTML]{C0C0C0} 
$t_{route}(c_j)$& Processing time to run the routing protocol by controller $c_j$\\
 $A_{f_i}$ & Set of switches assigned to controller $c_i$\\
  \rowcolor[HTML]{C0C0C0}
 $Cost(c)$ &  Cost of   controller installation \\
$Cost(E^{P_i}_c)$ & \begin{tabular}[c]{@{}l@{}} Connection establishment cost between controller to controller\\ and controller to switch for a placement $P_i$\end{tabular}\\
 \rowcolor[HTML]{C0C0C0}
 $E^{P_{i}}_{c}$ & Set of control paths for a given controller placement $P_i$ 
\end{tabular}} \quad
 }
\caption{Symbols and Descriptions}
\label{tab:symbols}
\end{table}

\begin{figure*}[!htb]
\centering
 \includegraphics[scale=0.65]{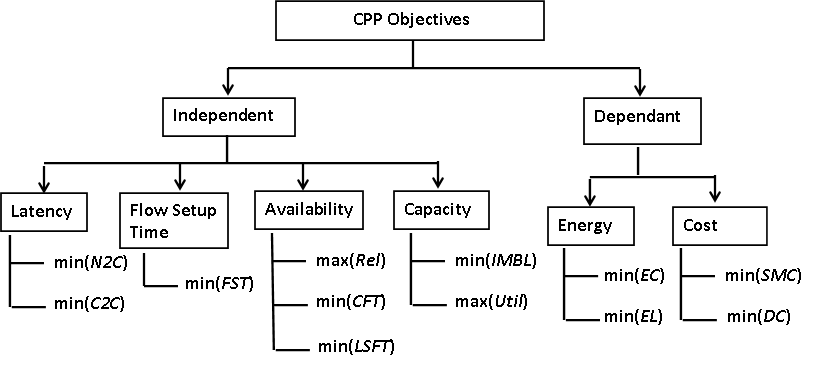}
     \caption{Classification of CPP Metrics}
     \label{fig:fig1}
\end{figure*}
\subsection{Objectives used in CPP}\label{sec:PM}
Researchers working on CPP use a number of metrics to measure performance. The number of controllers and their placement depends on the objective(s) one wants to optimize. We classify the performance metrics into two subclasses, one is independent, where the objectives can be considered in isolation and the other is dependent, where the metrics need to be jointly considered with the independent metrics. The performance metrics that come under independent objectives are  latency, flow setup time, availability and capacity. The latency metric can be further subclassified depending on the type of nodes between which the latency is measured. The availability metric can also be further sub classified depending on the availability of the network in the event of a failure or how the network recovers from failures. Load balancing is the most common capacity aware metric. The dependent objectives are energy aware and cost aware. The two main objectives that come under energy aware are the elastic controller and elastic link. The cost aware metrics that are mainly used is the deployment cost metric and switch migration cost metric.  A list of the metrics used by different CPP approaches is given in  Table \ref{tab:sCPP}.  The categorization of the metrics used in CPP is shown Figure \ref{fig:fig1}.

\begin{table}[]
\centering
\small
{\renewcommand{\arraystretch}{1.45}%
\begin{tabular}{@{}c|c@{}}
\rowcolor[HTML]{C0C0C0} 
\textbf{Metrics } & \textbf {Notations} \\
 Switch to Controller latency & $N2C$  \\
\rowcolor[HTML]{C0C0C0} 
 Controller to Controller latency&$C2C$ \\
 Flow Setup Time& $FST$\\
\rowcolor[HTML]{C0C0C0} 
 Controller Failure Tolerance& $CFT$ \\
Reliability& $Rel$    \\
\rowcolor[HTML]{C0C0C0} 
Elastic Controller & $EC$ \\
Elastic Link& $EL$\\
\rowcolor[HTML]{C0C0C0} 
Deployment Cost&$DC$ \\
 Switch Migration Cost& $SMC$\\
\rowcolor[HTML]{C0C0C0}
Link/Switch Failure Tolerance & $LSFT$\\
Imbalance  & $IMBL$\\
\rowcolor[HTML]{C0C0C0}
Utilization  & $Util$
\end{tabular}
}
\caption{Metrics and Notations}
\label{tab:sCPP}
\end{table}

\subsubsection{CPP with Independent Objective}
In CPP using independent objective(s), one or more objectives may be used. In CPP with multiple independent objectives, each of the objectives is formulated independently and they are solved as a multi-objective optimization problem. These objectives may constructively influence one another or may be conflicting in nature. 
\begin{itemize}
\item \textit {\textbf{Latency:}}
One of the most frequently used performance metrics is latency or delay. The overall latency consists of transmission, propagation, queuing and processing delay. Latency is measured between two nodes and depending on the type of nodes considered, latency can be of two types - switch to controller latency (also called node to controller latency) and controller to controller latency. 
\begin{itemize}
\item Node to Controller latency ($N2C$): This latency measures the delay between a node and its assigned controller\cite{heller2012controller},\cite{wang2018effective},\cite{zhao2017towards}. In order to minimize the $N2C$, each node will compute the minimum delay with all the controllers. The $N2C$ metric that will be minimized can be computed by computing the average across all these latencies. Let $d(u,v)$ be the delay between the nodes $u$ and $v$, where $u,v \in V$. The node to controller latency ($N2C$) for a particular placement $P_i$ is given as
\begin{equation}
N2C(P_i)=\frac{1}{|S|}\sum_{s \in S} \min_{p_j \in P_i} d(s,p_j)
\label{eqn:n2c}
\end{equation}
The objective is to find from amongst all the placements the one which minimizes the value of $N2C$, which is given as 
\begin{equation*}
\arg\min_{P_i \in P}[ N2C(P_i)]
\end{equation*}
  
The other way of computing $N2C$ is by minimizing the maximum value of all the latencies.
\item Controller to Controller latency ($C2C$): This latency measures the delay between two controllers\cite{hock2013pareto},\cite{lange2015heuristic},\cite{ksentini2016using1}. Analogous to $N2C$, the $C2C$ for a particular placement $P_k$ can be computed as: 
\begin{equation*}
C2C(P_k)=\frac{1}{|C|}\sum_{p_i,p_j \in P_k}d(p_i,p_j)
\end{equation*}
\end{itemize}

The objective to minimize the overall controller to controller latency can be defined as
\begin{equation*}
\arg\min_{P_k \in P}[ C2C(P_k)]
\end{equation*}

\item  \textit {\textbf{Flow Setup Time (FST):}} In reactive SDN controllers, the first packet of a flow arriving at a switch requires the controller to install flow rules in all the involved switches of its control domain.  The end-to-end flow setup time \cite{hemodeling} is defined as the duration for the first packet of a flow to travel from the source to its destination. Thus, end-to-end flow setup time includes the time required to setup forwarding rules in all the involved switches as well as the time taken to forward the packet along the data plane.  

In a control domain, the flow setup time\cite{zeng2015flow} will consist of the delay for the \textit{PACKET\_IN} message to travel from the first switch where the packet lands to controller; the processing time to run the routing protocol ($t_{route}$) at the controller; and the latency for  \textit{FLOW\_MOD} message to travel from the controller to  all the affected switches. The controller will also send a \textit{PACKET\_OUT} message to the switch. The \textit{FLOW\_MOD} and \textit{PACKET\_OUT} message will be sent out by the controller almost simultaneously, so we can consider a single delay for these two messages. The flow setup time of a switch $s$ for a particular controller placement $P_i$ can be written as 
\begin{equation*}
fst(s) =\min_{p_j \in P_i} d(s,p_j) + t_{route}(c_j) +\max_{s' \in S'}\{d(p_j,s),d(p_j,s')\}
\end{equation*}
where the set $S'$ denote the number of switches affected by the flow. The average flow setup time of the entire network for a given controller placement is given in the following equation.
\begin{equation*}
FST(P_i)= \frac{1}{|S|}\sum_{s \in S}  fst(s)
\end{equation*}
The objective is to find a placement which optimizes the overall flow setup time.
\begin{equation*}
\arg\min_{P_k \in P}[FST(P_k)]
\end{equation*}

\item \textit {\textbf{Availability:}} Availability is a measure of the duration a system will remain operational in the event of failures. The objective of this metric is to investigate the impact of the network component failure on the availability of SDN. The availability of a system may be either with reduced functionality or reduced performance. The first case gives a measure of the reliability of the system while the latter gives a measure of the fault tolerance of the system.  

\begin{itemize}
\item  \textit {\textbf{Reliability:}}
Reliability gives a quantitative analysis of the performance of a system. It is a measure of the availability of network in the face of the failure of physical network components. The probabilistic reliability metric assumes that those components may fail with a certain probability \cite{hu2014reliability}. The two commonly used metric to measure the reliability of a system is \textit{connectivity-based} and \textit{traffic-based}. F. Ros et al.\cite{ros2014five} aim to provide reliability of at least five nines. In SDN, failure of a network component will result in disruption of either the control traffic or the data traffic (or both). The traffic-based metric considers the traffic carrying capacity of the network. Failure of network components will result in degradation of the traffic carrying capacity.


Let ${x_{f}}$ be the probability of failure of a component ${x}$ and $\mathbb{F}$ is the set of all such failure scenarios $f$.  In this work, as we are dealing with CPP, we consider control path failures between switches and controllers as well as between controllers. Let $E^{P_i}_c$ represent the set of all control paths for a given controller placement $P_i$ and let ${\tilde{E}}^{P_i}_c(x_f)$ be the total number of control paths directly or indirectly affected path due to failure probability $x_f$.  The percentage of control path failure for a given controller placement scenario $P_i$ can be expressed as 
 \begin{equation*}
Rel(P_i)=[\sum_{f \in \mathbb{F}}(x_{f}\times \{\dfrac{\tilde{E}^{P_i}_c}{|E^{P_i}_c|}\})]
\end{equation*}

The objective is to find a controller placement combination which maximizes the reliability.
 \begin{equation*}
\arg\max_{P_i \in P} [Rel(P_i)]
 \end{equation*}

\item  \textit {\textbf{Fault Tolerance:}}
Fault tolerance measures the resilience of the system in recovering from any component failure such as controller, link, switch etc \cite{muller2014survivor}. In this survey, as our focus is on CPP, we consider the resilience of the system to controller failures.  In Equation \ref{eq:placement},  we defined the set $P_i$ as one possible placement of $k$ controllers. In a controller failure tolerance scenario, in the worst case up to $k-1$ of these controllers can fail. Thus the set ($F_i$) which considers one possible placement in a controller failure scenario  is
\begin{equation*}
F_i=\{p_i \in 2^k| p_i \in P_i\}
\label{eq:placement_k}
\end{equation*} 

The set ($F$) of all possible controller placements in a fault tolerance scenario is $F=\{F_i \in 2^V\}$. During a controller failure, switches may get reassigned resulting in increased latencies. The latency for one failure scenario $F_i$ is 
\begin{equation*}
CFT(F_i)= \frac{1}{|F_i|}\sum_{f_{p} \in F_i}N2C(f_{p})
\end{equation*}

The objective of  controller failure tolerance\cite{hock2013pareto},\cite{lange2015heuristic} is to minimize node to controller ($N2C$) latency during failure scenarios.
\begin{equation*}
\arg\min_{F_i\in F}[CFT(F_i)]
\end{equation*}
\textit {Link/Switch Failure Tolerance: ($LSFT$)} Failure of switches or links may partition the network. Due to the partitioning in the network, disconnection may occur between the switches and controllers or between controllers. Those partitions,  in which switches are not connected to a controller, the nodes are called controller-less. The objective of $LSFT$ is to  minimize the controller-less nodes. 

\end{itemize}
\item \textit {\textbf{Capacity Aware:}} The number of nodes a server can handle is limited  by its processing power, memory, access bandwidth and other resources. The problem of placement of controllers considering the load and capacity of controllers is called as capacitated CPP.  The objective of a capacity aware metric is to restrain the load on a controller to exceed its processing threshold.  
\begin{itemize}
\item  \textit {\textbf{Load Balance:}}
The load on a controller primarily consists of three components - (i) number of flow setup request processed (ii) managing the nodes in its domain and (iii) synchronization messages exchanged with other controllers to form a global view. As the capacity of SDN controller is limited in term of resources, so distributing the load equally on the controllers is another important measure to improve the performance of SDN \cite{lange2015heuristic}\cite{xiao2014sdn}. Let $A_{f_i}$ denote the set of switches assigned to controller $c_i$ when a maximum of $k-1$ out of $k$ controllers can fail. The imbalance metric $IMBL$ for a failure scenario is the difference between the maximum and minimum of nodes that can be assigned to a controller.    
 \begin{equation*}
   IMBL(F_i)=\max_{f_i \in F_i} |A_{f_i}| -  \min_{f_j \in F_i} |A_{f_j}|
  \end{equation*}
   The objective is to balance the load on the controllers or in other words to minimize the imbalance metric. 
  \begin{equation*}
\arg\min_{F_i\in F}[IMBL(F_i)]
\end{equation*}

\end{itemize}
\end{itemize}
\subsubsection{CPP with Dependent Objective}
In this section, we consider those objectives which cannot be considered in isolation but needs to be jointly considered with at least an independent metric. For example,  energy consumed is directly proportional to the number of active controllers. Although decreasing the number of active controllers will reduce energy consumption, it may have an adverse effect on the other performance metrics like latency and controller load. 

 \begin{itemize}
\item  \textit {\textbf{Energy Aware:}} 
Energy plays a vital role in the network. Based on real data center measurements,\cite{benson2010network}, it has been shown that the ratio of peak flow arrival rate to the median is one to two orders of magnitude.  In an energy-aware SDN, the objective is to switch off some controllers and links during off-peak hours.

 A. Ruiz-Riveria et al. \cite{ruiz2015greco}, Y. Hu et al.\cite{hu2017energy} propose an energy-aware SDN while ensuring that the node to controller latency is within a given delay bound and load on the controllers is balanced. 
\begin{itemize}
\item Elastic Link ($EL$): The elastic link metric considers switching off some links during off-peak hour. Let the binary variable $a_{e}$, denote whether the edge $e$ is active or not. The objective of the elastic link metric is to minimize the number of links, subject to certain constraints.
\begin{equation*}
EL = \min_{e \in E} \sum a_e
\end{equation*}

The constraints may be that there exists exactly one control path between the nodes, link utilization does not exceed its capacity and node to controller latency is minimized\cite{ruiz2015greco}.
\item Elastic Controller($EC$): The elastic controller metric considers switching off some controllers during the off-peak hours\cite{zeng2015flow},\cite{rath2014optimal}, \cite{bari2013dynamic}. The number of controllers active is a function of the network traffic. Let the binary variable $a_{c}$, denote whether the controller $c$ is active or not.  The objective is to minimize the number of controllers as shown in the following equation subject to some constraints.
\end{itemize}
\begin{equation*}
EC = \min_{c \in C} \sum a_c
\end{equation*}
 
\item  \textit {\textbf{Cost Aware:}}
The goal of cost-aware metric is to minimize the cost of setting up and managing SDN, while considering constraints such as controller capacity, latency and so on. 
\begin{itemize}
\item Deployment Cost ($DC$): The objective of this metric is to optimally plan\cite{sallahi2015optimal}  a SDN, including number of controllers, their placement, as well as interconnections between them. Given $C$ and  $E^{P_i}_c$ be the set of controllers and the set of control paths for a given control placement $P_i$ respectively.  On similar lines, let $Cost(c)$ and $Cost(E^{P_i}_c)$ be the cost of   controller installation and the cost of connection establishment cost between controller to controller and controller to switch respectively. The deployment cost $DC$ will be given as 
\begin{equation*}
DC = min \sum_{c \in C)} Cost(c)+\sum_{e\in E^{P_i}_c} Cost(e)
\end{equation*}
with  constraints similar to those of energy aware.\\
\end{itemize}

Switch migration is necessitated to balance the load among the controllers, recover from controller failure and reduce flow setup time as can be seen in the work by A. Dixit et al.\cite{dixit2014elasticon}, Cheng et al.\cite{cheng2016dynamic}. As switch migration involves some cost, the decision to whether to migrate a switch or not can be difficult at times. Moreover, we need to choose a target switch for migration from amongst multiple switches. Here, switch migration cost (SMC)\cite{wang2017switch}\cite{hu2018easm} can help out to choose one of the possible migration scenarios which can be the trade-off between objective and overhead. 
\begin{itemize}
\item Switch Migration Cost ($SMC$): The process of migrating a switch, involves the cost in terms of the control message to be exchanged between the controllers and the difference in delay encountered by flows associated due to the migration.  Let us consider a case where switch $s_k$ needs to be migrated from the controller $c_i$ to the target controller $c_j$. Let $t_c$ is the total control messages that need to be exchanged between the controllers to affect the migration and let $l_{m}$ be the load or number of flows associated with switch $s_k$. Thus the switch migration cost metric can be defined as
\begin{equation*}
SMC = min\{ t_c \times d(c_i,c_j) + l_m\times( |d(s,c_i)-d(s,c_j)|)\}
\end{equation*}
\end{itemize}
\end{itemize}




\section{ Static CPP Solutions} \label{CPP_Solutions}

In this section, we review the techniques and approaches of representative papers proposed to solve the static CPP. The papers are grouped on the basis of metrics (detailed in Table \ref{tab:sCPP}). They propose to optimize. During our study, we found that most researchers focused on minimizing latency as one of their objectives.
The other thing to be noted is that researchers are usually concerned with optimizing more than one objective.

\subsection{Latency}
Heller et al.\cite{heller2012controller} address the problem of controller placement as a facility location problem and report  that the answer to the problem depends on metric choice(s), network topology itself and desired reaction bound. They try to minimize the average propagation latency called the $k$-median or $k$-mean problem and worst case propagation latency called the $k$-centers problem. The authors  established that both these latencies cannot be minimized simultaneously. They experimented over the Internet2 OS3E\cite{OS3E} topology and the Internet topology zoo datasets\cite{6027859}. They measured the optimal average or worst case latency by trying out all possible combinations of the controllers (brute-force approach). They showed that latencies  reported for optimal placement of the controllers are significantly better than random placement of controllers. The approach  by Heller et al. other than being computationally exhaustive also  assumes that the number of controllers required is known beforehand. 

Zhao et al. \cite{zhao2017towards} extended the work of Heller et. al. \cite{heller2012controller} to find the optimal number of controllers using the affinity propagation clustering algorithm\cite{dueck2009affinity}. Their objective is to  form clusters and estimate the number of controllers. The authors proposed a modified version of the affinity propagation clustering algorithm since the algorithm cannot be applied directly as it uses the Euclidean distance to measure node similarity. In SDN, the Euclidean distance cannot be used since nodes are generally not directly reachable in a real network topology.

The authors in ~\cite{wang2016k} propose to minimize the worst case latency using a modified \textit{$k$-mean clustering} technique. Again the $k$-mean clustering technique cannot be directly used as it uses Euclidean distance to measure the distance between nodes. The $k$-mean algorithm also requires initialization of the $k$ cluster centers which makes the algorithm sensitive to the initialization parameters.  They compared their work with the standard $k$-mean approach and showed an improvement by a factor of greater than 2.

Wang et al.\cite{wang2018effective} also use latency as a performance metric to place the controllers. In addition to propagation latency, the authors also considered switch processing latency, transmission latency, and queuing latency. They partitioned the network into sub-networks using a clustering technique and proposed to place one or more controllers in each of these subdomains based on node density. The clustering technique proposed uses Dijkstra's algorithm \cite{johnson1973note} to measure the shortest distance between nodes and randomly selects cluster centers. Although the authors include the different latencies in their problem formulation, in their proposed solution, they make the simplifying assumption that the shortest distance incorporates the end-to-end latency. 

In \cite{su2015mdcp},  the objective is to minimize both the synchronization cost between controllers ($C2C$) as well as communication cost between controller and switch ($N2C$). The synchronization cost is defined as the measure of hops between a controller pair and communication cost is the sum of hops between a controller-switch pair. The authors formulate the problem as integer linear programming (ILP). Due to the high computational complexity of QIP, the authors propose greedy approximation solution and a heuristic approach. The approximation solution decomposes the initial problem as several classical facility location problems and selects the minimum cost among them as the final result. The heuristic approach sorts the nodes in decreasing order of their degree and returns a placement such that the synchronization cost and communication cost is below a threshold. 


\subsection{Latency and Capacity}
The SDN controller is implemented in software which runs either in a physical or a virtual server. Due to  server capacity limitations and higher failure probability of heavily loaded controllers, a controller placement strategy should also take into account controller capacity.  Yao et al.\cite{yao2014capacitated} extend Heller's\cite{heller2012controller} work of minimizing the worst case latency by incorporating the controller capacity as a constraint. The problem is defined as capacitated  controller placement problem and it is shown to be NP-hard as well as different from the classical capacitated $k$-center problem. The authors use integer programming to find the minimal number of controllers within a specified radius, where the radius denote the worst case latency. The authors report that their result in terms of the number of controllers and latency  is less than that of the $k$-center problem.
 
Gao et. al.  \cite{gao2015particle} propose to minimize the average node to the controller ($N2C$) and controller to controller($C2C$) latency subject to the condition that the load on the controllers does not exceed their capacity.  They solve the problem using   \textit{PSO} (Particle Swarm Optimization) algorithm where each particle represents a placement instance of the $k$ controllers. They compare their approach in terms of latency and computational time with a greedy algorithm and integer linear programming. 

Liu et. al. \cite{liu2015ncpso} also use PSO to computer clusters, where the particles represent a placement instance of the $k$ controllers like in \cite{gao2015particle}. However, elements of the particle also include the number of switches associated with each of the controllers. Their objectives include minimizing the node to controller latency, load balance the controllers such that the load on the controllers do not exceed their capacity. The authors introduce a new metric, switch to switch latency, however, the authors do not mention how this metric is different from node to node latency.
 
Sanner et al.\cite{sanner2016hierarchical} propose to find the number of controllers $k$  and place them to minimize the latency as well as balance load on the controllers in terms of the switch attached to the controllers. They proposed two flavors of the solution - adapted $k$-mean clustering and hierarchical clustering. In the heuristic hierarchical clustering technique  initially, each node (switch) is considered as a cluster with a controller. The clusters are then merged without violating the bound on latency and load.

In \cite{ksentini2016using1}, the authors distinctively formulate the problem of $N2C$ and $C2C$. Next, they use Nash bargaining to model the trade-off between these two latencies while considering controller load balancing as a constraint. In their proposed approach, the $N2C$ latency and $C2C$ latency are considered as two players. To use Nash bargaining, the threat point needs to be found first which in their case is  $N2C_{worst}$ and $C2C_{worst}$. The trade-off between the two latencies is modeled as an optimization problem and the threat point is computed using linear optimization. The authors claim that their proposed solution ensures better trade-off in comparison to the mono objective CPP. 
 
Obtaining Pareto optimal solutions for problems with multiple conflicting objectives is known to be a hard problem, thus researchers resort to multi-objective evolutionary algorithms (\textit{MOEAs})\cite{deb2003fast}. These evolutionary algorithms have the capability to obtain  Pareto optimal solutions in relatively few iterations. A number of researchers \cite{ahmadi2015hybrid}, \cite{jalili2017optimal},\cite{jalili2015controller}, \cite{ahmadi2017adaptive}   have tried to solve the problem of minimizing latency and capacity management using a modified version of  \textit{NSGA-II}\cite{deb2002fast}(non-dominating sorting genetic algorithm). The modification is in terms of greedy initialization among others. The convergence analysis is performed using the metric  {\em IGD (Inverted generational distance)} which compare the performance of algorithms in terms of  real Pareto front generated.

In traditional  \textit{MOGA} (Multi-Objective Genetic Algorithm), the mutation process is random and thus the algorithms do not necessarily converge in each iteration. In PSO, the search direction is guided which leads to reduced convergence time. Liao\cite{liao2017genetic} modified the traditional \textit{MOGA} by adding \textit{PSO} based mutation and show that it leads to reduced convergence time as compared to the classical genetic algorithm.

\subsection{Latency and Availability}
In SDN, while latency is undoubtedly an important design metric, many researchers argue that availability is crucial for operational SDNs. Guo et al.\cite{guo2013controller} study the problem of controller placement to improve latency as well as network resilience. The authors consider the SDN architecture as two interdependent networks - a switch to switch network ($SS$) for data forwarding and a controller to switch ($CS$) network for network control.  They model these two networks as an interdependence graph. Failure of a switch will affect its entire associated links in $SS$ and due to mutual dependence between the two networks, the effect of the failure will propagate to the associated links in $CS$. The authors examine the effect of cascading failure between the $SS$ and $CS$ network. Their objective is to make the network resilient to such failures. The solution proposed is  a greedy algorithm, where given $k$ controllers,  the algorithm partitions the network into $k$ clusters and place the controller in each cluster	 with maximal closeness centrality. The authors introduce a \textit{resilience} metric for cascading failure analysis.

Hu et. al. \cite{hu2013reliability},\cite{hu2014reliability}  propose a metric \textit{expected control path loss} which measures the reliability of the control path between a controller and switch or between controllers. Their objective is to minimize the metric. The problem is shown to be NP-hard by reducing it to the dominated set problem. The authors propose a {\em l-w-}greedy algorithm which ranks potential locations of switches based on increasing switch failure probability. From among these locations, the algorithm chooses the first $w$ locations. For non-zero values of $l$, the controller allows $l$ steps of backtracking. The authors also use simulated annealing\cite{kirkpatrick1983optimization} to solve the problem. They show that the optimal reliability can be obtained if the number of controllers lies in the range   [$0.035n, 0.117n$], where $n$ is the number of nodes. 

The objective in \cite{ros2014five} is to achieve at least five nines reliability in the southbound interface. The problem is formulated to minimize the dual cost of deploying a controller and the cost of serving a node by the controller. The network topology in addition to network nodes proposes facility locations where controllers can be deployed. The objective is to choose for every node a subset of deployed controllers such that the probability of having an operational path between them is higher than a given threshold. The heuristic proposes to solve the problem by selecting facilities on disjoint paths which allows the system to avoid controllers at facilities that do not contribute to increased reliability.  Further to reduce the number of deployed controllers, the facilities are ranked according to their expected contribution to southbound reliability. The rationale is to give higher priority to facilities that are useful to larger number of nodes.

The work in \cite{guo2015towards} is motivated by the observation that most of the network failures affect a single link at a time. Thus instead of considering controller placement under \textit{comprehensive network state} where any number of nodes and  links can fail, they consider single link failure case only. The problem is shown to be NP-hard and a greedy algorithm is proposed. In each iteration, the algorithm chooses a potential controller, computes the worst case latency under single link failure scenario for the new placement scheme. If the computed worst case latency improves upon the previously computed value, it is accepted as a better placement scheme. 

Zhong et. al. \cite{zhong2016min} also consider the availability of a SDN network in the event of a single link failure. They define two reliability metric which measures the average number of disconnected switches caused either by a single link failure in the network or a single link failure  in the control network.  Their objective is to find a minimum cover of the network, which is equivalent to find the least number of controllers, that also minimizes their proposed reliability metric. The authors propose a PSO inspired heuristic to compute initial the random controller placements in each iteration, and choose from among them the most optimized solution.

In \cite{aoki2016controller}, the authors in addition to latency propose a \textit{survivability} metric, which is  a measure of the number of the edge-disjoint path between a switch and controller. The objective is to maximize the survivability metric while partitioning the network to decide controller domains. The authors, however, do not propose any solution approach in their paper.

Ishigaki et al.\cite{ishigaki2016controller} define a new metric \textit{burden} of a node which is defined as the number of shortest paths that use the node as a relay point. The objective is to find a placement which minimizes the total burdens on all the nodes in the graph. Computing the burden of a node requires finding the shortest path. The authors propose to use the Dijkstra's algorithm \cite{johnson1973note} to find the shortest path, when the number of controllers is small as compared to the number of nodes and the Floyd-Warshall algorithm \cite{cormen1990floyd} if the two sets are comparable. They further show that from an aspect of latency and availability, their results are comparable to that of the closeness centrality placement but they achieve it with much less computational complexity.
                                                        

In this paper\cite{sanner2017evolutionary}, the objective is to enhance reliability by maximizing the  average connectivity between clusters. Further, they also propose to reduce the latency between controllers by minimizing the imbalance of control traffic between clusters and reduce the number of controllers. The problem is formulated as a linear programming model  using the max-flow min-cut theorem \cite{sakarovitchoptimisation} since the minimum cut represents the connectivity between two clusters. As the objectives are conflicting and uncorrelated, the solution 
proposed uses the NSGA-II framework \cite{deb2002fast}.

\subsection{Latency, Capacity and Availability}
Deploying multiple controllers not only reduces latency but also makes the network resilient to controller failures. However, during a controller outage, nodes need to be re-assigned to the second best controller. Researchers \cite{hock2013pareto}, \cite{lange2015heuristic}, \cite{lange2015specialized}  in addition to minimizing latency and balancing of controller load have also proposed to handle link failures. Their objective is to minimize the latencies not only during failure-free routing but also during controller failures. They consider up to $k-1$ controller failures. 

Hock et al.\cite{hock2013pareto} proposed a resilient \textit{POCO} (Pareto-based Optimal COntroller-placement) framework which consider the Pareto-optimal controller placements out of all the placements. The POCO framework does not give any recommendation and leaves it to the network administrator to choose any one of the Pareto-optimal placement that best meet their needs. In order to find the set of Pareto-optimal placement of the controllers, the POCO framework performs an exhaustive search.  Thus the {POCO} framework will not scale well in case of  large size networks. To overcome the computational overhead, Lange et al.\cite{lange2015heuristic} extended the POCO framework and proposed to find the Pareto-optimal front using  a  heuristic technique based on Pareto simulated annealing\cite{czyzzak1998pareto}. 

Perrot et. al.\cite{perrot2016optimal} extend the POCO framework to find the minimum number of controllers required. The authors formulate the problem as an integer programming problem where the main objective is to minimize the number of active controllers and all other objectives are considered as constraints. They solved their integer programming model using CPLEX\cite{ilog2012cplex} and used the POCO framework to evaluate their solution.

In \cite{yao2013cascading}, the authors inspired by the Motter model\cite{motter2002cascade} discuss the possibility of cascading failure in SDN. In a SDN, when a controller fails, its load will be distributed among the remaining controllers. Due to this load redistribution, the load on a controller may exceed its capacity and may trigger a cascading failure. The authors outline three optimal strategies to prevent cascading failures. The optimal strategies must fulfill the following conditions (i) the system's ability to handle the load of the failed controller (ii) controller load must be initially balanced to avoid failure of an overloaded controller and (iii) load distribution after a controller failure, must not result in a controller exceeding its capacity. They, however, do not specify the number of controller failures that can be tolerated.

Jimenez et al.\cite{jimenez2014controller} propose to place the controllers such that the delay between node to controller as well as the delay between controllers is bounded. The other objective is to build a robust control plane by proposing to select shortest branches with optimal connectivity so as to minimize the impact of node failures. To solve the CPP, the authors introduce an  algorithm, \textit{$k$-critical} broadly based on the \textit{$k$-center} and \textit{$k$-median} algorithm. The algorithm computes a parameter to select candidate nodes which is a function of node connectivity and path weight. From among the candidate nodes, the nodes that can become a controller are those nodes that satisfy delay requirement of a switch which is not yet assigned to any controller.

Muller et. al. \cite{muller2014survivor} proposed \textit{Survivor}, a framework that exploits path diversity to handle peak-time load failure while balancing the load on the controllers. The failover mechanism is further enhanced by preparing a list of backup controllers. The problem is formulated as integer linear programming (ILP) with the objective of maximizing connectivity and controller capacity as constraints. The backup controller is computed using a heuristic which selects the closest controller instances in terms of delay or \textit{residual} capacity.

Killi et. al. \cite{killi2016optimal}, similar to the idea of backup controllers, propose \textit{reference} controllers for every switch which will be used as  failover controllers. They consider the worst case scenario where  up to $k-1$ controllers can fail and their objective is to minimize the delay from a switch to its $k^{th}$ reference controller. An assumption made by the authors is that the switches have \textit{failure foresight}, that is the switches know the status of all the controllers which is unrealistic. The authors in a subsequent work \cite{killi2017capacitated} handle the issue of failure foresight by minimizing the maximum  sum of the $C2C$ latency from a switch to its first reference controller. In other words, the objective is to minimize the maximum, sum of latency from a switch to its first reference controller  and from the first reference controller to the second reference controller and so on.

Xiao et al.~\cite{xiao2014sdn} handle the problem of deployment of SDN in WAN (wide area network). They propose to partition a WAN using \textit{spectral clustering} called \textit{SDN domains} where each domain is controlled by a SDN controller. The \textit{min-max cut function}\cite{ding2001min} is used as a metric for spectral clustering which minimizes interdomain similarity but maximizes intradomain similarity. The authors propose to place a SDN controller in each of the SDN domain so as to optimize latency, capacity, and reliability. The authors in a subsequent work \cite{xiao2016ak} extend their work to compute the optimal number of SDN domains by exploiting the structure of the eigenvectors. The \textit{eigengap} gives a measure of the graph connectivity and it is used to reveal the number of clusters. Spectral clustering technique has time complexity $ O(n^3)$ which is greater than other clustering technique. 
The work by Liao et al.\cite{liao2017density} is similar to the spectral clustering of SDN \cite{xiao2014sdn} described above in the sense that they propose to split the network into sub-networks, each of which will be controlled by a controller, however, the clustering is density-based. The network is split into clusters by computing for each switch its local density as well as its distance to switches with higher density. In the second part of their work, the authors propose that in addition to density, the capacity of the controllers can also be used during cluster formation, the so-called capacitated density-based clustering. In this type of clustering, if the load of a cluster cannot be expanded the switches are assigned to another cluster or form a new cluster. The controller placement in each sub-network is done considering $N2C$ latency, $C2C$ latency, and reliability.

\subsection{Latency, Capacity, and Energy}
Rivera et al.\cite{ruiz2015greco} focus on minimizing the energy consumption by reducing the number of active control link, while ensuring that all switches have path to a controller within a given delay bound and the load of the controllers  is balanced. The problem is formulated as a $BIP$ (Binary integer programming) which being exhaustive for large-scale network, a heuristic method is proposed. Given the demand between a pair of nodes, the algorithm computes $k$ paths for each demand. The authors then find the \textit{surviving} links, which is the smallest connected set of links which route all demands and other links are switched off. The authors report that their framework is able to reduce energy consumption by up to  $55\%$ during the off-peak period and only $20\%$ more links are used compared to the optimal solution. 

In \cite{ruiz2015greco}, a simplifying assumption made was that placement of the controllers is  known. Hu et al.\cite{hu2017energy} extend their work by showing that controller placement also affects energy consumption. Their objective is to find a controller placement that minimizes the power consumption of the control network while ensuring that delay and capacity constraints are not exceeded. The problem is formulated as a binary integer program with the aim of minimizing the number of active control links. The authors propose a  genetic algorithm where a chromosome represents a placement solution of the controllers. The fitness function computes the number of links in the solution using GreCo\cite{ruiz2015greco}.

\subsection{Latency, Capacity and Cost}
Sallahi et al.\cite{sallahi2015optimal} propose a mathematical model  to minimize the network cost while considering other constraints. The problem is formulated as an ILP where the objective is to minimize the cost of controller installation, linking of controllers to switches and interlinking of controllers, while ensuring constraints like controller and link capacity is not exceeded.   The authors in \cite{sallahi2017expansion}, extend their work with the objective of minimizing the cost of expanding an existing SDN network with constraints similar to that of their previous work. In both cases, the authors use CPLEX \cite{ilog2012cplex} to compute the solutions and as such their solution will work only for small-scale networks.

Zhao et al.\cite{zhao2017scalable} define the CPP as the problem of minimizing the number of clusters and cluster the controllers into sets called controller sets. Each controller set involves some \textit{basic costs} like warehouse rent and power supply cost which need to be minimized. The problem is formulated as an ILP that aims to minimize the controller cost, basic costs, $N2C$, $C2C$ and manpower costs. The proposed solution adopts the divide-and-conquer philosophy by partitioning the network into SDN domains and places a controller set in each such domain. 
A greedy algorithm is proposed to partition the network which exploits the property of  minimum nonzero eigenvalue of Laplacian matrix  to cluster closer nodes to a SDN domain by considering the minimum non-zero eigenvalue.


\subsection{Flow setup time}
One of the primary tasks of a SDN controller is setting up flows. In SDN, the first packet of a flow that arrives at the egress switch is forwarded to the controller for setting up the flow table entries. The response time of a request sent from a switch is defined as the duration the request is sent from the switch until the response is received from the controller. Cheng et al.\cite{cheng2015qos} propose to minimize the response time. The problem is formulated by assuming each controller as a M/M/1 queuing model.  The objective is to minimize the mean response time of the queuing model.  The authors propose three heuristics-  incremental greedy algorithm, primal-dual-based algorithm and network-partition-based algorithm to solve this problem. In the greedy approach, in each iteration, the controllers that can serve the maximum number of switches are chosen. The process is repeated until every switch is served by a controller. The authors report that the greedy algorithm performs better than the other two approaches.

Zeng et. al. \cite{zeng2015flow} assume that from the perspective of a controller, the flow setup request queue at the controller. They also use a M/M/1 queuing model to model the SDN controller.  The authors propose to minimize the number of controllers while ensuring that the flow setup time does not exceed a threshold. The problem is modeled as  an integer programming model and  a greedy solution is proposed. In their solution approach, a switch is  assigned to a controller with the smallest transmission latency provided the flow setup time constraint is not violated. 

A comparison of the different objectives and summary of the solution approaches of static CPP discussed in this paper is given in Table \ref{tab:static_cpp_solns}.


\begin{center}
\onecolumn
\begin{longtable}{|l|p{1cm}|l|p{8.0cm}|}
\caption{Comparison of solution approaches on static CPP
\label{tab:static_cpp_solns}}\\
\hline
\hline
\rowcolor[HTML]{EFEFEF} 
\textbf{\begin{tabular}[c]{@{}l@{}}OPTIMIZE\\ OBJECTIVE\end{tabular}} & \textbf{REF.} & \textbf{\begin{tabular}[c]{@{}l@{}}PROBLEM\\ FORMULATION\end{tabular}} & \textbf{SHORT DESCRIPTION OF SOLUTION APPROACH} \\ \hline \hline \hline
\endfirsthead
\multicolumn{4}{c}%
{{\bfseries Table \thetable\ continued from previous page}} \\
\endhead
\rowcolor[HTML]{9B9B9B} 
\cellcolor[HTML]{9B9B9B} & \cite{heller2012controller} & \textit{\begin{tabular}[c]{@{}l@{}}Facility location problem\\ $k$-median and $k$-center problem\end{tabular}} & Used exhaustive search to provide optimal solution. \\ \cline{2-4} 

\cellcolor[HTML]{9B9B9B} &\cite{su2015mdcp}  & Integer linear programming & Decomposed the problem to facility location problem and used greedy solution approach based on synchronization cost and connectivity.  \\ \cline{2-4} 

\rowcolor[HTML]{9B9B9B} 
\cellcolor[HTML]{9B9B9B} &\cite{wang2016k}  & \textit{\begin{tabular}[c]{@{}l@{}}Network partitioning\\ with linear programming\end{tabular}} & Modified $k$-mean clustering technique is used,  where similarity is a function of $N2C$. \\ \cline{2-4} 

\cellcolor[HTML]{9B9B9B} &\cite{wang2018effective}  & \textit{\begin{tabular}[c]{@{}l@{}}Network partitioning\\ with linear programming\end{tabular}} & Clustering-based network partition algorithm is proposed, where similarity is a function of latency(End2End and queuing latency). \\ \cline{2-4} 
\rowcolor[HTML]{9B9B9B}
\cellcolor[HTML]{9B9B9B} & \cite{zhao2017towards} & Linear programming & \begin{tabular}[c]{@{}l@{}}Modified Affinity propagation clustering is used,\\ where similarity is a function of negative $N2C$.\end{tabular}\\\cline{2-4}

\cellcolor[HTML]{9B9B9B} & \cite{sahoo2016optimal},\cite{sahoo2017metaheuristic},\cite{sahoo2017placement}& Facility location problem & Spectral clustering technique is used to minimize the $N2C$. And PSO and firefly meta heuristic are also used to solve the problem.
 \\ \cline{2-4}
 \rowcolor[HTML]{9B9B9B}
 \cellcolor[HTML]{9B9B9B} & \cite{ishigaki2017cluster}&\begin{tabular}[c]{@{}l@{}} Weighted minimum set\\ cover problem (WMSCP)\end{tabular}& Greedy algorithm.
  \\ \cline{2-4}
  \multirow{-6}{*}{\cellcolor[HTML]{9B9B9B}LATENCY}& \cite{killi2018placement}&
          Integer linear programming  & CPLEX is used for optimal solution. Joint placement of hpervisiors and controllers in virtual SDN .  
 \\ \hline

\rowcolor[HTML]{EFEFEF} 
\cellcolor[HTML]{EFEFEF} &\cite{yao2014capacitated}  & Capacited facility location problem & \begin{tabular}[c]{@{}l@{}}\\Used exact algorithm \cite{ozsoy2006exact} to get optimal solution.\end{tabular} \\ \cline{2-4} 
\cellcolor[HTML]{EFEFEF} &\cite{gao2015particle}  & Linear programming & PSO-based meta-heuristic technique is used to solve global latency ($N2C$ \& $C2C$) with capacity constraint. \\ \cline{2-4} 
\rowcolor[HTML]{EFEFEF} 
\cellcolor[HTML]{EFEFEF} &\cite{ahmadi2015hybrid},\cite{ahmadi2017adaptive}  & \begin{tabular}[c]{@{}l@{}}Multi-objective combinatorial\\ optimizing model (MOCO)\end{tabular} & Used meta-heuristic approach (NSGA-II) with greedy  initialization to solve objectives ($N2C,$ $C2C,$ \& $IMBL$) and \textit{Inverted generational distance} as fitness function.  \\ \cline{2-4} 
\cellcolor[HTML]{EFEFEF} &\cite{jalili2015controller},\cite{jalili2017optimal}  & \begin{tabular}[c]{@{}l@{}}Multi-objective combinatorial\\ optimizing model(MOCO)\end{tabular} & Used meta-heuristic approach (NSGA-II) with greedy  initialization to solve objectives ($N2C,$ $C2C,$ \& $IMBL$) and \textit{Inverted generational distance} as fitness function.  \\ \cline{2-4} 
\rowcolor[HTML]{EFEFEF} 
\cellcolor[HTML]{EFEFEF} &\cite{liu2015ncpso}  & Linear programming &   Computed the number of controllers that need to be deployed using PSO based meta-heuristic algorithm. \\ \cline{2-4} 

\cellcolor[HTML]{EFEFEF} &\cite{sanner2016hierarchical}  &Set Covering Problem(ILP)  & Hierarchical based clustering algorithm is used.  \\ \cline{2-4} 
\rowcolor[HTML]{EFEFEF} 
\cellcolor[HTML]{EFEFEF} &\cite{ksentini2016using1}  & Integer linear programming & Co-operative game theory (Nash Bargaining algorithm ) is used to solve the multi-objective problem. \\ \cline{2-4}
\cellcolor[HTML]{EFEFEF}
&\cite{liao2017genetic}  & \begin{tabular}[c]{@{}l@{}}Multi-objective combinatorial\\ optimization model (MOCO)\end{tabular} & Used meta-heuristic approach(MOGA),and modified the mutation process using PSO.\\ \cline{2-4} 
\rowcolor[HTML]{EFEFEF} 
\cellcolor[HTML]{EFEFEF} &\cite{han2016minimum} & Partitioning& Proposed greedy based controlling pattern algorithm to minimize the number of controller with control latency, where control latency is calculated using Kalman filter. \\ \cline{2-4}
\cellcolor[HTML]{EFEFEF}
&\cite{hu2016load}  & linear programming& CPLEX is used to minimize both $IMLB$ and $N2C$.\\ \cline{2-4}
\rowcolor[HTML]{EFEFEF}
 \cellcolor[HTML]{EFEFEF}&\cite{bo2016controller} & Partitioning & Used meta-heuristic approach(MOGA),and fitness function is formulated based on load diversity and minimum spanning tree (Prims is used  to spanning tree). \\ \cline{2-4} 
 \cellcolor[HTML]{EFEFEF}&\cite{kuang2018hierarchical} & Partitioning & Used $k*$-mean algorithm\cite{qi2016k} (hierarchical $k$-means algorithm
initializes more than $k$ clusters and merges top-n nearest clusters to get $k$ clusters), where similarity is a function of $N2C$ and load balance. \\ \cline{2-4}
\rowcolor[HTML]{EFEFEF}
 \cellcolor[HTML]{EFEFEF}&\cite{zhu2017control}& Integer linear programming &A heuristic algorithm is proposed where $k$-mean clustering technique is used to find first cluster. \\ \cline{2-4}
 
\multirow{-11}{*}{\cellcolor[HTML]{EFEFEF}\begin{tabular}[c]{@{}l@{}}LATENCY\\ \\     AND\\ CAPACITY\end{tabular}}& \cite{killi2018cooperative}&
       Partitioning   & Proposed clustering method utilizing cooperative game theory approach.  \\ \hline
\rowcolor[HTML]{9B9B9B} 
\cellcolor[HTML]{9B9B9B} & \cite{hock2013pareto} & \begin{tabular}[c]{@{}l@{}}Multi-objective combinatorial\\ optimization model (MOCO)\end{tabular} & Proposed a Pareto optimal solution using exhaustive search. Developed the framework (POCO) in MATLAB.   \\ \cline{2-4} 
\cellcolor[HTML]{9B9B9B} & \cite{yao2013cascading} & Theoretical & Used Motter  based solution proposed for complex network\cite{motter2002cascade} to avoid cascading failure.  \\ \cline{2-4} 
\rowcolor[HTML]{9B9B9B} 
\cellcolor[HTML]{9B9B9B} &\cite{muller2014survivor} & Integer linear programming & Proposed a framework to find node-disjoint, capacity sensitive path; selected backup controller based on heuristic approach i.e., delay or capacity residual based algorithm.  \\ \cline{2-4} 
\cellcolor[HTML]{9B9B9B} &\cite{jimenez2014controller}   & $k-$ center and $k-$ median problem & Proposed $k-$critical algorithm to find out  minimum controllers and robustness to failures. \\ \cline{2-4} 
\rowcolor[HTML]{9B9B9B} 
\cellcolor[HTML]{9B9B9B} &\cite{xiao2014sdn},\cite{xiao2016ak} & Network partitioning  & Spectral clustering is used, where min-cut and normalized-cut is used to partition.   \\ \cline{2-4} 
\cellcolor[HTML]{9B9B9B} & \cite{lange2015heuristic},\cite{lange2015specialized} & \begin{tabular}[c]{@{}l@{}}Multi-objective combinatorial\\ optimization model (MOCO)\end{tabular}  & Pareto based meta and specific heuristic (PSA \& PCKM) is proposed to avoid exhaustive evaluation in terms of time and memory.   \\ \cline{2-4} 
\rowcolor[HTML]{9B9B9B} 
\cellcolor[HTML]{9B9B9B} &\cite{perrot2016optimal}   & Integer linear programming & Used CPLEX to get optimal solution. \\ \cline{2-4} 

\cellcolor[HTML]{9B9B9B} &\cite{killi2016optimal},\cite{killi2017capacitated} & Mixed integer linear programming  & Used CPLEX as well as meta-heuristic approach(simulated annealing) to solve the problem.  \\ \cline{2-4} 
\rowcolor[HTML]{9B9B9B} 
\cellcolor[HTML]{9B9B9B} &\cite{liao2017density}  &  Network Partitioning  & Used density based clustering algorithm with capacity constraint.  \\ \cline{2-4} 
\cellcolor[HTML]{9B9B9B}& \cite{tanha2016enduring}&\begin{tabular}[c]{@{}l@{}} Capacitated reliable fixed-charge\\ location problem
\\(Integer linear programming)\end{tabular}
  & Gurobi is used  to provide optimal solution.   \\ \cline{2-4}
 \rowcolor[HTML]{9B9B9B}
 \cellcolor[HTML]{9B9B9B}& \cite{liu2016heuristics}&\begin{tabular}[c]{@{}l@{}} Partitioning \\(Integer linear programming)\end{tabular}
  & Proposed heuristic algorithm to solve the problem in linear time that is based on community detection algorithm(Capacitated label propagation algorithm).  \\ \cline{2-4}
  \cellcolor[HTML]{9B9B9B}& \cite{liu2016reliability}&\begin{tabular}[c]{@{}l@{}} Mathematical formulation \end{tabular}
    & $k$-mean clustering technique is used to provide optimal solution while greedy approach to sub optimal solution.  \\ \cline{2-4}
  \rowcolor[HTML]{9B9B9B}
  \cellcolor[HTML]{9B9B9B}& \cite{naning2016sdn}&\begin{tabular}[c]{@{}l@{}}Multi-objective combinatorial\\ optimization model (MOCO)\end{tabular}
   & POCO framework is used for production network.  \\ \cline{2-4}
  
  \cellcolor[HTML]{9B9B9B}& \cite{borcoci2016multi}&\begin{tabular}[c]{@{}l@{}}Multi-objective combinatorial\\ optimization model (MOCO)\end{tabular}
     & Exhaustive search is used to obtain optimal solution.  \\ \cline{2-4}
 
 \rowcolor[HTML]{9B9B9B}
  \cellcolor[HTML]{9B9B9B}&\cite{bannour2017scalability}& \begin{tabular}[c]{@{}l@{}}Multi-objective combinatorial\\ optimization model (MOCO)\end{tabular}& A clustering technique (PAM-partitioning around medoids) and NSGA-$II$ heuristic approach are used to solve the problem.  \\ \cline{2-4}
 
  \cellcolor[HTML]{9B9B9B}&\cite{killi2018link}& Mixed integer linear programming& CPLEX is used to provide optimal solution in case of single link failure.  \\ \cline{2-4}
  \rowcolor[HTML]{9B9B9B}
  \cellcolor[HTML]{9B9B9B}&\cite{tanha2018capacity}& Linear programming& Two heuristic algorithms are proposed based on  clique graph theory technique i.e using all maximal clique and  single maximal clique respectively,.  \\ \cline{2-4}
\multirow{-11}{*}{\cellcolor[HTML]{9B9B9B}\begin{tabular}[c]{@{}l@{}}LATENCY,\\ CAPACITY,\\    AND\\ AVAILABILITY\end{tabular}} &\cite{zhang2016optimal}  &\begin{tabular}[c]{@{}l@{}}Multi-objective combinatorial\\ optimization model (MOCO)\end{tabular}  & Meta-heuristic (adaptive bacterial foraging optimization) is used. \\ \hline
\rowcolor[HTML]{EFEFEF} 
\cellcolor[HTML]{EFEFEF} & \cite{guo2013controller} & \begin{tabular}[c]{@{}l@{}}Graph theoretical approach\\ (Interdependence Graph)\end{tabular} & Used hierarchal agglomeration algorithm to partition the network; controller placement chosen using maximal closeness centrality. \\ \cline{2-4} 
\cellcolor[HTML]{EFEFEF} & \cite{hu2012placement},\cite{hu2013reliability},\cite{hu2014reliability} &Integer linear programming  & Proposed two approaches: $l-w-$greedy algorithm and simulated annealing (SA).
 The greedy algorithm allow  $l$ backtrack steps , $w$ list of ranked potential controller location. \\ \cline{2-4} 
\rowcolor[HTML]{EFEFEF} 
\cellcolor[HTML]{EFEFEF} &\cite{ros2014five},\cite{ros2016reliable}  & \begin{tabular}[c]{@{}l@{}}Fault tolerance facility\\ location problem\end{tabular} & To meet reliability constraints, a greedy algorithm was proposed where nodes are ranked in terms of their degrees. \\ \cline{2-4} 
\cellcolor[HTML]{EFEFEF} & \cite{guo2015towards} & $k-$center problem  & Proposed optimal solution based on traversal algorithm (state-first search algorithm is used )as well as a greedy based solution. \\ \cline{2-4} 
\rowcolor[HTML]{EFEFEF} 
\cellcolor[HTML]{EFEFEF} &\cite{ishigaki2016controller}  & \begin{tabular}[c]{@{}l@{}}Graph theoretical approach\\ (Stress centrality)\end{tabular} &  Computed burden of each node, which is the number of time the node appears on the shortest path. Proposed to minimize node burden. \\ \cline{2-4} 
\cellcolor[HTML]{EFEFEF} &\cite{zhong2016min}  & Min cover problem & The min cover problem was reduced to finding the least number of controllers. A PSO inspired heuristic was proposed. \\ \cline{2-4} 
\rowcolor[HTML]{EFEFEF} 
\cellcolor[HTML]{EFEFEF} &\cite{aoki2016controller}  & Graph theoretical approach & Used spectral clustering technique to partition the network, where similarity is a function $N2C$ and survivability metric. \\ \cline{2-4} 
\cellcolor[HTML]{EFEFEF} &\cite{sanner2017evolutionary}  & Mixed integer programming &Proposed  modified version of NSGA-II to find a tradeoff between average connectivity and $IMBL$ as well as minimize the active controllers. \\ \cline{2-4} 
\rowcolor[HTML]{EFEFEF}
   \cellcolor[HTML]{EFEFEF}&\cite{vizarreta2016controller}& Mixed integer linear programming& Solver is used to minimize the latency between switch to assigned  controller and its backup controller.
 \\ \cline{2-4}
 
   \cellcolor[HTML]{EFEFEF}&\cite{zhang2017survivability}& Integer linear programming& Used heuristic approach to handle single link failure.  \\ \cline{2-4}
   \rowcolor[HTML]{EFEFEF}
  \cellcolor[HTML]{EFEFEF}&\cite{li2017sharing}&\begin{tabular}[c]{@{}l@{}}Multi-objective combinatorial\\ optimization model (MOCO)\end{tabular}& PSO and GA based meta heuristic is proposed to find optimal placement of backup controllers. \\ \cline{2-4}
   \cellcolor[HTML]{EFEFEF}&\cite{beheshti2012fast}& Theoretical& Greedy approach is used to place a single controller to make system reliable. \\ \cline{2-4}
   \rowcolor[HTML]{EFEFEF}
\multirow{-8}{*}{\cellcolor[HTML]{EFEFEF}\begin{tabular}[c]{@{}l@{}}LATENCY,\\    AND\\ AVAILABILITY\end{tabular}} &\cite{li2017efficient}  & Facility location problem & Using weight random early detection(function of latency and rate of unreachability ) a modified SA algorithm is implemented. \\ \hline
\rowcolor[HTML]{9B9B9B} 
\cellcolor[HTML]{9B9B9B} &\cite{ruiz2015greco}  & Binary integer programming & The classical heuristic approach (GreCo) along with Yen's algorithm is used to find the $k-$-shortest path that meet the capacity and latency constraints. \\ \cline{2-4} 
\multirow{-4}{*}{\cellcolor[HTML]{9B9B9B}\begin{tabular}[c]{@{}l@{}}LATENCY,\\ CAPACITY,\\    AND\\ ENERGY\end{tabular}} &\cite{hu2017energy}  & Binary integer programming & CPLEX is used to get optimal solution. Proposed a modified genetic algorithm where new generation operator is used in place of the crossover operator and GreCo is used as fitness functions. \\ \hline
\rowcolor[HTML]{EFEFEF} 
\cellcolor[HTML]{EFEFEF} & \cite{sallahi2015optimal},\cite{sallahi2017expansion} &\begin{tabular}[c]{@{}l@{}}\\\\ Linear programming\end{tabular} & CPLEX is used to provide optimal solution but exhaustive for large network. \\ \cline{2-4} 
\multirow{-4}{*}{\cellcolor[HTML]{EFEFEF}\begin{tabular}[c]{@{}l@{}}COST,\\    AND\\ CAPACITY\end{tabular}} &\cite{zhao2017scalable}  & \begin{tabular}[c]{@{}l@{}}\\Integer linear programming\end{tabular} &  Proposed a greedy algorithm which chunk the network into small domain based on minimum non-zero eigenvalue  of Laplacian matrix. \\ \hline
\rowcolor[HTML]{9B9B9B} 
\cellcolor[HTML]{9B9B9B} &\cite{cheng2015qos}  & Queuing model & Proposed three heuristic algorithms to guarantee; QoS; (1) Incremental greedy algorithm that iteratively add switches to controller (2) Primal-dual algorithm which claims to provide near optimal solution (3) Network partition algorithm which provides balanced placement. \\ \cline{2-4} 
\multirow{-5}{*}{\cellcolor[HTML]{9B9B9B}\begin{tabular}[c]{@{}l@{}}FLOW SETUP\\  TIME\end{tabular}}&\cite{zeng2015flow} & \begin{tabular}[c]{@{}l@{}}Queuing model\\    Integer programming\end{tabular} & Used Gurobi\cite{gurobi2015gurobi} to find optimal solution and  also proposed a heuristic algorithm. SDN sub-domains are merged to minimize controller without violating the constraints or flow setup time.\\ 
 \hline

\caption{Comparison of solution approaches on static CPP}

\end{longtable}

\end{center}
\twocolumn


\section{Adaptive CPP Solutions} \label{DCPP_Solutions}
In the CPP solutions that we have seen so far the controller pool, as well as the mapping between a switch and a controller, is fixed based on some initial goals. However, as traffic conditions change dynamically, the initial configuration may no longer fulfill the intended objectives.      The adaptive CPP solutions propose a framework to dynamically vary the number of controllers and their switch mappings so as to  adapt to changing network conditions.

\subsection{Capacity}
Dixit et al.\cite{dixit2013towards} \cite{dixit2014elasticon} observed that real networks exhibit both temporal and spatial traffic variations. The temporal variations will arise due to traffic conditions depending on the time of day as well as in smaller time scales due to application characteristics. Spatial traffic variations will occur due to the flows generated by applications connected to the different switches. To handle such variations, the authors propose an \textit{elastic distributed controller framework} to balance controller load by migrating switches from overloaded controllers to lightly loaded ones. To address disruption of ongoing flows during switch migration, they proposed to use the \textit{equal}  controller mode (specified in \textit{OpenFlow} $v1.2$) while transitioning a controller from master to  slave. Further, they propose to grow or shrink the controller pool  if the controller load exceeds the upper or lower threshold. 

Cheng et. al.  \cite{cheng2015dha} \cite{cheng2016dynamic} use game theory to handle switch migration policies. The authors define \textit{network utility} as the number of events a controller can handle under available resources. Their objective is to maximize the overall network utility, which put differently means maximizing the resource utilization of each controller. They aim to load balance the controllers by re-assigning switches instead of adding/deleting controllers. The problem is formulated as a 0-1 integer linear program.  The authors initially propose to approximate the optimal value using the log-sum-exp function. In a latter work\cite{cheng2016dynamic}, the authors propose a non-cooperative game theoretic  approach. The controllers are envisioned as players and switches as commodities. The game is to trade switches among the controllers so as to maximize their profit (utility). Events emitted by a switch are broadcasted to all involved controllers of the events. Each of these controller computes its utilization changes if the switch is accepted. The controller whose change in utilization will be maximum wins and the switch is accordingly assigned to the controller.

Cello et al.\cite{cello2017balcon} propose to reduce the load imbalance among SDN controllers using switch migration. The SDN network is modeled as a vertex-weighted and edge-weighted graph where vertices represent SDN switches. The vertex weights and edge weights represent new flow arrivals at the switch and flow arrivals from other SDN domains respectively. The controller load balancing problem is reduced to a graph partitioning problem where the controller load is the sum of weight of the vertices in the partition plus the sum of weight of the edges directed towards the partition. The partition problem being NP-complete, the authors propose \textit{BalCon}, a heuristic algorithm. The algorithm monitors congestion at each controller and list potential candidate switches for migrations. The traffic pattern of these candidate switches are analyzed to form clusters of heavily connected switches and the best cluster is selected for migration. 

The controller load balancing techniques that we have discussed so far achieve their objective by reassignment of switches from an overloaded controller to a less loaded controller. Kyung et al\cite{kyung2015load} propose a load distribution paradigm based on a per-flow basis. In this approach, a switch forwards a new packet to its default controller. Depending on the load of the controller, the packet is either processed or forwarded to another controller. The authors claim that their approach outperforms the conventional switch migration technique in terms of blocking probability and controller capacity utilization ratio. The approach was evaluated using a Markov chain model. Although the per-flow load balancing approach looks attractive, it will incur additional cost in terms of controller-to-controller traffic which has been overlooked by the authors.

In \cite{sridharan2017multiple}, a fraction of the flows is distributed to multiple controllers instead of per-flow load balancing. The authors assume a switch to be mapped to multiple controllers, a so-called \textit{multiple mapping} approach. The mapping is determined based on observation of the network traffic over a period of time. The objective is to find the optimal multiple mapping for each switch such that the flow setup time with respect to resilience constraint is minimized. This approach provides better fairness to the switches as well as resilience to each switch in terms of controller failure. However, the main drawback of this scheme is that the switch needs to keep track of flows and accordingly distribute the flows to the multiple controllers.

In \cite{zhou2017load}, the authors introduce the problem of load oscillation where target controllers used to offload the load of overloaded controllers themselves becomes overloaded quickly. The authors identify that this is because the switch  migration strategy only considers the load of the switch but not the network status. A group of switches is chosen for migration such that the load of the target controller is close to the network average load. They also show that this process improves the time complexity of the balancing process. 

Although data center networks (DCNs) like Internet traffic have an unpredictable traffic pattern, their volume of traffic is considerably high. Thus load balancing the controllers by duplicating the database at each controller will not solve the problem of degradation in response time. Tam et. al. \cite{tam2011}, therefore, introduced the concept of \textit{devolved} controllers for DCNs, where each controller covers only a part of the total network. However, together they can respond to any request from any node. The authors show that the optimal allocation problem of the devolved controller is NP-hard and propose an approximate solution. 

A number of researchers\cite{liang2014balancing} have shown that the devolved controllers too can suffer from the problem of traffic imbalance. Gao et al.\cite{gao2017traffic} handle the problem of traffic load imbalance in devolved controllers by formulating it as a problem of balancing traffic load among $m$ partitions. The problem of migrating switches among the partitions is solved using integer programming. The authors also show that the load balancing problem in devolved controllers is NP-complete and design an $f$-approximation algorithm. The authors further show that their solution can be integrated with OpenFlow.

The work by Kim et. al. \cite{kim2018hes} is also targeted towards DCN. Their primary aim is to distribute the control traffic load of a SDN network among multiple controllers so as to reduce packet delay in the data plane as well as reduce the flow setup time. 
 
The authors propose two policies, forward and backward algorithm, to change the master controller of a switch and which of these policies will be activated depends on the CPU load. The forward algorithm changes the master controller of a switch generating the highest traffic while the backward algorithm changes the master controller of a switch generating lowest traffic. The authors observe that the time to change the master role of a switch depends on the controller load. Thus when the controller load is high the backward algorithm is used and when the controller load is low it is possible to change the master role of any switch and so the forward algorithm is used.

\subsection{Flow Setup Time}
Bari et al.\cite{bari2013dynamic} proposed a dynamic controller placement framework which consists of three modules - monitoring, reassignment, and provisioning. The monitoring module periodically pulls relevant statistics from the controllers, which is used by the reassignment module to decide whether to reassign a switch. The provision module provisions the controllers and makes the necessary switch to controller mappings. Their objective is to minimize the cost incurred during statistics collection, flow setup, synchronization and switch reassignment. The problem is formulated as an ILP and two heuristics are proposed. The first solution is a greedy approach based on the knapsack problem and the second solution is based on simulated annealing. In the objective function, the authors have not considered the inter-domain communication cost. 

He et al.\cite{hemodeling} build their work on top of \cite{bari2013dynamic}. The authors formulate the end-to-end flow setup time which basically consists of three components. These components include the flow setup time when the switches are in the same SDN domain when the switches are in different control domain and forwarding latency between the source and destination nodes. Their objective is to minimize the average flow setup time. The problem is formulated as an ILP where either the placement of the controllers or the mappings between switches and  controllers can change  or both the placement and assignments can change synchronously. The problem is solved using generic optimizers like CPLEX \cite{ilog2012cplex}.
    
\subsection{Capacity and Latency}
Rath et al.~\cite{rath2014optimal} capture the problem of addition/deletion of controllers so as to minimize the number of controllers with bounds on the
controller load and latency as a centralized non-linear function. As a centralized solution will not be scalable, the problem is reformulated as a  distributed individual optimization problem per controller. These individual problems being non-strictly competitive, it is solved as a non-zero sum game. The authors further, reason that to make the approach adaptive, the equations need to be solved iteratively. They, however, do not mention the frequency of the iterations.

Ul Haque et al.\cite{ul2015revisiting} propose to handle the variation in network traffic by dynamically changing the placement of controllers. The authors propose two search techniques to find the placement of controllers. In the \textit{open} search technique, given the location of the switches the entire region is searched to find the optimal placement of controllers. Although the open search technique results in maximum utilization of the controllers, the approach is not practical as controllers cannot be placed at any location. Thus a \textit{restricted} search problem is introduced, where placement of the controllers is restricted to a set of selected locations. The authors find the location of the controllers using a \textit{location searching algorithm} ($LSA$).  The algorithm partitions the network into sub-regions using latency bounds and it uses a heuristic to find the smallest enclosing circle, given the location of switches. The heuristic is based on Welzl's algorithm\cite{welzl1991smallest} with the controller being placed at the center of the circle.  

Zhou et al.~\cite{zhou2018elastic} formulate the problem of switch migration using the 3-D earth mover's distance model (EMD)\cite{rubner2000earth}, a well-known algorithm to measure difference between images. In EMD, images are represented using signatures and the objective is to find the optimal flow of earth between images subject to a number of constraints. The authors formulate the switch migration problem as a signature matching problem and extend the traditional EMD algorithm to reduce the difference in traffic load between controllers. The authors also consider capacity, bandwidth, and latency constraint during switch migration. A heuristic model is also proposed to reduce the time and computational complexity. The authors further claim that the results achieved outperform the results reported in one of their previous work \cite{wang2017switch} on the same problem. 

\subsection{Capacity and Cost}
An efficient switch migration technique should not only discuss the process to choose target switch and target controllers to effect the migration  but it should also consider switch migration cost.  Switch migration-based decision-making (\textit{SMDM})\cite{wang2017switch}  is a framework to build a tradeoff between migration costs and load imbalance among the controllers. The tradeoff is measured using a metric \textit{migration efficiency} which is a ratio of the difference in controller load before and after migration to the migration cost.  The switch migration problem is shown to be NP-hard and a greedy approach is proposed. In the proposed approach, a controller prefers to migrate a switch with less load and higher efficiency. The switch migration algorithm is triggered when the load diversity between two controllers exceeds a threshold. 

Hu et al.\cite{hu2018easm} propose Efficiency-Aware Switch Migration (EASM), a dynamic switch migration framework whose objectives are similar to that of SMDM\cite{wang2017switch}. However, unlike SMDM the migration efficiency is the  ratio of load balancing rate to switch migration cost. The migration cost is computed in terms of the average size of \textit{PACKET\_IN} message and migration hops between the switch and controller. The authors introduce \textit{load difference matrix} and \textit{trigger factor} which is a measure of the load variance of the controllers. If the trigger factor exceeds a threshold the switch migration process is invoked. 

Table \ref{tab:dynamic_cpp_solns} list the solution approaches on dynamic CPP discussed in this paper.

\begin{table*}[]
\centering
\small
\caption{Comparison of solution approaches on adaptive CPP}
\label{tab:dynamic_cpp_solns}
\begin{tabular}{|l|p{1cm}|p{4cm}|p{6cm}|l|}
\hline
\textbf{OBJECTIVE} & \textbf{REF.} & \textbf{\begin{tabular}[c]{@{}l@{}}PROBLEM\\ FORMULATION\end{tabular}} & \textbf{\begin{tabular}[c]{@{}l@{}}SHORT DESCRIPTION OF SOLUTION\\ APPROACH\end{tabular}} &\textbf{\begin{tabular}[c]{@{}l@{}} DYNAMIC\\ TECHNIQUE\end{tabular}} \\ \hline
\rowcolor[HTML]{C0C0C0}
\cellcolor[HTML]{C0C0C0} & \cite{dixit2013towards},\cite{dixit2014elasticon} &  Decision model(ElastiCon framework) and partitioning&  Heuristic algorithms are proposed to expand/shrink the control plane and change switch migration accordingly.&\begin{tabular}[c]{@{}l@{}} Add/delete controller\\ as well as\\ Switch reassignment\end{tabular} \\ \cline{2-5} 

\cellcolor[HTML]{C0C0C0} & \cite{cheng2015dha}& $0-1$ Integer linear programming to maximize network utility & Distributed Hopping algorithm is proposed to give approximate result (Markov approximation using log-sum-exp).& Switch reassignment \\ \cline{2-5}
 \rowcolor[HTML]{C0C0C0}
 \cellcolor[HTML]{C0C0C0}
  & \cite{cheng2016dynamic} &  Network utility maximization as partitioning problem & Used non-cooperative game theory approach to maximize profit resource utilization. & Switch reassignment \\ \cline{2-5}
  
 \cellcolor[HTML]{C0C0C0} & \cite{cello2017balcon} & \begin{tabular}[c]{@{}l@{}}Min-max problem\\Partitioning \end{tabular} & Reduced problem to a graph partitioning problem, and proposed a heuristic algorithm(BalCon). & Switch reassignment   \\ \cline{2-5}
 \rowcolor[HTML]{C0C0C0}
 \cellcolor[HTML]{C0C0C0} 
 & \cite{zhou2017load} & Mathematical model to dampen load oscillation & Switches and target controller are selected  taking into consideration the network status to avoid the load oscillation.  & Switch reassignment   \\ \cline{2-5}
 
  \cellcolor[HTML]{C0C0C0} & \cite{kyung2015load} & Markov chain model & Proposed approximation solution by reducing 2-D Markov chain to 1-D using normalization to the service rate. & Flow reassignment   \\ \cline{2-5} 
  \rowcolor[HTML]{C0C0C0}
   \cellcolor[HTML]{C0C0C0} 
  & \cite{sridharan2017multiple} &\begin{tabular}[c]{@{}l@{}} Queuing technique\\Linear programming \end{tabular} & Presented an analysis of the problem without proposing any specific solution.  & Flows reassignment  \\ \cline{2-5}
 \cellcolor[HTML]{C0C0C0}  & \cite{gao2017traffic} & Integer linear programming & Proposed multiple solution approaches and compared with different scenarios; 1)linear programming relaxation algorithm; 2) centralized greedy algorithm; 3) decentralized greedy algorithm & Switch reassignment. \\ \cline{2-5} 
  \rowcolor[HTML]{C0C0C0}
  \multirow{-15}{*}{\cellcolor[HTML]{C0C0C0}} & \cite{kim2018hes} & $0/1$-knapsack problem & Proposed classical heuristic approach for switch migration(Greedy based forward-backward algorithm). & Switch reassignment \\
 \multirow{-15}{*}{\cellcolor[HTML]{C0C0C0} CAPACITY} & \cite{mattos2016profiling} & Markov chain model& Proposed greedy algorithm based on the betweenness centrality. & Add/delete controller\\ \hline
 \rowcolor[HTML]{EFEFEF} 
 \cellcolor[HTML]{EFEFEF}
  
    &\cite{bari2013dynamic} &\begin{tabular}[c]{@{}l@{}} Integer linear programming\\Single
    source unsplittable flow problem\end{tabular} & Classical and meta- heuristic solutions based on greedy knapsack algorithm and simulated annealing. &\begin{tabular}[c]{@{}l@{}} Add/delete controller\\ \end{tabular}  \\ \cline{2-5} 
    
 \cellcolor[HTML]{EFEFEF} & \cite{hemodeling} & Mixed Integer linear programming & Generic optimizer solver (Gurobi\cite{gurobi2015gurobi}) is used to minimize FST. & \begin{tabular}[c]{@{}l@{}} Switch reassignment,\\ as well as\\ Controller reallocation\end{tabular} \\ \cline{2-5}

  \rowcolor[HTML]{EFEFEF} 
  \cellcolor[HTML]{EFEFEF} & \cite{wang2017efficient} & Assignment problem & Provided a solution for dynamically adjustment of controllers: using randomized fixed horizon control (RFHC) algorithm\cite{zhang2013moving} and modified stable matching with coalition game theory. & Add/delete controller \\ \cline{2-5}
   \multirow{-3}{*}{\cellcolor[HTML]{EFEFEF} {\begin{tabular}[c]{@{}l@{}}FLOW\\ SETUP\\ TIME\end{tabular}}}  &\cite{kumari2018optimizing} & linear programming & --- & Switch reassignment\\ \hline
 \rowcolor[HTML]{C0C0C0}
   \cellcolor[HTML]{C0C0C0}     
 & \cite{rath2014optimal} & \begin{tabular}[c]{@{}l@{}}Non linear problem and \\simplify to linear problem\end{tabular}  & Non zero sum game theory are applied (finding payoff of conflicting objectives).  &Add/delete controller \\ \cline{2-5} 
 \cellcolor[HTML]{C0C0C0}  
 & \cite{ul2015revisiting},\cite{ul2017large} & Network partitioning and mathematical model & Proposed two search technique to find the controller ; without restriction (open search) and restricted to location of switches (restricted search).  & Add/delete controller \\ \cline{2-5}
 
  \rowcolor[HTML]{C0C0C0} 
  \multirow{-4}{*}{  \cellcolor[HTML]{C0C0C0} \begin{tabular}[c]{@{}l@{}}CAPACITY\\AND\\ LATENCY\end{tabular}}& \cite{zhou2018elastic} &  Signature matching problem 3-D EMD model i.e., linear programming transportation problem & CPLEX is used to solve EMD Problem; A heuristic model is also proposed.& Switch reassignment   \\ \hline
    \rowcolor[HTML]{EFEFEF}
    \cellcolor[HTML]{EFEFEF}
 & \cite{wang2017switch} & Bin packing/Decision model; Switch Migration-based Decision-Making(SMDM)  & Propose greedy approach to switch selection based on load and efficiency. & Switch reassignment  \\ \cline{2-5} 

\multirow{-3}{*}{\cellcolor[HTML]{EFEFEF}\begin{tabular}[c]{@{}l@{}}CAPACITY\\                AND\\ COST\end{tabular}} & \cite{hu2018easm} & Decision model )Efficiency-Aware Switch Migration)
(EASM)& Used meta-heuristic(SA) to select the target controller and greedy approach to select switch  i.e., based on load and efficiency. &  Switch reassignment \\ \hline

\end{tabular}

\end{table*}
\section{CPP solutions for wireless networks}\label{sec:CPP_sol_Wireless}
SDN has found applications in different areas of wireless networks like resource sharing, scalability, and traffic engineering. Slicing is a popular technique used to slice traffic flow into separate subspaces. Resource sharing can be improved in SDN network by using the SDN controller to optimize the allocation of channels to the slices\cite{chaudet2013wireless}.  Scalability in terms of reduced turnaround time and robustness to link/node failures\cite{detti2013wireless},\cite{jin2013softcell} can be improved by using a distributed SDN control plane in wireless networks. Wireless network operators use SDN to extend support to traffic load balancing and energy-aware routing. In this survey, however, we limit our scope to those research works that propose solutions to the control placement problem in wireless networks.

Ahmad et al. \cite{crowd} put forward the concept of \textit{densenets}, dense and heterogeneous  wireless networks, to meet the increasing traffic demand in wireless networks. However, increasing the density of networks also considerably increase the load on the backhaul networks and existing network protocols thus requiring massive deployment of network devices. To address these issues introduced by densenets, the authors proposed a dynamic two-tier SDN controller hierarchy. In~\cite{auroux2014flow,auroux2015efficient}, the authors formulate the problem of optimizing the flow setup time in densenets, taking into consideration the stringent constraints of cellular networks. The problem is formulated as a mixed integer program and is shown to be NP-hard. The authors propose a heuristic, where nodes are \textit{greedily} assigned to \textit{regional} controllers depending on the number of unassigned neighbor nodes or number of \textit{unsatisfied} flows.

In wireless networks, the channel state of a link evolves over time. In order to activate the correct subset of links, a SDN controller obtains the channel state information (CSI) of each link in the network. Theoretically maintaining a single centralized controller will yield high performance since the controller can compute a globally optimal schedule but it is impractical due to the latency involved in obtaining the network-wide CSIs. Johnston et. al. \cite{johnston2015controller} analyze the impact of static placement of multiple controllers so as to optimize the throughput of a wireless network. However, as the channel state is time-varying, the optimal controller placement will depend on the channel transmission probabilities. The authors subsequently pose the problem as a dynamic controller placement problem and characterize the problem using linear programming. 

There is a growing interest in integrating SDN in the envisaged 5G network\cite{akyildiz2015softair} by separating the control plane and data plane functions within the  
Serving  GateWay (SGW) and Packet data GateWay (PGW). This separation enables the creation of S/PGW-C (control plane) and S/PGW-U  (user/data plane) responsible for managing the user plane and forwarding traffic respectively. Assuming that a mobile operator has deployed one S/PGW-C per region when a user equipment (UE) moves from one region to another, the UE context needs to be moved from the first S/PGW-C to the second one to maintain UE connectivity. This relocation of the UE context is costly and ideally, we would like to limit to a single S/PGW-C. However, limiting the number of S/PGW-C would result in its overloading, which in turn will increase the flow setup time. Ksentini et. al. \cite{ksentini2016using} formulate the problem of minimizing the number of S/PGW-C locations while maintaining an acceptable load on each of these control planes as an integer linear problem. Given the two conflicting objectives, the authors argue that a Pareto-optimal solution needs to be derived and they use Nash bargaining game theory approach to compute the Pareto-optimal solution. It may be noted that the authors have also proposed a game theoretic solution to the controller placement problem for wired networks in \cite{ksentini2016using1}.

Abdel et al.~\cite{abdel2017stochastic} address the controller placement problem in a cellular network controlled by a set of SDN controllers. The authors initially assume a wired connection between the controller and their controlled elements. Later they assume these connections to be wireless. For the wireless model, path loss and shadowing model are combined, retransmission is included and delay is considered a combination of propagation latency and transmission latency.  The authors propose two problem formulations. In the first, the average response time of the controllers is constrained to a predefined value. In the second formulation, the response time of each controller is restricted to be less than a pre-defined value. The problems are formulated as an integer linear programming. 

Abdel et al. ~\cite{abdel2017robust}, extend the controller placement problem in the cellular network to LTE networks. The system model is a set of evolved nodeBs (eNBs) forming a cellular network and a set of SDN controllers deployed to control the eNBs. The objective is to find the minimum number of controllers and their optimal locations to fulfill the eNB's delay requirements. The authors consider two flavors of the problem, a static and adaptive eNB-controller assignment. The static assignment problem is formulated using a stochastic model where the objective is to minimize the number of controllers while ensuring that the response time of each eNB does not exceed a threshold with a certain pre-defined probability. In the adaptive assignment problem, the eNB-controller assignment adapts to the variations in eNB request rates. The problem is formulated using as a two-stage stochastic problem, wherein the first stage  decision is static and in the second stage the distribution of the stochastic variables are assumed to be known and the assignment is done accordingly. These stochastic problems are converted to deterministic programs using sample average approximation and formulated as  mixed integer linear programming(MILP).

A summary of the solution approaches on wireless CPP discussed in this paper is given in Table \ref{tab:wireless_cpp_solns}.

. 

\begin{table*}[]
\centering
\small
\caption{Comparison of solution approaches on CPP for wireless networks}
\label{tab:wireless_cpp_solns}
\begin{adjustbox}{max width=\textwidth}
\begin{tabular}{|p{3.5cm}|p{1cm}|l|p{4.7 cm}|p{3cm}|}
\hline
 \rowcolor[HTML]{C0C0C0}
\textbf{Objective} & \textbf{Ref.} & \textbf{\begin{tabular}[c]{@{}l@{}}Problem \\ Formulation\end{tabular}} & \textbf{\begin{tabular}[c]{@{}l@{}}Short Description of \\Solution methodology\end{tabular}} & \textbf{\begin{tabular}[c]{@{}l@{}}Network\\ Type \end{tabular}}  \\ \hline
 Minimize number of controllers based on data flow processing(Data rate, $N2C$, Capacity)
  &  \cite{auroux2014flow} \cite{auroux2015efficient} & \begin{tabular}[c]{@{}l@{}}  MIQCP  \\(Mixed integer\\ programming with quadratic\\ terms in the constraints)\end{tabular}  &  LP Optimizer(Gurobi) is only applicable for prototype of LTE network.  Classical heuristic(greedy algorithm) is proposed for practical domain & Assumed unlimited data rate and zero latency based backbone infrastructure\\   \hline
 \rowcolor[HTML]{C0C0C0}
Proposed adaptive CPP  to optimize throughput& \cite{johnston2015controller} & Linear programming &  Used node queue length  for controller placement  and channel state information  for dynamic channel allocation  & Wireless Network  \\ \hline
Minimize the frequency of SGW-C relocations and $IMBL$ to ensure given $FST$ & \cite{ksentini2016using} & Integer linear programming & Propose {\em {Nash bargaining}} with {\em Threat points} to tradeoff between {\emph{conflicting nature}} of objectives, and provide {\em Pareto optimal}  solution & 5G network  \\ \hline
 \rowcolor[HTML]{C0C0C0}
 Minimize average $FST$ and per controller $FST$& \cite{abdel2017stochastic} & \begin{tabular}[c]{@{}l@{}}1)Mixed integer program\\ 2)Chance-constrained \\stochastic programs\end{tabular} & CPLEX is used  to get optimal solution & 1)Link between $N2C$ is wireless
2) Wireless network model based on combined path loss and  shadowing channel model \\ \hline
\begin{tabular}[c]{@{}l@{}}Minimize the active\\ controller with $FST$ \end{tabular}  & \cite{abdel2017robust} &\begin{tabular}[c]{@{}l@{}}1)Queuing model\\2)Chance-constrained \\stochastic programs(CCSP) \\ 3)Mixed integer programming\end{tabular} & Convert chance-constrained stochastic programs (intractable to solve)to Mixed integer programming and use CPLEX to solve this problem & LTE network  \\
\hline
 \rowcolor[HTML]{C0C0C0}
 Minimize the average latency and the average outage(link failure probability)
 probability&\cite{dvir2018wireless}  & Linear programming & Used Matlab $k$-Medoid function to find out $k$ clusters and brute force algorithm is applied to place the controller in clusters & Assumed that channel is based on a Rayleigh fading 
 with no line of sight \\ \hline
 Maximize reliability to meet latency constraint for joint placement of satellite and controller &\cite{liu2018joint} &Linear programming & Proposed meta-heuristics algorithm which adopt cluster based
 approximation approach in each iteration, (simulated annealing and clustering hybrid algorithm) & 5-G satellite integrated network \\ \hline
  \rowcolor[HTML]{C0C0C0}
  Minimize interference  &\cite{ashrafplacing} & ---& Proposed a greedy algorithm based on ranking function& Wireless mesh network \\ \hline
\end{tabular}
\end{adjustbox}
\end{table*}

\section{Characteristics of CPP Solutions}\label{sec:Ch_CPP_Sol}	

\subsection{Static CPP}
After reviewing the CPP solutions available in literature, we observed that the most general way of formulating CPP is to model the problem as a linear program (LP). Typically one of the CPP metrics (described in Section \ref{sec:PM}) is considered as an objective, and one or more of the other metrics are regarded as constraints. Latency is one of the metrics which is commonly optimized. A variant of the LP model is to consider multiple objectives by formulating it is a weighted objective function. The CPP problem is also formulated as a facility location problem\cite{heller2012controller}, which is an instance of mixed integer programming. Two variants of the facility location problem has been encountered, capacitated and uncapacitated. The capacitated form assume that the maximum number of switches that can be serviced by a controller is bounded by the processing capability of the controller. In the uncapaciated version, no such restriction is assumed. These linear programming problems are solved optimally by using well know \textit{optimizer tools} like CPLEX\cite{ilog2012cplex}, Gurobi\cite{gurobi2015gurobi} etc or by exploring the entire solution space. Solutions using LP solvers and exhaustive search cannot scale for large networks, thus a number of researchers have proposed solutions based on classical heuristic (greedy) or meta-heuristic approach. 

The second most popular way of formulating static CPP is to explore the entire solution space by defining the problem as a multi-objective, combinatorial optimization problem (MOCO). The solution approach here is to design a framework that returns the Pareto-optimal fronts. As finding these optimal fronts is known to be NP-hard, the strategy is thus to find an approximation using heuristics or evolutionary (meta-heuristic) algorithms. However, evolutionary algorithms suffer from the problem of getting stuck in local Pareto-optimal solutions, thus researchers have suggested modifications to the evolutionary algorithms to find accurate enough approximations. Another approach used to solve the multi-objective problem is using game theory, where the conflicting objectives are looked upon as players. 

 \begin{figure*}[thb]
    \centering
  
    \includegraphics[scale=0.7]{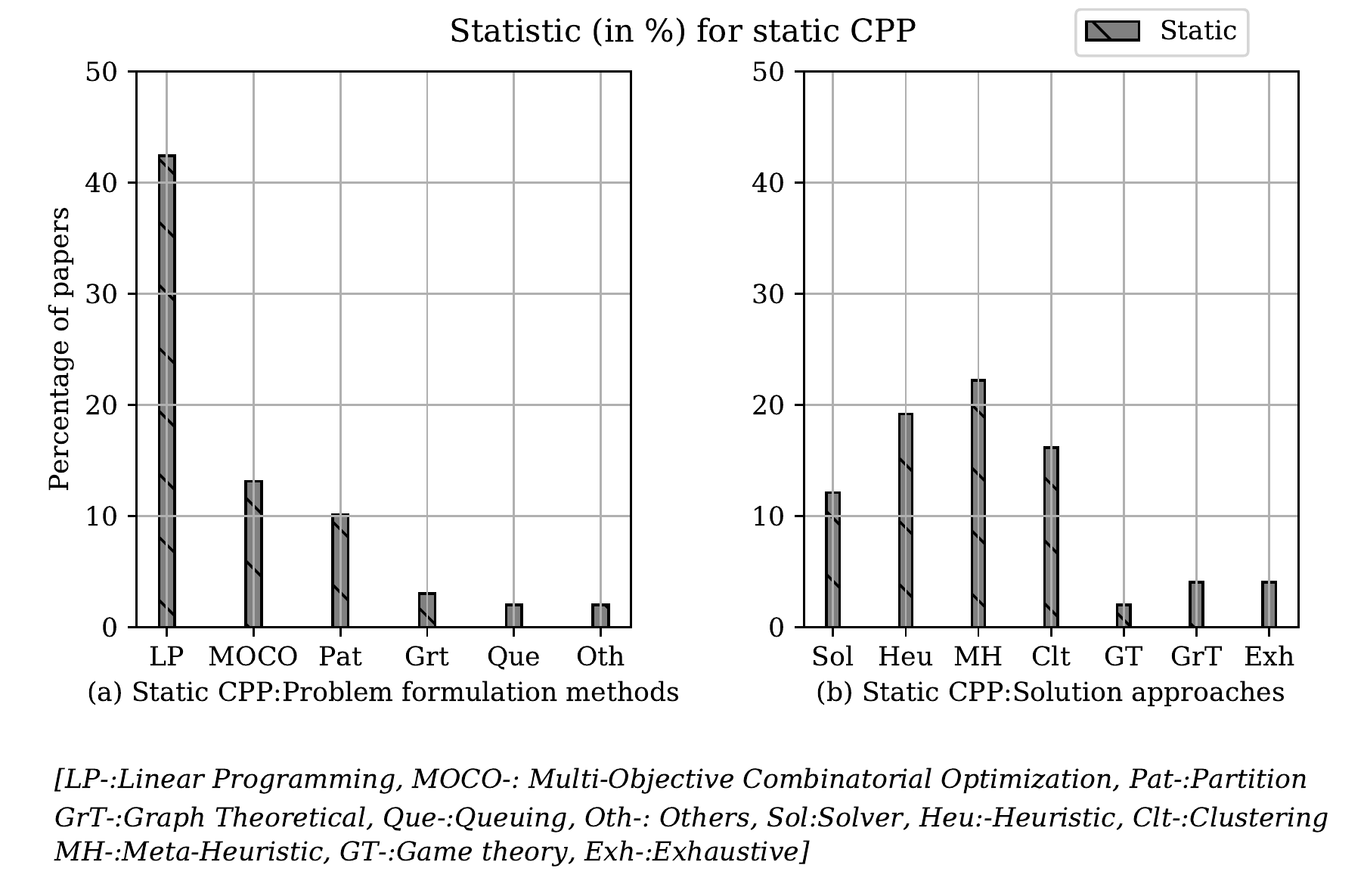}
    \caption{ Problem formulation and solution approaches for static controller placement problem}
    \label{fig:static_cpp}
  \end{figure*}

Other than the traditional way of formulating CPP, another popular approach is to reduce the complexity of the controller placement problem by partitioning the network into sub-networks.  The objectives are separately implemented in each sub-network such that the network as a whole achieves the prescribed objectives.  The solution to this problem formulation leverages on the inherent partitioning capability of clustering algorithms. A sizeable number of researcher work have proposed to use the $k$-means algorithm, a popular and effective clustering approach. The $k$-means algorithm, however, cannot be directly implemented in CPP, since it requires choosing initial random centers. The second concern with this algorithm is that it measures separation between the nodes using Euclidean distance. However, the use of Euclidean distance may not be always feasible in practical networks due to the absence of directly connected paths. Thus the CPP solutions based on $k$-means algorithm have proposed its variant to overcome these two issues. The other clustering techniques that are used to solve the network partitioning formulation are based on  affinity propagation, density, hierarchical and spectral clustering. The clustering techniques can also be studied based on the assumption of whether the numbers of clusters are known beforehand or not. Affinity propagation is a graph-based technique in which the number of partitions is not known a priori.  

There is also a class of static CPP approach where the problem is posed as a graph analysis problem. The solutions to these approaches are based on complex network analysis such as using centrality measures for controller placement or examination of cliques in a graph. There is also a small number of work which formulate the problem as a queueing theory problem. The controllers are regarded as servers and request from switches as clients demanding service. These problems are finally solved using heuristics.

Figure \ref{fig:static_cpp} show the different problem formulation methods and solution approaches used in static CPP. The percentage of papers is computed over the total number of 99 CPP papers that we have reviewed in this survey. 


  \begin{figure*}[thb]
     \centering
   
     \includegraphics[scale=0.7]{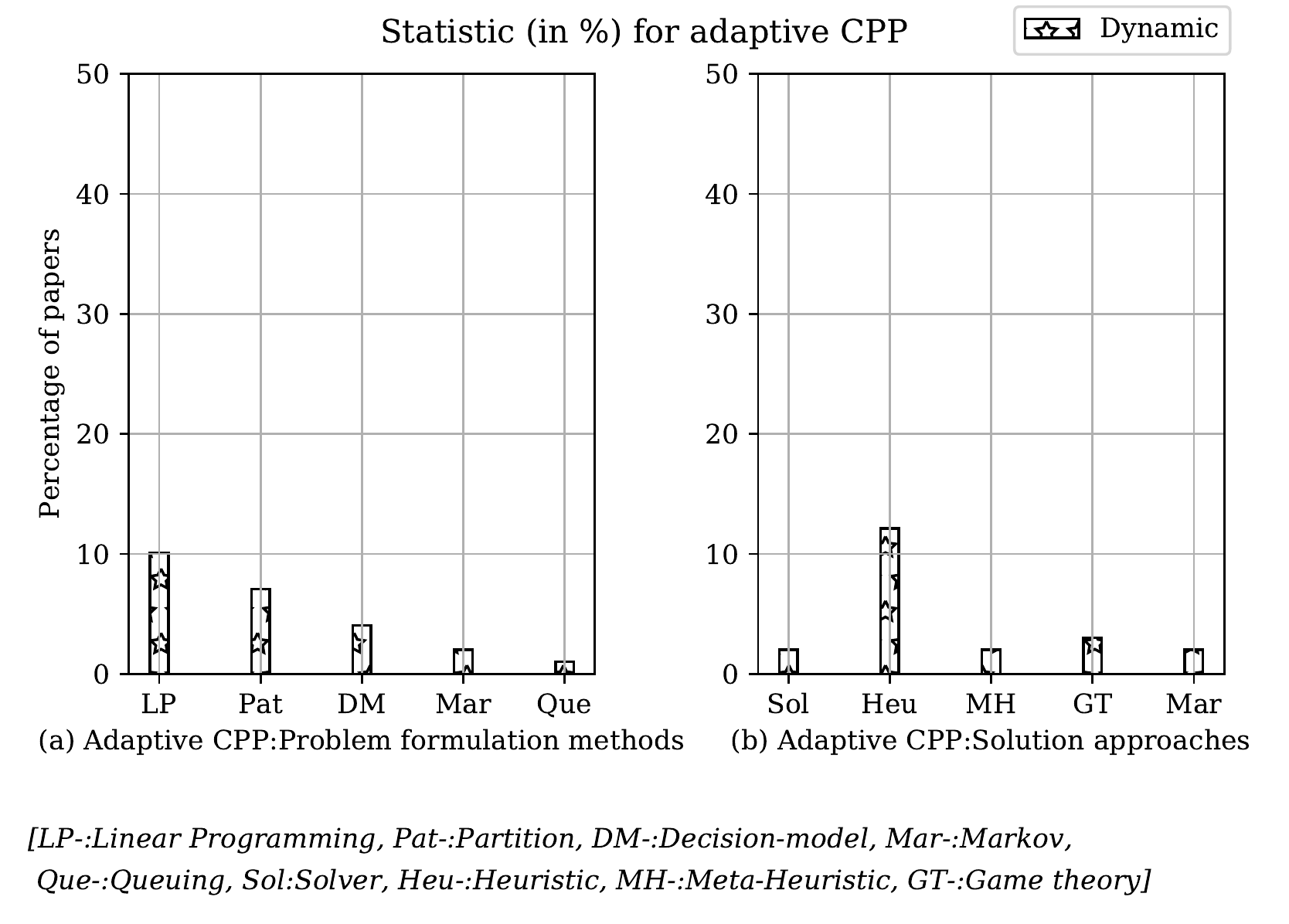}
     \caption{Problem formulation and solution approaches for dynamic controller placement problem}
     \label{fig:dynamic_cpp}
   \end{figure*}

 \begin{figure*}[thb]
     \centering
   
     \includegraphics[scale=0.7]{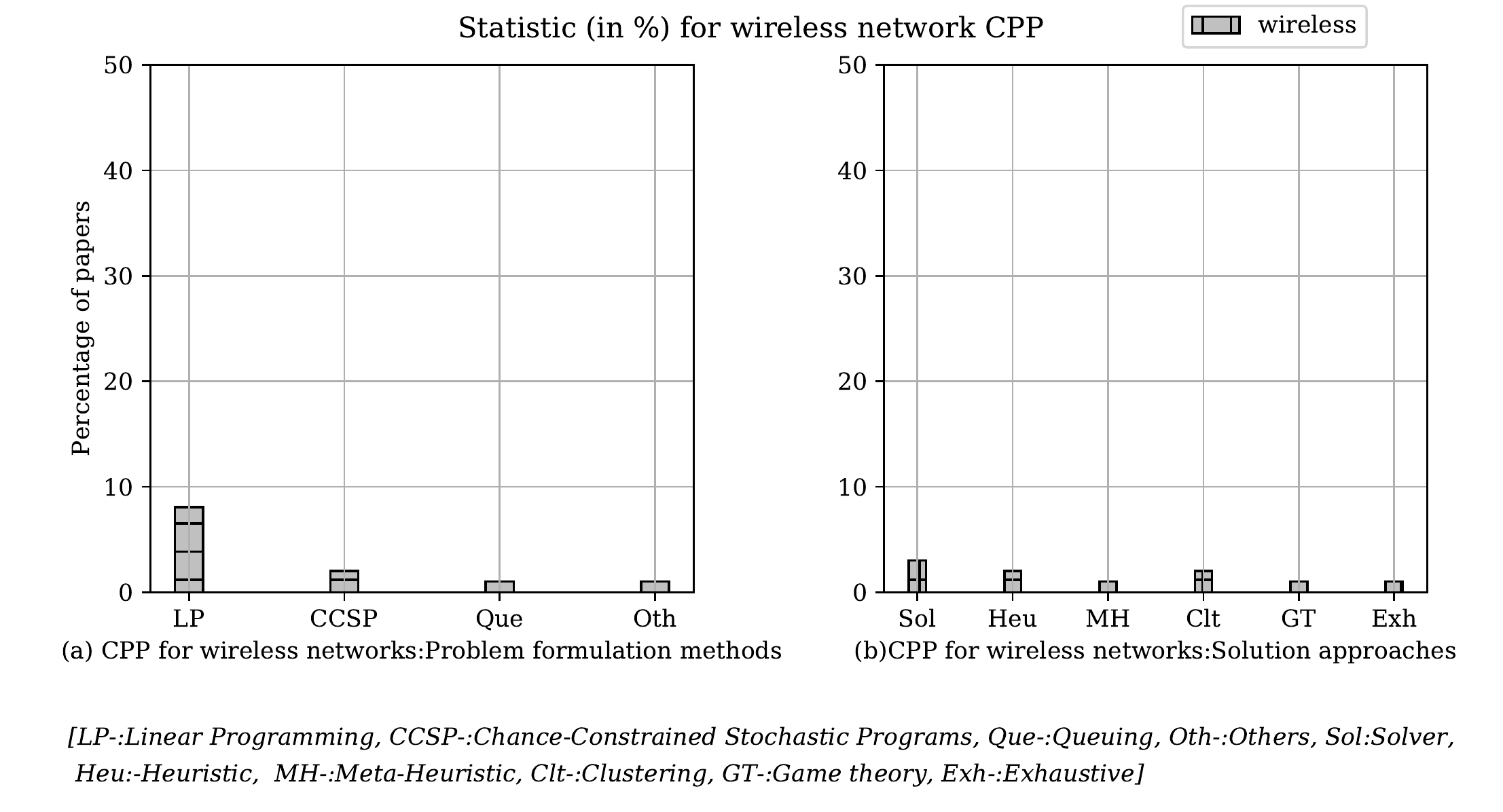}
     \caption{Problem formulation and solution approaches for controller placement problem in wireless domain}
     \label{fig:wireless_cpp}
   \end{figure*}
\subsection{Adaptive CPP}
Researchers realized that static placement of controller and static \textit{switch to controller} mapping cannot adapt to real networs where the traffic condition is highly dynamic. In adaptive CPP, the main focus is on handling the traffic load imbalance in SDN controllers called the \textit{controller load balancing} problem. Figure \ref{fig:dynamic_cpp} show the problem formulation and solution approaches used in adaptive CPP. The problem formulation, as well as the proposed solution used in adaptive CPP is to a large extent similar to those of static CPP. In addition to the problem formulation approach used in static CPP, two new problem formulation approaches are used in dynamic, \textit{decision} and Markov model. The adaptive CPP formulations using \textit{decision model} are those where the framework iteratively computes or monitor a parameter and based on the measured values makes a decision to trigger the solution process. For instance, a decision on whether to migrate a switch or not may be based on if the difference in measured load between controllers exceed a threshold. The different states of the controller like under-utilized, over-utilized or stable and the transition between these states are modeled using a Markov model. 

The main distinction of adaptive CPP solution approaches over static CPP is in the way the former approaches handle the varying traffic conditions. Researchers have proposed to handle the dynamism in SDN traffic by reassigning the flows. Flow rearrangement can be achieved by per-flow reassignment, reassigning fraction of flows, or all flows of a switch. The last approach, that is reassigning all of flows of a switch basically translates to migrating a switch from one controller to another. In such switch migrations, the solution approach should incorporate an approach to choose the target switch that will be reallocated and a target controller which will act as the new primary controller of the target switch. The different approaches to select the target switch and controller are random, utilization-based, delay sensitive, or reducing migration cost.

A second way of handling dynamic traffic in CPP is by relocating the controllers. In this case, we need a method to select the target controller as well as a target switch which will accommodate the controller.  

A third way of handling dynamism is by adding or deleting controllers based on requirement. A controller can be deleted if either the controller utilization is below a threshold or no switch is assigned to the controller. Likewise a new controller  may be added if the current set of controllers cannot handle the overall traffic in the network. In the case of controller addition, it is usually added close to an overloaded controller so that the cost of migrating flows is reduced.

The solution approaches employed in adaptive CPP are same as that of static CPP. An interesting observation is that although some researchers have modeled the problem using the partitioning method the solution approach do not feature clustering. Such a measure can be attributed to the fact that in adaptive CPP the primary concern of the solution approach is to consider when a migration procedure needs to be affected and not how to create clusters.


\subsection{CPP for wireless domain}
Figure \ref{fig:wireless_cpp} shows the various problem formulation method and solution approaches used in wireless domain. Problem formulation in wireless CPP need to consider varying traffic conditions similar to adaptive CPP. However, unlike adaptive CPP where the variation in traffic condition is primarily due to user behavior, in wireless networks the dynamic traffic conditions are largely due to node mobility. The second reason that contribute to variation in traffic conditions in wireless domain is due to the unpredictable channel conditions. Thus other than formulating the problem as a linear programming model, the second most popular CPP problem formulation approach in wireless networks is to model the uncertainty  using CCS (chance constrained stochastic) program. The CCS model is however characterized using linear programming. 

Researchers also proposed to handle the channel uncertainty based on queue length information and transmission probability. Yet another approach to handle varying channel conditions is by allowing  switches to periodically transmit the channel state information (CSI) to the controller.



\section{Discussion \& Future Research Direction }\label{sec:Disc}
\begin{figure}[thb]
   \centering
 
   \includegraphics[scale=0.58]{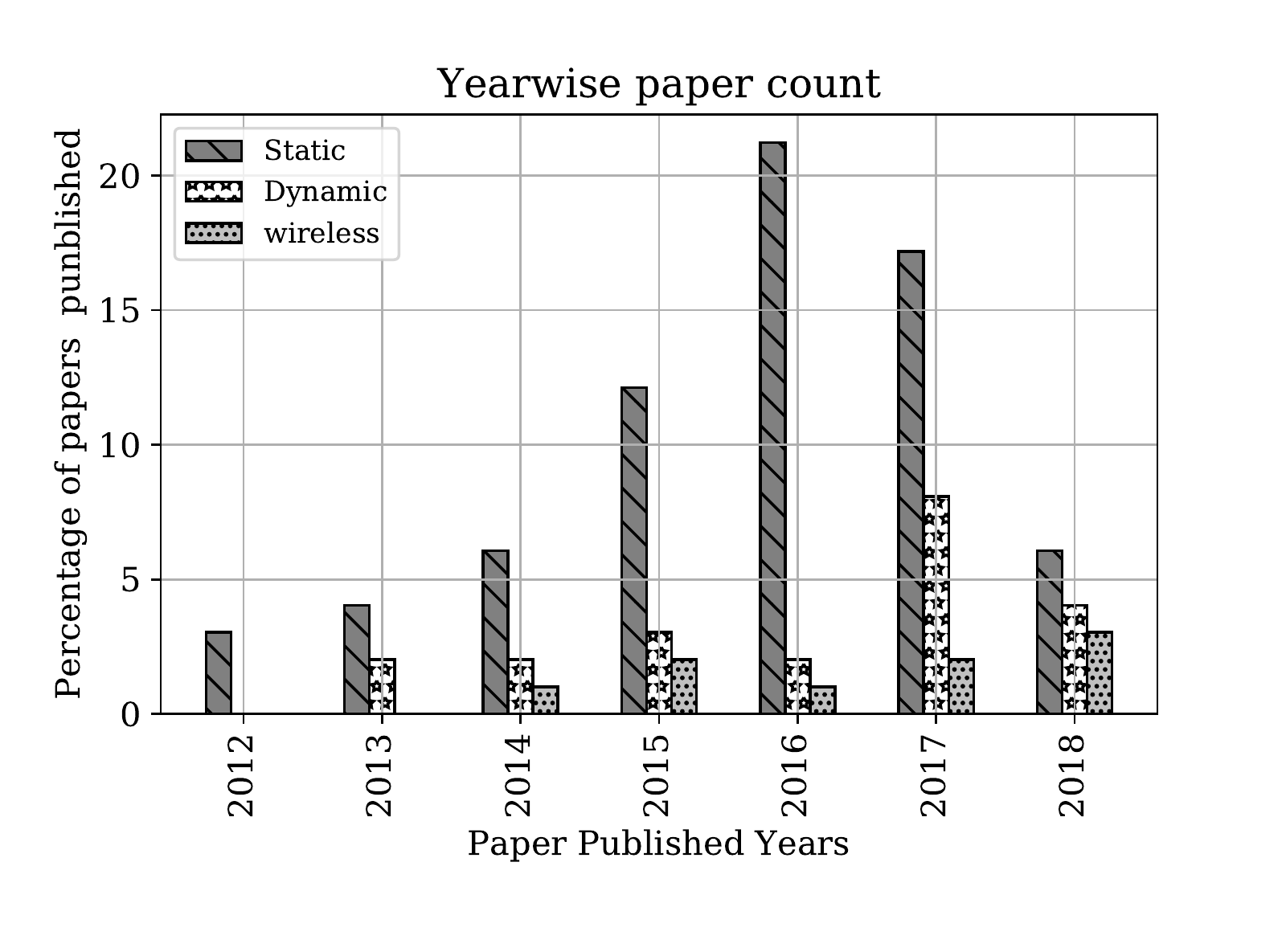}
   \caption{ Year-wise papers detail of Controller Placement Problem (CPP)}
   \label{fig:fig7}
 \end{figure}

In this survey, we have reviewed 99 research papers on CPP published between 2012 to mid of 2018, spread across three domains. Figure \ref{fig:fig7} provides a domain-wise statistics of these papers. The bulk of these papers is on static CPP but there is  a clear trend that researchers are now focusing more on adaptive CPP. The research effort on placement of controllers in wireless domain is the least. Similar to adaptive CPP, work in this domain is also gathering momentum. An insight from this survey, is that formulating the problem as a linear programming model is the most common approach across all the three domains. We found that greedy heuristics is the most popular solution approach followed by meta-heuristic approaches. 

A clear research challenge in CPP, is to find a traffic agnostic controller placement. Considering the real world dynamics, we feel that adaptive CPP is the way forward. The issues of wireless CPP are comparable to those of adaptive CPP. We identify the possible research directions in these two domains:

\begin{itemize}
\item \textbf{Controller relocation:} In the course of our survey, we found that controller relocation technique is used to optimize a number of parameters. However, we feel that relocating the controller within its own domain will not help in improving traffic load imbalance across the SDN domains. Controller relocation at most can improve the flow setup time. Inter-domain relocation of controller may help in improving the load imbalance metric provided the controllers are not of the same capacity. However, this will require a centralized approach where an oracle is aware of the capacity of all the controllers. Further controller load balancing will only be effective, if in-band signaling is used. Hence, we feel that controller relocation alone will not be an effective approach.

\item \textbf{Switch migration overhead}
Switch reassignment strategy is a popular and effective technique to deal with dynamic network traffic. This approach has the dual effect of limiting the flow setup time and improve load imbalance. However, the reassignment comes with a cost and we found that very few work have considered the overhead involved in switch migration. The other issue is that if the traffic has a seasonal trend then there is a possibility that a switch may toggle between two controllers, a so called ping-pong effect observed in cellular networks.   

\item 	\textbf{Target switch/controller selection:} An additional issue with switch migration that most authors have not considered is a systematic way of selecting the target switch and controller. The target switch is selected randomly and the target controller from the neighbor. Some authors have suggested  to select an \textit{under-utilized} controller but in case multiple such controllers exist there is no clear strategy. There is a limited number of work that have proposed to consider  migration efficiency during switch/controller reassignment. The efficiency is computed based on the presumption that traffic conditions will remain same in the next epoch notwithstanding the fact that future traffic does not necessarily depend on past and current traffic condition.  

\item \textbf{Traffic prediction based CPP} 
During our survey, we found that decision on switch migration is basically based on current or past traffic conditions. Researchers have shown that though Internet traffic show a seasonal trend for large time scales (hours, weeks, etc) it is highly chaotic for smaller time scales (seconds or minutes). Hence developing methods to predict traffic behavior in the next epoch and then deciding on switch reassignment will prove to be much more beneficial. The challenge here would be in predicting the traffic of real SDN networks. Modeling Internet traffic is still an open research challenge but in case of SDN networks, we can do in on a network-by-network basis by collecting real traffic and developing suitable learning models.

\item \textbf{Flow relocation:} We have observed that a few researchers instead of migrating a switch have proposed to distribute a fraction of the flows to other controllers. There are two crucial issues that need to be considered in such an approach. The first is keeping track of the flows and discover the correct division of flows is a non-trivial task. Moreover, the switch needs to keep track of the flows which means some intelligence is being pushed back to the infrastructure layer which is against the SDN paradigm. The second issue is that in case flow relocation is implemented, a switch will need to have multiple masters simultaneously.

\item \textbf{Addition/deletion of controller(s):} Researchers have proposed addition or deletion of controllers to enhance the capacity or reduce the energy usage of a SDN network. In such cases, an issue that has been overlooked is the cost of turning on or switching off a controller in terms of time and energy. It may be less productive to switch off controllers for short duration, if they need to be powered up again.
 
\item \textbf{Mobility aware based CPP:} SDN has been proposed for the next-generation future network like 5G where user mobility impacts the performance of the network. Literature have revealed that traditional mobility model may not be valid for such networks since in these networks femto cells will handle majority of the data transfer. Thus mobility-aware placement of controllers is an open research issue.


\item \textbf{CPP for other network domains}
Researchers have been contemplating deployment of SDN architecture in other network domains such as IoT, sensor network etc. These networks have their own typical characteristics in terms of range, topology, energy limitations etc. Thus there is a possibility to explore the challenges in placement of SDN controllers in such networks.
\end{itemize}

\section{Conclusion}\label{Con}
This paper presents a quantitative as well as qualitative perspective of controller placement problem (CPP) in software-defined networks (SDN). The aim of this survey is to provide a comprehensive view of CPP, identify the research gaps and provide future research directions. In order to provide a complete coverage on the subject, the survey begins with a brief introduction on SDN, shortcomings of a single SDN controller and  evolution of a multi-controller(distributed) SDN architecture. The paper highlights the issues related to a distributed SDN architecture and introduces the problem of controller placement. A two pronged strategy is adopted to examine the problem of controller placement - CPP objectives and CPP solutions. The state-of-the-art on CPP solutions is discussed based on whether the solution proposed is for wired or wireless network. In regard to the nature of mapping between a switch and its controller, the CPP solutions for wired networks are further sub-divided into static and adaptive. Finally, we generalize the characteristics of the different CPP solutions. With regard to future research direction, we suggest examining the issues with controller relocation, controller addition/deletion, challenges related to switch migration, examining issues related to mobility and CPP for other network domains such as IoT and sensor networks.  

\ifCLASSOPTIONcaptionsoff
  \newpage
\fi

\bibliographystyle{IEEEtran}
\bibliography{bare_adv}

\begin{thebibliography}{100}
\providecommand{\url}[1]{#1}
\csname url@samestyle\endcsname
\providecommand{\newblock}{\relax}
\providecommand{\bibinfo}[2]{#2}
\providecommand{\BIBentrySTDinterwordspacing}{\spaceskip=0pt\relax}
\providecommand{\BIBentryALTinterwordstretchfactor}{4}
\providecommand{\BIBentryALTinterwordspacing}{\spaceskip=\fontdimen2\font plus
\BIBentryALTinterwordstretchfactor\fontdimen3\font minus
  \fontdimen4\font\relax}
\providecommand{\BIBforeignlanguage}[2]{{%
\expandafter\ifx\csname l@#1\endcsname\relax
\typeout{** WARNING: IEEEtran.bst: No hyphenation pattern has been}%
\typeout{** loaded for the language `#1'. Using the pattern for}%
\typeout{** the default language instead.}%
\else
\language=\csname l@#1\endcsname
\fi
#2}}
\providecommand{\BIBdecl}{\relax}
\BIBdecl

\bibitem{jain2013b4}
S.~Jain, A.~Kumar, S.~Mandal, J.~Ong, L.~Poutievski, A.~Singh, S.~Venkata,
  J.~Wanderer, J.~Zhou, M.~Zhu \emph{et~al.}, ``B4: Experience with a
  globally-deployed software defined wan,'' \emph{ACM SIGCOMM Computer
  Communication Review}, vol.~43, no.~4, pp. 3--14, 2013.

\bibitem{211243}
\BIBentryALTinterwordspacing
M.~Dalton, D.~Schultz, J.~Adriaens, A.~Arefin, A.~Gupta, B.~Fahs,
  D.~Rubinstein, E.~C. Zermeno, E.~Rubow, J.~A. Docauer, J.~Alpert, J.~Ai,
  J.~Olson, K.~DeCabooter, M.~de~Kruijf, N.~Hua, N.~Lewis, N.~Kasinadhuni,
  R.~Crepaldi, S.~Krishnan, S.~Venkata, Y.~Richter, U.~Naik, and A.~Vahdat,
  ``Andromeda: Performance, isolation, and velocity at scale in cloud network
  virtualization,'' in \emph{15th {USENIX} Symposium on Networked Systems
  Design and Implementation ({NSDI} 18)}.\hskip 1em plus 0.5em minus
  0.4em\relax Renton, WA: {USENIX} Association, 2018, pp. 373--387. [Online].
  Available: \url{https://www.usenix.org/conference/nsdi18/presentation/dalton}
\BIBentrySTDinterwordspacing

\bibitem{hong2013achieving}
C.-Y. Hong, S.~Kandula, R.~Mahajan, M.~Zhang, V.~Gill, M.~Nanduri, and
  R.~Wattenhofer, ``Achieving high utilization with software-driven wan,'' in
  \emph{ACM SIGCOMM Computer Communication Review}, vol.~43, no.~4.\hskip 1em
  plus 0.5em minus 0.4em\relax ACM, 2013, pp. 15--26.

\bibitem{WindowsServer2016}
Windows Server 2016 [accessed on 29/05/2018] available
  at:\url{https://www.microsoft.com/en-in/cloud-platform/software-defined-networking}.

\bibitem{ACI}
ACI (Application-Centric-Infrastructure [accessed on 05/6/2018] available
  at:\url{
  https://www.cisco.com/c/en_in/solutions/data-center-virtualization/application-centric-infrastructure/index.html#~stickynav=2}.

\bibitem{beheshti2012fast}
N.~Beheshti and Y.~Zhang, ``Fast failover for control traffic in
  software-defined networks,'' in \emph{Global Communications Conference
  (GLOBECOM), 2012 IEEE}.\hskip 1em plus 0.5em minus 0.4em\relax IEEE, 2012,
  pp. 2665--2670.

\bibitem{heller2012controller}
B.~Heller, R.~Sherwood, and N.~McKeown, ``The controller placement problem,''
  in \emph{Proceedings of the first workshop on Hot topics in software defined
  networks}.\hskip 1em plus 0.5em minus 0.4em\relax ACM, 2012, pp. 7--12.

\bibitem{nunes2014survey}
B.~A.~A. Nunes, M.~Mendonca, X.-N. Nguyen, K.~Obraczka, and T.~Turletti, ``A
  survey of software-defined networking: Past, present, and future of
  programmable networks,'' \emph{IEEE Communications Surveys \& Tutorials},
  vol.~16, no.~3, pp. 1617--1634, 2014.

\bibitem{hu2014survey}
F.~Hu, Q.~Hao, and K.~Bao, ``A survey on software-defined network and openflow:
  From concept to implementation,'' \emph{IEEE Communications Surveys \&
  Tutorials}, vol.~16, no.~4, pp. 2181--2206, 2014.

\bibitem{jarraya2014survey}
Y.~Jarraya, T.~Madi, and M.~Debbabi, ``A survey and a layered taxonomy of
  software-defined networking,'' \emph{IEEE Communications Surveys \&
  Tutorials}, vol.~16, no.~4, pp. 1955--1980, 2014.

\bibitem{xia2015survey}
W.~Xia, Y.~Wen, C.~H. Foh, D.~Niyato, and H.~Xie, ``A survey on
  software-defined networking,'' \emph{IEEE Communications Surveys \&
  Tutorials}, vol.~17, no.~1, pp. 27--51, 2015.

\bibitem{farhady2015software}
H.~Farhady, H.~Lee, and A.~Nakao, ``Software-defined networking: A survey,''
  \emph{Computer Networks}, vol.~81, pp. 79--95, 2015.

\bibitem{kreutz2015software}
D.~Kreutz, F.~M. Ramos, P.~E. Verissimo, C.~E. Rothenberg, S.~Azodolmolky, and
  S.~Uhlig, ``Software-defined networking: A comprehensive survey,''
  \emph{Proceedings of the IEEE}, vol. 103, no.~1, pp. 14--76, 2015.

\bibitem{xie2015control}
J.~Xie, D.~Guo, Z.~Hu, T.~Qu, and P.~Lv, ``Control plane of software defined
  networks: A survey,'' \emph{Computer communications}, vol.~67, pp. 1--10,
  2015.

\bibitem{oktian2017distributed}
Y.~E. Oktian, S.~Lee, H.~Lee, and J.~Lam, ``Distributed sdn controller system:
  A survey on design choice,'' \emph{Computer Networks}, vol. 121, pp.
  100--111, 2017.

\bibitem{murat2017distributed}
\BIBentryALTinterwordspacing
M.~Karakus and A.~Durresi, ``A survey: Control plane scalability issues and
  approaches in software-defined networking (sdn),'' \emph{Computer Networks},
  vol. 112, pp. 279 -- 293, 2017. [Online]. Available:
  \url{http://www.sciencedirect.com/science/article/pii/S138912861630411X}
\BIBentrySTDinterwordspacing

\bibitem{bannour2017distributed}
F.~Bannour, S.~Souihi, and A.~Mellouk, ``Distributed sdn control: Survey,
  taxonomy and challenges,'' \emph{IEEE Communications Surveys \& Tutorials},
  2017.

\bibitem{trois2016survey}
C.~Trois, M.~D. Del~Fabro, L.~C. de~Bona, and M.~Martinello, ``A survey on sdn
  programming languages: Toward a taxonomy,'' \emph{IEEE Communications Surveys
  \& Tutorials}, vol.~18, no.~4, pp. 2687--2712, 2016.

\bibitem{fonseca2017survey}
P.~Fonseca and E.~Mota, ``A survey on fault management in software-defined
  networks,'' \emph{IEEE Communications Surveys \& Tutorials}, 2017.

\bibitem{scott2013sdn}
S.~Scott-Hayward, G.~O'Callaghan, and S.~Sezer, ``Sdn security: A survey,'' in
  \emph{Future Networks and Services (SDN4FNS), 2013 IEEE SDN For}.\hskip 1em
  plus 0.5em minus 0.4em\relax IEEE, 2013, pp. 1--7.

\bibitem{yan2016software}
Q.~Yan, F.~R. Yu, Q.~Gong, and J.~Li, ``Software-defined networking (sdn) and
  distributed denial of service (ddos) attacks in cloud computing environments:
  A survey, some research issues, and challenges,'' \emph{IEEE Communications
  Surveys \& Tutorials}, vol.~18, no.~1, pp. 602--622, 2016.

\bibitem{ahmad2015security}
I.~Ahmad, S.~Namal, M.~Ylianttila, and A.~Gurtov, ``Security in software
  defined networks: A survey,'' \emph{IEEE Communications Surveys \&
  Tutorials}, vol.~17, no.~4, pp. 2317--2346, 2015.

\bibitem{karakus2017quality}
M.~Karakus and A.~Durresi, ``Quality of service (qos) in software defined
  networking (sdn): A survey,'' \emph{Journal of Network and Computer
  Applications}, vol.~80, pp. 200--218, 2017.

\bibitem{badirzadeh2018survey}
A.~Badirzadeh and S.~Jamali, ``A survey on load balancing methods in software
  defined network,'' \emph{Networking and Communication Engineering}, vol.~10,
  no.~2, pp. 21--27, 2018.

\bibitem{li2017load}
L.~Li and Q.~Xu, ``Load balancing researches in sdn: A survey,'' in
  \emph{Electronics Information and Emergency Communication (ICEIEC), 2017 7th
  IEEE International Conference on}.\hskip 1em plus 0.5em minus 0.4em\relax
  IEEE, 2017, pp. 403--408.

\bibitem{mendiola2016survey}
A.~Mendiola, J.~Astorga, E.~Jacob, and M.~Higuero, ``A survey on the
  contributions of software-defined networking to traffic engineering,''
  \emph{IEEE Communications Surveys \& Tutorials}, vol.~19, no.~2, pp.
  918--953, 2016.

\bibitem{zehra2017survey}
U.~Zehra and M.~A. Shah, ``A survey on resource allocation in software defined
  networks (sdn),'' in \emph{Automation and Computing (ICAC), 2017 23rd
  International Conference on}.\hskip 1em plus 0.5em minus 0.4em\relax IEEE,
  2017, pp. 1--6.

\bibitem{huang2016survey}
T.~Huang, F.~R. Yu, C.~Zhang, J.~Liu, J.~Zhang, and Y.~Liu, ``A survey on
  large-scale software defined networking (sdn) testbeds: Approaches and
  challenges,'' \emph{IEEE Communications Surveys \& Tutorials}, vol.~19,
  no.~2, pp. 891--917, 2016.

\bibitem{tuysuz2017survey}
M.~F. Tuysuz, Z.~K. Ankarali, and D.~G{\"o}z{\"u}pek, ``A survey on energy
  efficiency in software defined networks,'' \emph{Computer Networks}, vol.
  113, pp. 188--204, 2017.

\bibitem{rawat2017software}
D.~B. Rawat and S.~R. Reddy, ``Software defined networking architecture,
  security and energy efficiency: A survey,'' \emph{IEEE Communications Surveys
  \& Tutorials}, vol.~19, no.~1, pp. 325--346, 2017.

\bibitem{liang2015wireless}
C.~Liang and F.~R. Yu, ``Wireless network virtualization: A survey, some
  research issues and challenges,'' \emph{IEEE Communications Surveys \&
  Tutorials}, vol.~17, no.~1, pp. 358--380, 2015.

\bibitem{haque2016wireless}
I.~T. Haque and N.~Abu-Ghazaleh, ``Wireless software defined networking: A
  survey and taxonomy,'' \emph{IEEE Communications Surveys \& Tutorials},
  vol.~18, no.~4, pp. 2713--2737, 2016.

\bibitem{thyagaturu2016software}
A.~S. Thyagaturu, A.~Mercian, M.~P. McGarry, M.~Reisslein, and W.~Kellerer,
  ``Software defined optical networks (sdons): A comprehensive survey,''
  \emph{IEEE Communications Surveys \& Tutorials}, vol.~18, no.~4, pp.
  2738--2786, 2016.

\bibitem{wang2017controller}
G.~Wang, Y.~Zhao, J.~Huang, and W.~Wang, ``The controller placement problem in
  software defined networking: a survey,'' \emph{IEEE Network}, vol.~31, no.~5,
  pp. 21--27, 2017.

\bibitem{yoon2017controller}
S.-K. Yoon, Z.~Khalib, N.~Yaakob, and A.~Amir, ``Controller placement
  algorithms in software defined network-a review of trends and challenges,''
  in \emph{MATEC Web of Conferences}, vol. 140.\hskip 1em plus 0.5em minus
  0.4em\relax EDP Sciences, 2017, p. 01014.

\bibitem{singh2018survey}
A.~K. Singh and S.~Srivastava, ``A survey and classification of controller
  placement problem in sdn,'' \emph{International Journal of Network
  Management}, p. e2018, 2018.

\bibitem{mckeown2008openflow}
N.~McKeown, T.~Anderson, H.~Balakrishnan, G.~Parulkar, L.~Peterson, J.~Rexford,
  S.~Shenker, and J.~Turner, ``Openflow: enabling innovation in campus
  networks,'' \emph{ACM SIGCOMM Computer Communication Review}, vol.~38, no.~2,
  pp. 69--74, 2008.

\bibitem{doria2010forwarding}
A.~Doria, J.~H. Salim, R.~Haas, H.~Khosravi, W.~Wang, L.~Dong, R.~Gopal, and
  J.~Halpern, ``Forwarding and control element separation (forces) protocol
  specification,'' Tech. Rep., 2010.

\bibitem{smith2014opflex}
M.~Smith, M.~Dvorkin, Y.~Laribi, V.~Pandey, P.~Garg, and N.~Weidenbacher,
  ``Opflex control protocol,'' \emph{IETF}, 2014.

\bibitem{saint2009xmpp}
P.~Saint-Andre, K.~Smith, and R.~Tron{\c{c}}on, \emph{XMPP: the definitive
  guide}.\hskip 1em plus 0.5em minus 0.4em\relax " O'Reilly Media, Inc.", 2009.

\bibitem{bakhshi2017state}
T.~Bakhshi, ``State of the art and recent research advances in software defined
  networking,'' \emph{Wireless Communications and Mobile Computing}, vol. 2017,
  2017.

\bibitem{zhou2014rest}
W.~Zhou, L.~Li, M.~Luo, and W.~Chou, ``Rest api design patterns for sdn
  northbound api,'' in \emph{Advanced Information Networking and Applications
  Workshops (WAINA), 2014 28th International Conference on}.\hskip 1em plus
  0.5em minus 0.4em\relax IEEE, 2014, pp. 358--365.

\bibitem{cummins2013enterprise}
H.~Cummins and T.~Ward, \emph{Enterprise OSGi in action: with examples using
  Apache Aries}.\hskip 1em plus 0.5em minus 0.4em\relax Manning Publications
  Co., 2013.

\bibitem{yin2012sdni}
H.~Yin, H.~Xie, T.~Tsou, D.~Lopez, P.~Aranda, and R.~Sidi, ``Sdni: A message
  exchange protocol for software defined networks (sdns) across multiple
  domains,'' \emph{IETF draft, work in progress}, 2012.

\bibitem{lin2015west}
P.~Lin, J.~Bi, S.~Wolff, Y.~Wang, A.~Xu, Z.~Chen, H.~Hu, and Y.~Lin, ``A
  west-east bridge based sdn inter-domain testbed,'' \emph{IEEE Communications
  Magazine}, vol.~53, no.~2, pp. 190--197, 2015.

\bibitem{benamrane2017east}
F.~Benamrane, R.~Benaini \emph{et~al.}, ``An east-west interface for
  distributed sdn control plane: Implementation and evaluation,''
  \emph{Computers \& Electrical Engineering}, vol.~57, pp. 162--175, 2017.

\bibitem{lin2016control}
S.-C. Lin, P.~Wang, and M.~Luo, ``Control traffic balancing in software defined
  networks,'' \emph{Computer Networks}, vol. 106, pp. 260--271, 2016.

\bibitem{gude2008nox}
N.~Gude, T.~Koponen, J.~Pettit, B.~Pfaff, M.~Casado, N.~McKeown, and
  S.~Shenker, ``Nox: towards an operating system for networks,'' \emph{ACM
  SIGCOMM Computer Communication Review}, vol.~38, no.~3, pp. 105--110, 2008.

\bibitem{erickson2013beacon}
D.~Erickson, ``The beacon openflow controller,'' in \emph{Proceedings of the
  second ACM SIGCOMM workshop on Hot topics in software defined
  networking}.\hskip 1em plus 0.5em minus 0.4em\relax ACM, 2013, pp. 13--18.

\bibitem{McNettle}
McNettle [accessed on 05/12/2017] available at:\url{
  haskell.cs.yale.edu/wp-content/uploads/2012/08/yale-tr-1468.pdf}.

\bibitem{ng2010maestro}
E.~Ng, Z.~Cai, and A.~Cox, ``Maestro: A system for scalable openflow control,''
  \emph{Rice University, Houston, TX, USA, TSEN Maestro-Techn. Rep, TR10-08},
  2010.

\bibitem{RouteFlow}
RouteFlow [accessed on 05/12/2017] available
  at:\url{https://github.com/routeflow}.

\bibitem{POX}
POX [accessed on 05/12/2017] available at: \url{
  http://www.noxrepo.org/pox/about-pox/}.

\bibitem{Ryu}
Ryu [accessed on 05/12/2017] available at: \url{http://osrg.github.com/ryu/}.

\bibitem{Beacon}
Beacon [accessed on 05/12/2017] available at:
  \url{https://openflow.stanford.edu/display/Beacon/Home}.

\bibitem{Floodlight}
Floodlight [accessed on 05/12/2017] available at: \url{
  http://Floodlight.openflowhub.org/}.

\bibitem{berde2014onos}
P.~Berde, M.~Gerola, J.~Hart, Y.~Higuchi, M.~Kobayashi, T.~Koide, B.~Lantz,
  B.~O'Connor, P.~Radoslavov, W.~Snow \emph{et~al.}, ``Onos: towards an open,
  distributed sdn os,'' in \emph{Proceedings of the third workshop on Hot
  topics in software defined networking}.\hskip 1em plus 0.5em minus
  0.4em\relax ACM, 2014, pp. 1--6.

\bibitem{OpenDaylight}
OpenDaylight [accessed on 05/12/2017] available at:
  \url{https://www.opendaylight.org/}.

\bibitem{Trema}
Trema [accessed on 05/12/2017] available at:\url{
  https://github.com/trema/trema}.

\bibitem{MUL}
MUL [accessed on 05/12/2017] available at:\url{ http://www.openmul.org/}.

\bibitem{opencontrail}
Opencontrail [accessed on 05/12/2017] available at:
  \url{http://www.opencontrail.org/}.

\bibitem{curtis2011devoflow}
A.~R. Curtis, J.~C. Mogul, J.~Tourrilhes, P.~Yalagandula, P.~Sharma, and
  S.~Banerjee, ``Devoflow: Scaling flow management for high-performance
  networks,'' \emph{ACM SIGCOMM Computer Communication Review}, vol.~41, no.~4,
  pp. 254--265, 2011.

\bibitem{tavakoli2009applying}
A.~Tavakoli, M.~Casado, T.~Koponen, and S.~Shenker, ``Applying nox to the
  datacenter.'' in \emph{HotNets}, 2009.

\bibitem{al2010hedera}
M.~Al-Fares, S.~Radhakrishnan, B.~Raghavan, N.~Huang, and A.~Vahdat, ``Hedera:
  Dynamic flow scheduling for data center networks.'' in \emph{NSDI}, vol.~10,
  2010, pp. 19--19.

\bibitem{pereini2014espres}
P.~Pere{\'\i}ni, M.~Kuzniar, M.~Canini, and D.~Kosti{\'c}, ``Espres:
  transparent sdn update scheduling,'' in \emph{Proceedings of the third
  workshop on Hot topics in software defined networking}.\hskip 1em plus 0.5em
  minus 0.4em\relax ACM, 2014, pp. 73--78.

\bibitem{qin2015bandwidth}
P.~Qin, B.~Dai, B.~Huang, and G.~Xu, ``Bandwidth-aware scheduling with sdn in
  hadoop: A new trend for big data,'' \emph{IEEE Systems Journal}, 2015.

\bibitem{shvachko2010hadoop}
K.~Shvachko, H.~Kuang, S.~Radia, and R.~Chansler, ``The hadoop distributed file
  system,'' in \emph{Mass storage systems and technologies (MSST), 2010 IEEE
  26th symposium on}.\hskip 1em plus 0.5em minus 0.4em\relax Ieee, 2010, pp.
  1--10.

\bibitem{blial2016overview}
O.~Blial, M.~Ben~Mamoun, and R.~Benaini, ``An overview on sdn architectures
  with multiple controllers,'' \emph{Journal of Computer Networks and
  Communications}, vol. 2016, 2016.

\bibitem{yu2010scalable}
M.~Yu, J.~Rexford, M.~J. Freedman, and J.~Wang, ``Scalable flow-based
  networking with difane,'' \emph{ACM SIGCOMM Computer Communication Review},
  vol.~40, no.~4, pp. 351--362, 2010.

\bibitem{ros2016reliable}
F.~J. Ros and P.~M. Ruiz, ``On reliable controller placements in
  software-defined networks,'' \emph{Computer Communications}, vol.~77, pp.
  41--51, 2016.

\bibitem{stribling2009flexible}
J.~Stribling, Y.~Sovran, I.~Zhang, X.~Pretzer, J.~Li, M.~F. Kaashoek, and
  R.~Morris, ``Flexible, wide-area storage for distributed systems with
  wheelfs.'' in \emph{NSDI}, vol.~9, 2009, pp. 43--58.

\bibitem{hows2014mongodb}
D.~Hows, P.~Membrey, and E.~Plugge, \emph{MongoDB Basics}.\hskip 1em plus 0.5em
  minus 0.4em\relax Apress, 2014.

\bibitem{wang2012nosql}
G.~Wang and J.~Tang, ``The nosql principles and basic application of cassandra
  model,'' in \emph{Computer Science \& Service System (CSSS), 2012
  International Conference on}.\hskip 1em plus 0.5em minus 0.4em\relax IEEE,
  2012, pp. 1332--1335.

\bibitem{Hazelcast}
Hazelcast [accessed on 05/12/2017] available at:\url{ https://hazelcast.org/}.

\bibitem{koponen2010onix}
T.~Koponen, M.~Casado, N.~Gude, J.~Stribling, L.~Poutievski, M.~Zhu,
  R.~Ramanathan, Y.~Iwata, H.~Inoue, T.~Hama \emph{et~al.}, ``Onix: A
  distributed control platform for large-scale production networks.'' in
  \emph{OSDI}, vol.~10, 2010, pp. 1--6.

\bibitem{tootoonchian2010hyperflow}
A.~Tootoonchian and Y.~Ganjali, ``Hyperflow: A distributed control plane for
  openflow,'' in \emph{Proceedings of the 2010 internet network management
  conference on Research on enterprise networking}, 2010, pp. 3--3.

\bibitem{phemius2014disco}
K.~Phemius, M.~Bouet, and J.~Leguay, ``Disco: Distributed sdn controllers in a
  multi-domain environment,'' in \emph{Network Operations and Management
  Symposium (NOMS), 2014 IEEE}.\hskip 1em plus 0.5em minus 0.4em\relax IEEE,
  2014, pp. 1--2.

\bibitem{hassas2012kandoo}
S.~Hassas~Yeganeh and Y.~Ganjali, ``Kandoo: a framework for efficient and
  scalable offloading of control applications,'' in \emph{Proceedings of the
  first workshop on Hot topics in software defined networks}.\hskip 1em plus
  0.5em minus 0.4em\relax ACM, 2012, pp. 19--24.

\bibitem{khattak2014performance}
Z.~K. Khattak, M.~Awais, and A.~Iqbal, ``Performance evaluation of opendaylight
  sdn controller,'' in \emph{Parallel and Distributed Systems (ICPADS), 2014
  20th IEEE International Conference on}.\hskip 1em plus 0.5em minus
  0.4em\relax IEEE, 2014, pp. 671--676.

\bibitem{RamCloud}
RamCloud [accessed on 05/12/2017] available at:\url{
  https://github.com/jdellithorpe/blueprints-ramcloud-graph}.

\bibitem{mccauley2013extending}
J.~McCauley, A.~Panda, M.~Casado, T.~Koponen, and S.~Shenker, ``Extending sdn
  to large-scale networks,'' \emph{Open Networking Summit}, pp. 1--2, 2013.

\bibitem{lee2014iris}
B.~Lee, S.~H. Park, J.~Shin, and S.~Yang, ``Iris: the openflow-based recursive
  sdn controller,'' in \emph{Advanced Communication Technology (ICACT), 2014
  16th International Conference on}.\hskip 1em plus 0.5em minus 0.4em\relax
  IEEE, 2014, pp. 1227--1231.

\bibitem{fu2014orion}
Y.~Fu, J.~Bi, K.~Gao, Z.~Chen, J.~Wu, and B.~Hao, ``Orion: A hybrid
  hierarchical control plane of software-defined networking for large-scale
  networks,'' in \emph{Network Protocols (ICNP), 2014 IEEE 22nd International
  Conference on}.\hskip 1em plus 0.5em minus 0.4em\relax IEEE, 2014, pp.
  569--576.

\bibitem{dixit2014elasticon}
A.~Dixit, F.~Hao, S.~Mukherjee, T.~Lakshman, and R.~R. Kompella, ``Elasticon;
  an elastic distributed sdn controller,'' in \emph{Architectures for
  Networking and Communications Systems (ANCS), 2014 ACM/IEEE Symposium
  on}.\hskip 1em plus 0.5em minus 0.4em\relax IEEE, 2014, pp. 17--27.

\bibitem{santos2014decentralizing}
M.~A. Santos, B.~A. Nunes, K.~Obraczka, T.~Turletti, B.~T. de~Oliveira, and
  C.~B. Margi, ``Decentralizing sdn's control plane,'' in \emph{Local Computer
  Networks (LCN), 2014 IEEE 39th Conference on}.\hskip 1em plus 0.5em minus
  0.4em\relax IEEE, 2014, pp. 402--405.

\bibitem{wang2015lazy}
L.~Wang, K.~Zheng, B.~Yang, Y.~Sun, Y.~Zhang, and S.~Uhlig, ``Lazy ctrl:
  Scalable network control for cloud data centers,'' in \emph{Distributed
  Computing Systems (ICDCS), 2015 IEEE 35th International Conference on}.\hskip
  1em plus 0.5em minus 0.4em\relax IEEE, 2015, pp. 788--789.

\bibitem{killi2017capacitated}
B.~P.~R. Killi and S.~V. Rao, ``Capacitated next controller placement in
  software defined networks,'' \emph{IEEE Transactions on Network and Service
  Management}, vol.~14, no.~3, pp. 514--527, 2017.

\bibitem{hu2017energy}
Y.~Hu, T.~Luo, N.~C. Beaulieu, and C.~Deng, ``The energy-aware controller
  placement problem in software defined networks,'' \emph{IEEE Communications
  Letters}, vol.~21, no.~4, pp. 741--744, 2017.

\bibitem{wang2018effective}
G.~Wang, Y.~Zhao, J.~Huang, and Y.~Wu, ``An effective approach to controller
  placement in software defined wide area networks,'' \emph{IEEE Transactions
  on Network and Service Management}, vol.~15, no.~1, pp. 344--355, 2018.

\bibitem{zhao2017towards}
J.~Zhao, H.~Qu, J.~Zhao, Z.~Luan, and Y.~Guo, ``Towards controller placement
  problem for software-defined network using affinity propagation,''
  \emph{Electronics Letters}, 2017.

\bibitem{hock2013pareto}
D.~Hock, M.~Hartmann, S.~Gebert, M.~Jarschel, T.~Zinner, and P.~Tran-Gia,
  ``Pareto-optimal resilient controller placement in sdn-based core networks,''
  in \emph{Teletraffic Congress (ITC), 2013 25th International}.\hskip 1em plus
  0.5em minus 0.4em\relax IEEE, 2013, pp. 1--9.

\bibitem{lange2015heuristic}
S.~Lange, S.~Gebert, T.~Zinner, P.~Tran-Gia, D.~Hock, M.~Jarschel, and
  M.~Hoffmann, ``Heuristic approaches to the controller placement problem in
  large scale sdn networks,'' \emph{IEEE Transactions on Network and Service
  Management}, vol.~12, no.~1, pp. 4--17, 2015.

\bibitem{ksentini2016using1}
A.~Ksentini, M.~Bagaa, T.~Taleb, and I.~Balasingham, ``On using bargaining game
  for optimal placement of sdn controllers,'' in \emph{Communications (ICC),
  2016 IEEE International Conference on}.\hskip 1em plus 0.5em minus
  0.4em\relax IEEE, 2016, pp. 1--6.

\bibitem{hemodeling}
M.~He, A.~Basta, A.~Blenk, and W.~Kellerer, ``Modeling flow setup time for
  controller placement in sdn: Evaluation for dynamic flows,'' 2017.

\bibitem{zeng2015flow}
D.~Zeng, C.~Teng, L.~Gu, H.~Yao, and Q.~Liang, ``Flow setup time aware minimum
  cost switch-controller association in software-defined networks,'' in
  \emph{Heterogeneous Networking for Quality, Reliability, Security and
  Robustness (QSHINE), 2015 11th International Conference on}.\hskip 1em plus
  0.5em minus 0.4em\relax IEEE, 2015, pp. 259--264.

\bibitem{hu2014reliability}
Y.~Hu, W.~Wang, X.~Gong, X.~Que, and S.~Cheng, ``On reliability-optimized
  controller placement for software-defined networks,'' \emph{China
  Communications}, vol.~11, no.~2, pp. 38--54, 2014.

\bibitem{ros2014five}
F.~J. Ros and P.~M. Ruiz, ``Five nines of southbound reliability in
  software-defined networks,'' in \emph{Proceedings of the third workshop on
  Hot topics in software defined networking}.\hskip 1em plus 0.5em minus
  0.4em\relax ACM, 2014, pp. 31--36.

\bibitem{muller2014survivor}
L.~F. M{\"u}ller, R.~R. Oliveira, M.~C. Luizelli, L.~P. Gaspary, and M.~P.
  Barcellos, ``Survivor: an enhanced controller placement strategy for
  improving sdn survivability,'' in \emph{Global Communications Conference
  (GLOBECOM), 2014 IEEE}.\hskip 1em plus 0.5em minus 0.4em\relax IEEE, 2014,
  pp. 1909--1915.

\bibitem{xiao2014sdn}
P.~Xiao, W.~Qu, H.~Qi, Z.~Li, and Y.~Xu, ``The sdn controller placement problem
  for wan,'' in \emph{Communications in China (ICCC), 2014 IEEE/CIC
  International Conference on}.\hskip 1em plus 0.5em minus 0.4em\relax IEEE,
  2014, pp. 220--224.

\bibitem{benson2010network}
T.~Benson, A.~Akella, and D.~A. Maltz, ``Network traffic characteristics of
  data centers in the wild,'' in \emph{Proceedings of the 10th ACM SIGCOMM
  conference on Internet measurement}.\hskip 1em plus 0.5em minus 0.4em\relax
  ACM, 2010, pp. 267--280.

\bibitem{ruiz2015greco}
A.~Ruiz-Rivera, K.-W. Chin, and S.~Soh, ``Greco: An energy aware controller
  association algorithm for software defined networks,'' \emph{IEEE
  Communications Letters}, vol.~19, no.~4, pp. 541--544, 2015.

\bibitem{rath2014optimal}
H.~K. Rath, V.~Revoori, S.~Nadaf, and A.~Simha, ``Optimal controller placement
  in software defined networks (sdn) using a non-zero-sum game,'' in
  \emph{World of Wireless, Mobile and Multimedia Networks (WoWMoM), 2014 IEEE
  15th International Symposium on a}.\hskip 1em plus 0.5em minus 0.4em\relax
  IEEE, 2014, pp. 1--6.

\bibitem{bari2013dynamic}
M.~F. Bari, A.~R. Roy, S.~R. Chowdhury, Q.~Zhang, M.~F. Zhani, R.~Ahmed, and
  R.~Boutaba, ``Dynamic controller provisioning in software defined networks,''
  in \emph{Network and Service Management (CNSM), 2013 9th International
  Conference on}.\hskip 1em plus 0.5em minus 0.4em\relax IEEE, 2013, pp.
  18--25.

\bibitem{sallahi2015optimal}
A.~Sallahi and M.~St-Hilaire, ``Optimal model for the controller placement
  problem in software defined networks,'' \emph{IEEE Communications Letters},
  vol.~19, no.~1, pp. 30--33, 2015.

\bibitem{cheng2016dynamic}
G.~Cheng, H.~Chen, H.~Hu, and J.~Lan, ``Dynamic switch migration towards a
  scalable sdn control plane,'' \emph{International Journal of Communication
  Systems}, vol.~29, no.~9, pp. 1482--1499, 2016.

\bibitem{wang2017switch}
C.~Wang, B.~Hu, S.~Chen, D.~Li, and B.~Liu, ``A switch migration-based
  decision-making scheme for balancing load in sdn,'' \emph{IEEE Access},
  vol.~5, pp. 4537--4544, 2017.

\bibitem{hu2018easm}
T.~Hu, J.~Lan, J.~Zhang, and W.~Zhao, ``Easm: Efficiency-aware switch migration
  for balancing controller loads in software-defined networking,''
  \emph{Peer-to-Peer Networking and Applications, Springer}, 2018.

\bibitem{OS3E}
OS3E [accessed on 05/12/2017] available at:\url{
  https://www.internet2.edu/products-services/advanced-networking/layer-2-services/}.

\bibitem{6027859}
S.~Knight, H.~Nguyen, N.~Falkner, R.~Bowden, and M.~Roughan, ``The internet
  topology zoo,'' \emph{Selected Areas in Communications, IEEE Journal on},
  vol.~29, no.~9, pp. 1765 --1775, october 2011.

\bibitem{dueck2009affinity}
D.~Dueck, \emph{Affinity propagation: clustering data by passing
  messages}.\hskip 1em plus 0.5em minus 0.4em\relax Citeseer, 2009.

\bibitem{wang2016k}
G.~Wang, Y.~Zhao, J.~Huang, Q.~Duan, and J.~Li, ``A k-means-based network
  partition algorithm for controller placement in software defined network,''
  in \emph{Communications (ICC), 2016 IEEE International Conference on}.\hskip
  1em plus 0.5em minus 0.4em\relax IEEE, 2016, pp. 1--6.

\bibitem{johnson1973note}
D.~B. Johnson, ``A note on dijkstra's shortest path algorithm,'' \emph{Journal
  of the ACM (JACM)}, vol.~20, no.~3, pp. 385--388, 1973.

\bibitem{su2015mdcp}
Z.~Su and M.~Hamdi, ``Mdcp: Measurement-aware distributed controller placement
  for software defined networks,'' in \emph{Parallel and Distributed Systems
  (ICPADS), 2015 IEEE 21st International Conference on}.\hskip 1em plus 0.5em
  minus 0.4em\relax IEEE, 2015, pp. 380--387.

\bibitem{yao2014capacitated}
G.~Yao, J.~Bi, Y.~Li, and L.~Guo, ``On the capacitated controller placement
  problem in software defined networks,'' \emph{IEEE Communications Letters},
  vol.~18, no.~8, pp. 1339--1342, 2014.

\bibitem{gao2015particle}
C.~Gao, H.~Wang, F.~Zhu, L.~Zhai, and S.~Yi, ``A particle swarm optimization
  algorithm for controller placement problem in software defined network,'' in
  \emph{International Conference on Algorithms and Architectures for Parallel
  Processing}.\hskip 1em plus 0.5em minus 0.4em\relax Springer, 2015, pp.
  44--54.

\bibitem{liu2015ncpso}
S.~Liu, H.~Wang, S.~Yi, and F.~Zhu, ``Ncpso: a solution of the controller
  placement problem in software defined networks,'' in \emph{International
  Conference on Algorithms and Architectures for Parallel Processing}.\hskip
  1em plus 0.5em minus 0.4em\relax Springer, 2015, pp. 213--225.

\bibitem{sanner2016hierarchical}
J.-M. Sanner, Y.~Hadjadj-Aoufi, M.~Ouzzif, and G.~Rubino, ``Hierarchical
  clustering for an efficient controllers' placement in software defined
  networks,'' in \emph{Global Information Infrastructure and Networking
  Symposium (GIIS), 2016}.\hskip 1em plus 0.5em minus 0.4em\relax IEEE, 2016,
  pp. 1--7.

\bibitem{deb2003fast}
K.~Deb, M.~Mohan, and S.~Mishra, ``A fast multi-objective evolutionary
  algorithm for finding well-spread pareto-optimal solutions,'' \emph{KanGAL
  report}, vol. 2003002, pp. 1--18, 2003.

\bibitem{ahmadi2015hybrid}
V.~Ahmadi, A.~Jalili, S.~M. Khorramizadeh, and M.~Keshtgari, ``A hybrid nsga-ii
  for solving multiobjective controller placement in sdn,'' in
  \emph{Knowledge-Based Engineering and Innovation (KBEI), 2015 2nd
  International Conference on}.\hskip 1em plus 0.5em minus 0.4em\relax IEEE,
  2015, pp. 663--669.

\bibitem{jalili2017optimal}
A.~Jalili, M.~Keshtgari, and R.~Akbari, ``Optimal controller placement in large
  scale software defined networks based on modified nsga-ii,'' \emph{Applied
  Intelligence}, pp. 1--15, 2017.

\bibitem{jalili2015controller}
A.~Jalili, V.~Ahmadi, M.~Keshtgari, and M.~Kazemi, ``Controller placement in
  software-defined wan using multi objective genetic algorithm,'' in
  \emph{Knowledge-Based Engineering and Innovation (KBEI), 2015 2nd
  International Conference on}.\hskip 1em plus 0.5em minus 0.4em\relax IEEE,
  2015, pp. 656--662.

\bibitem{ahmadi2017adaptive}
V.~Ahmadi and M.~Khorramizadeh, ``An adaptive heuristic for multi-objective
  controller placement in software-defined networks,'' \emph{Computers \&
  Electrical Engineering}, 2017.

\bibitem{deb2002fast}
K.~Deb, A.~Pratap, S.~Agarwal, and T.~Meyarivan, ``A fast and elitist
  multiobjective genetic algorithm: Nsga-ii,'' \emph{IEEE transactions on
  evolutionary computation}, vol.~6, no.~2, pp. 182--197, 2002.

\bibitem{liao2017genetic}
L.~Liao and V.~C. Leung, ``Genetic algorithms with particle swarm optimization
  based mutation for distributed controller placement in sdns,'' in
  \emph{Network Function Virtualization and Software Defined Networks
  (NFV-SDN), 2017 IEEE Conference on}.\hskip 1em plus 0.5em minus 0.4em\relax
  IEEE, 2017, pp. 1--6.

\bibitem{guo2013controller}
M.~Guo and P.~Bhattacharya, ``Controller placement for improving resilience of
  software-defined networks,'' in \emph{Networking and Distributed Computing
  (ICNDC), 2013 Fourth International Conference on}.\hskip 1em plus 0.5em minus
  0.4em\relax IEEE, 2013, pp. 23--27.

\bibitem{hu2013reliability}
Y.~Hu, W.~Wendong, X.~Gong, X.~Que, and C.~Shiduan, ``Reliability-aware
  controller placement for software-defined networks,'' in \emph{Integrated
  Network Management (IM 2013), 2013 IFIP/IEEE International Symposium
  on}.\hskip 1em plus 0.5em minus 0.4em\relax IEEE, 2013, pp. 672--675.

\bibitem{kirkpatrick1983optimization}
S.~Kirkpatrick, C.~D. Gelatt, and M.~P. Vecchi, ``Optimization by simulated
  annealing,'' \emph{science}, vol. 220, no. 4598, pp. 671--680, 1983.

\bibitem{guo2015towards}
S.~Guo, S.~Yang, Q.~Li, and Y.~Jiang, ``Towards controller placement for robust
  software-defined networks,'' in \emph{Computing and Communications Conference
  (IPCCC), 2015 IEEE 34th International Performance}.\hskip 1em plus 0.5em
  minus 0.4em\relax IEEE, 2015, pp. 1--8.

\bibitem{zhong2016min}
Q.~Zhong, Y.~Wang, W.~Li, and X.~Qiu, ``A min-cover based controller placement
  approach to build reliable control network in sdn,'' in \emph{Network
  Operations and Management Symposium (NOMS), 2016 IEEE/IFIP}.\hskip 1em plus
  0.5em minus 0.4em\relax IEEE, 2016, pp. 481--487.

\bibitem{aoki2016controller}
H.~Aoki and N.~Shinomiya, ``Controller placement problem to enhance performance
  in multi-domain sdn networks,'' in \emph{ICN}, vol. 120, 2016, p. 2016.

\bibitem{ishigaki2016controller}
G.~Ishigaki and N.~Shinomiya, ``Controller placement algorithm to alleviate
  burdens on communication nodes,'' in \emph{Computing, Networking and
  Communications (ICNC), 2016 International Conference on}.\hskip 1em plus
  0.5em minus 0.4em\relax IEEE, 2016, pp. 1--5.

\bibitem{cormen1990floyd}
T.~H. Cormen, C.~E. Leiserson, and R.~L. Rivest, ``The floyd-warshall
  algorithm,'' \emph{Introduction to Algorithms}, pp. 558--565, 1990.

\bibitem{sanner2017evolutionary}
J.-M. Sanner, Y.~Hadjadj-Aoul, M.~Ouzzif, and G.~Rubino, ``An evolutionary
  controllers' placement algorithm for reliable sdn networks,'' in \emph{IFIP
  International Workshop on Management of SDN and NFV Systems
  (ManSDNNFV'2017)}, 2017.

\bibitem{sakarovitchoptimisation}
M.~Sakarovitch, ``Optimisation combinatoire: Programmation discrete, ser.
  collection enseignement des sciences. hermann, 1984.''

\bibitem{lange2015specialized}
S.~Lange, S.~Gebert, J.~Spoerhase, P.~Rygielski, T.~Zinner, S.~Kounev, and
  P.~Tran-Gia, ``Specialized heuristics for the controller placement problem in
  large scale sdn networks,'' in \emph{Teletraffic Congress (ITC 27), 2015 27th
  International}.\hskip 1em plus 0.5em minus 0.4em\relax IEEE, 2015, pp.
  210--218.

\bibitem{czyzzak1998pareto}
P.~Czyz{\.z}ak and A.~Jaszkiewicz, ``Pareto simulated annealing—a
  metaheuristic technique for multiple-objective combinatorial optimization,''
  \emph{Journal of Multi-Criteria Decision Analysis}, vol.~7, no.~1, pp.
  34--47, 1998.

\bibitem{perrot2016optimal}
N.~Perrot and T.~Reynaud, ``Optimal placement of controllers in a resilient sdn
  architecture,'' in \emph{Design of Reliable Communication Networks (DRCN),
  2016 12th International Conference on the}.\hskip 1em plus 0.5em minus
  0.4em\relax IEEE, 2016, pp. 145--151.

\bibitem{ilog2012cplex}
I.~ILOG, ``Cplex optimizer,'' \emph{En ligne]. Available: http://www-01. ibm.
  com/software/commerce/optimization/cplex-optimizer}, 2012.

\bibitem{yao2013cascading}
G.~Yao, J.~Bi, and L.~Guo, ``On the cascading failures of multi-controllers in
  software defined networks,'' in \emph{Network Protocols (ICNP), 2013 21st
  IEEE International Conference on}.\hskip 1em plus 0.5em minus 0.4em\relax
  IEEE, 2013, pp. 1--2.

\bibitem{motter2002cascade}
A.~E. Motter and Y.-C. Lai, ``Cascade-based attacks on complex networks,''
  \emph{Physical Review E}, vol.~66, no.~6, p. 065102, 2002.

\bibitem{jimenez2014controller}
Y.~Jimenez, C.~Cervello-Pastor, and A.~J. Garcia, ``On the controller placement
  for designing a distributed sdn control layer,'' in \emph{Networking
  Conference, 2014 IFIP}.\hskip 1em plus 0.5em minus 0.4em\relax IEEE, 2014,
  pp. 1--9.

\bibitem{killi2016optimal}
B.~P.~R. Killi and S.~V. Rao, ``Optimal model for failure foresight capacitated
  controller placement in software-defined networks,'' \emph{IEEE
  Communications Letters}, vol.~20, no.~6, pp. 1108--1111, 2016.

\bibitem{ding2001min}
C.~H. Ding, X.~He, H.~Zha, M.~Gu, and H.~D. Simon, ``A min-max cut algorithm
  for graph partitioning and data clustering,'' in \emph{Data Mining, 2001.
  ICDM 2001, Proceedings IEEE International Conference on}.\hskip 1em plus
  0.5em minus 0.4em\relax IEEE, 2001, pp. 107--114.

\bibitem{xiao2016ak}
P.~Xiao, Z.-y. Li, S.~Guo, H.~Qi, W.-y. Qu, and H.-s. Yu, ``Ak self-adaptive
  sdn controller placement for wide area networks,'' \emph{Frontiers of
  Information Technology \& Electronic Engineering}, vol.~17, no.~7, pp.
  620--633, 2016.

\bibitem{liao2017density}
J.~Liao, H.~Sun, J.~Wang, Q.~Qi, K.~Li, and T.~Li, ``Density cluster based
  approach for controller placement problem in large-scale software defined
  networkings,'' \emph{Computer Networks}, vol. 112, pp. 24--35, 2017.

\bibitem{sallahi2017expansion}
A.~Sallahi and M.~St-Hilaire, ``Expansion model for the controller placement
  problem in software defined networks,'' \emph{IEEE Communications Letters},
  vol.~21, no.~2, pp. 274--277, 2017.

\bibitem{zhao2017scalable}
Z.~Zhao and B.~Wu, ``Scalable sdn architecture with distributed placement of
  controllers for wan,'' \emph{Concurrency and Computation: Practice and
  Experience}, vol.~29, no.~16, 2017.

\bibitem{cheng2015qos}
T.~Y. Cheng, M.~Wang, and X.~Jia, ``Qos-guaranteed controller placement in
  sdn,'' in \emph{Global Communications Conference (GLOBECOM), 2015
  IEEE}.\hskip 1em plus 0.5em minus 0.4em\relax IEEE, 2015, pp. 1--6.

\bibitem{sahoo2016optimal}
K.~S. Sahoo, B.~Sahoo, R.~Dash, and N.~Jena, ``Optimal controller selection in
  software defined network using a greedy-sa algorithm,'' in \emph{Computing
  for Sustainable Global Development (INDIACom), 2016 3rd International
  Conference on}.\hskip 1em plus 0.5em minus 0.4em\relax IEEE, 2016, pp.
  2342--2346.

\bibitem{sahoo2017metaheuristic}
K.~S. Sahoo, A.~Sarkar, S.~K. Mishra, B.~Sahoo, D.~Puthal, M.~S. Obaidat, and
  B.~Sadun, ``Metaheuristic solutions for solving controller placement problem
  in sdn-based wan architecture,'' 2017.

\bibitem{sahoo2017placement}
K.~S. Sahoo, S.~Sahoo, A.~Sarkar, B.~Sahoo, and R.~Dash, ``On the placement of
  controllers for designing a wide area software defined networks,'' in
  \emph{Region 10 Conference, TENCON 2017-2017 IEEE}.\hskip 1em plus 0.5em
  minus 0.4em\relax IEEE, 2017, pp. 3123--3128.

\bibitem{ishigaki2017cluster}
G.~Ishigaki, R.~Gour, A.~Yousefpour, N.~Shinomiya, and J.~P. Jue, ``Cluster
  leader election problem for distributed controller placement in sdn,'' in
  \emph{GLOBECOM 2017-2017 IEEE Global Communications Conference}.\hskip 1em
  plus 0.5em minus 0.4em\relax IEEE, 2017, pp. 1--6.

\bibitem{killi2018placement}
B.~P.~R. Killi and S.~V. Rao, ``On placement of hypervisors and controllers in
  virtualized software defined network,'' \emph{IEEE Transactions on Network
  and Service Management}, vol.~15, no.~2, pp. 840--853, 2018.

\bibitem{ozsoy2006exact}
F.~A. {\"O}zsoy and M.~{\c{C}}. P{\i}nar, ``An exact algorithm for the
  capacitated vertex p-center problem,'' \emph{Computers \& Operations
  Research}, vol.~33, no.~5, pp. 1420--1436, 2006.

\bibitem{han2016minimum}
L.~Han, Z.~Li, W.~Liu, K.~Dai, and W.~Qu, ``Minimum control latency of sdn
  controller placement,'' in \emph{Trustcom/BigDataSE/I​ SPA, 2016
  IEEE}.\hskip 1em plus 0.5em minus 0.4em\relax IEEE, 2016, pp. 2175--2180.

\bibitem{hu2016load}
Y.~Hu, T.~Luo, W.~Wang, and C.~Deng, ``On the load balanced controller
  placement problem in software defined networks,'' in \emph{Computer and
  Communications (ICCC), 2016 2nd IEEE International Conference on}.\hskip 1em
  plus 0.5em minus 0.4em\relax IEEE, 2016, pp. 2430--2434.

\bibitem{bo2016controller}
H.~Bo, W.~Youke, W.~Chuan'an, and W.~Ying, ``The controller placement problem
  for software-defined networks,'' in \emph{Computer and Communications (ICCC),
  2016 2nd IEEE International Conference on}.\hskip 1em plus 0.5em minus
  0.4em\relax IEEE, 2016, pp. 2435--2439.

\bibitem{kuang2018hierarchical}
H.~Kuang, Y.~Qiu, R.~Li, and X.~Liu, ``A hierarchical k-means algorithm for
  controller placement in sdn-based wan architecture,'' in \emph{Measuring
  Technology and Mechatronics Automation (ICMTMA), 2018 10th International
  Conference on}.\hskip 1em plus 0.5em minus 0.4em\relax IEEE, 2018, pp.
  263--267.

\bibitem{qi2016k}
J.~Qi, Y.~Yu, L.~Wang, and J.~Liu, ``K*-means: An effective and efficient
  k-means clustering algorithm,'' in \emph{Big Data and Cloud Computing
  (BDCloud), Social Computing and Networking (SocialCom), Sustainable Computing
  and Communications (SustainCom)(BDCloud-SocialCom-SustainCom), 2016 IEEE
  International Conferences on}.\hskip 1em plus 0.5em minus 0.4em\relax IEEE,
  2016, pp. 242--249.

\bibitem{zhu2017control}
L.~Zhu, R.~Chai, and Q.~Chen, ``Control plane delay minimization based sdn
  controller placement scheme,'' in \emph{Wireless Communications and Signal
  Processing (WCSP), 2017 9th International Conference on}.\hskip 1em plus
  0.5em minus 0.4em\relax IEEE, 2017, pp. 1--6.

\bibitem{killi2018cooperative}
B.~P.~R. Killi, E.~A. Reddy, and S.~V. Rao, ``Cooperative game theory based
  network partitioning for controller placement in sdn,'' in
  \emph{Communication Systems \& Networks (COMSNETS), 2018 10th International
  Conference on}.\hskip 1em plus 0.5em minus 0.4em\relax IEEE, 2018, pp.
  105--112.

\bibitem{tanha2016enduring}
M.~Tanha, D.~Sajjadi, and J.~Pan, ``Enduring node failures through resilient
  controller placement for software defined networks,'' in \emph{Global
  Communications Conference (GLOBECOM), 2016 IEEE}.\hskip 1em plus 0.5em minus
  0.4em\relax IEEE, 2016, pp. 1--7.

\bibitem{liu2016heuristics}
B.~Liu, B.~Wang, and X.~Xi, ``Heuristics for sdn controller deployment using
  community detection algorithm,'' in \emph{Software Engineering and Service
  Science (ICSESS), 2016 7th IEEE International Conference on}.\hskip 1em plus
  0.5em minus 0.4em\relax IEEE, 2016, pp. 253--258.

\bibitem{liu2016reliability}
J.~Liu, J.~Liu, and R.~Xie, ``Reliability-based controller placement algorithm
  in software defined networking,'' \emph{Computer Science and Information
  Systems}, no.~00, pp. 14--14, 2016.

\bibitem{naning2016sdn}
H.~S. Naning, R.~Munadi, and M.~Z. Effendy, ``Sdn controller placement design:
  For large scale production network,'' in \emph{Wireless and Mobile (APWiMob),
  2016 IEEE Asia Pacific Conference on}.\hskip 1em plus 0.5em minus 0.4em\relax
  IEEE, 2016, pp. 74--79.

\bibitem{borcoci2016multi}
E.~Borcoci, T.~Ambarus, and M.~Vochin, ``Multi-criteria based optimization of
  placement for software defined networking controllers and forwarding nodes,''
  \emph{ICN 2016}, p. 114, 2016.

\bibitem{bannour2017scalability}
F.~Bannour, S.~Souihi, and A.~Mellouk, ``Scalability and reliability aware sdn
  controller placement strategies,'' in \emph{2017 13th International
  Conference on Network and Service Management (CNSM)}.\hskip 1em plus 0.5em
  minus 0.4em\relax IEEE, 2017, pp. 1--4.

\bibitem{killi2018link}
B.~P.~R. Killi and S.~V. Rao, ``Link failure aware capacitated controller
  placement in software defined networks,'' in \emph{Information Networking
  (ICOIN), 2018 International Conference on}.\hskip 1em plus 0.5em minus
  0.4em\relax IEEE, 2018, pp. 292--297.

\bibitem{tanha2018capacity}
M.~Tanha, D.~Sajjadi, R.~Ruby, and J.~Pan, ``Capacity-aware and
  delay-guaranteed resilient controller placement for software-defined wans,''
  \emph{IEEE Transactions on Network and Service Management}, 2018.

\bibitem{zhang2016optimal}
B.~Zhang, X.~Wang, L.~Ma, and M.~Huang, ``Optimal controller placement problem
  in internet-oriented software defined network,'' in \emph{Cyber-Enabled
  Distributed Computing and Knowledge Discovery (CyberC), 2016 International
  Conference on}.\hskip 1em plus 0.5em minus 0.4em\relax IEEE, 2016, pp.
  481--488.

\bibitem{hu2012placement}
Y.-n. HU, W.-d. WANG, X.-y. GONG, X.-r. QUE, and S.-d. CHENG, ``On the
  placement of controllers in software-defined networks,'' \emph{The Journal of
  China Universities of Posts and Telecommunications}, vol.~19, pp.
  92\,171--97, 2012.

\bibitem{vizarreta2016controller}
P.~Vizarreta, C.~M. Machuca, and W.~Kellerer, ``Controller placement strategies
  for a resilient sdn control plane,'' in \emph{Resilient Networks Design and
  Modeling (RNDM), 2016 8th International Workshop on}.\hskip 1em plus 0.5em
  minus 0.4em\relax IEEE, 2016, pp. 253--259.

\bibitem{zhang2017survivability}
L.~Zhang, Y.~Wang, W.~Li, X.~Qiu, and Q.~Zhong, ``A survivability-based backup
  approach for controllers in multi-controller sdn against failures,'' in
  \emph{Network Operations and Management Symposium (APNOMS), 2017 19th
  Asia-Pacific}.\hskip 1em plus 0.5em minus 0.4em\relax IEEE, 2017, pp.
  100--105.

\bibitem{li2017sharing}
J.~Li, Y.~Wang, W.~Li, and X.~Qiu, ``Sharing data store and backup controllers
  for resilient control plane in multi-domain sdn,'' in \emph{Integrated
  Network and Service Management (IM), 2017 IFIP/IEEE Symposium on}.\hskip 1em
  plus 0.5em minus 0.4em\relax IEEE, 2017, pp. 476--482.

\bibitem{li2017efficient}
H.~Li, R.~E. De~Grande, and A.~Boukerche, ``An efficient cpp solution for
  resilience-oriented sdn controller deployment,'' in \emph{Parallel and
  Distributed Processing Symposium Workshops (IPDPSW), 2017 IEEE
  International}.\hskip 1em plus 0.5em minus 0.4em\relax IEEE, 2017, pp.
  540--549.

\bibitem{gurobi2015gurobi}
I.~Gurobi~Optimization, ``Gurobi optimizer reference manual,'' \emph{URL
  http://www. gurobi. com}, 2015.

\bibitem{dixit2013towards}
A.~Dixit, F.~Hao, S.~Mukherjee, T.~Lakshman, and R.~Kompella, ``Towards an
  elastic distributed sdn controller,'' in \emph{ACM SIGCOMM Computer
  Communication Review}, vol.~43, no.~4.\hskip 1em plus 0.5em minus 0.4em\relax
  ACM, 2013, pp. 7--12.

\bibitem{cheng2015dha}
G.~Cheng, H.~Chen, Z.~Wang, and S.~Chen, ``Dha: Distributed decisions on the
  switch migration toward a scalable sdn control plane,'' in \emph{IFIP
  Networking Conference (IFIP Networking), 2015}.\hskip 1em plus 0.5em minus
  0.4em\relax IEEE, 2015, pp. 1--9.

\bibitem{cello2017balcon}
M.~Cello, Y.~Xu, A.~Walid, G.~Wilfong, H.~J. Chao, and M.~Marchese, ``Balcon: A
  distributed elastic sdn control via efficient switch migration,'' in
  \emph{Cloud Engineering (IC2E), 2017 IEEE International Conference on}.\hskip
  1em plus 0.5em minus 0.4em\relax IEEE, 2017, pp. 40--50.

\bibitem{kyung2015load}
Y.~Kyung, K.~Hong, T.~M. Nguyen, S.~Park, and J.~Park, ``A load distribution
  scheme over multiple controllers for scalable sdn,'' in \emph{Ubiquitous and
  Future Networks (ICUFN), 2015 Seventh International Conference on}.\hskip 1em
  plus 0.5em minus 0.4em\relax IEEE, 2015, pp. 808--810.

\bibitem{sridharan2017multiple}
V.~Sridharan, M.~Gurusamy, and T.~Truong-Huu, ``On multiple controller mapping
  in software defined networks with resilience constraints,'' \emph{IEEE
  Communications Letters}, vol.~21, no.~8, pp. 1763--1766, 2017.

\bibitem{zhou2017load}
Y.~Zhou, Y.~Wang, J.~Yu, J.~Ba, and S.~Zhang, ``Load balancing for multiple
  controllers in sdn based on switches group,'' in \emph{Network Operations and
  Management Symposium (APNOMS), 2017 19th Asia-Pacific}.\hskip 1em plus 0.5em
  minus 0.4em\relax IEEE, 2017, pp. 227--230.

\bibitem{tam2011}
A.~S.~W. Tam, K.~Xi, and H.~J. Chao, ``Use of devolved controllers in data
  center networks,'' in \emph{2011 IEEE Conference on Computer Communications
  Workshops (INFOCOM WKSHPS)}.\hskip 1em plus 0.5em minus 0.4em\relax IEEE,
  2011, pp. 596--601.

\bibitem{liang2014balancing}
W.~Liang, X.~Gao, F.~Wu, G.~Clien, and W.~Wei, ``Balancing traffic load for
  devolved controllers in data center networks,'' in \emph{Global
  Communications Conference (GLOBECOM), 2014 IEEE}.\hskip 1em plus 0.5em minus
  0.4em\relax IEEE, 2014, pp. 2258--2263.

\bibitem{gao2017traffic}
X.~Gao, L.~Kong, W.~Li, W.~Liang, Y.~Chen, and G.~Chen, ``Traffic load
  balancing schemes for devolved controllers in mega data centers,'' \emph{IEEE
  Transactions on Parallel and Distributed Systems}, vol.~28, no.~2, pp.
  572--585, 2017.

\bibitem{kim2018hes}
W.~Kim, J.~Li, J.~W.-K. Hong, and Y.-J. Suh, ``Hes-cop: Heuristic
  switch-controller placement scheme for distributed sdn controllers in data
  center networks,'' \emph{International Journal of Network Management},
  vol.~28, no.~3, p. e2015, 2018.

\bibitem{ul2015revisiting}
M.~T.~I. ul~Huque, G.~Jourjon, and V.~Gramoli, ``Revisiting the controller
  placement problem,'' in \emph{Local Computer Networks (LCN), 2015 IEEE 40th
  Conference on}.\hskip 1em plus 0.5em minus 0.4em\relax IEEE, 2015, pp.
  450--453.

\bibitem{welzl1991smallest}
E.~Welzl, ``Smallest enclosing disks (balls and ellipsoids),'' in \emph{New
  results and new trends in computer science}.\hskip 1em plus 0.5em minus
  0.4em\relax Springer, 1991, pp. 359--370.

\bibitem{zhou2018elastic}
Y.~Zhou, K.~Zheng, W.~Ni, and R.~P. Liu, ``Elastic switch migration for control
  plane load balancing in sdn,'' \emph{IEEE Access}, 2018.

\bibitem{rubner2000earth}
Y.~Rubner, C.~Tomasi, and L.~J. Guibas, ``The earth mover's distance as a
  metric for image retrieval,'' \emph{International journal of computer
  vision}, vol.~40, no.~2, pp. 99--121, 2000.

\bibitem{mattos2016profiling}
D.~M. Mattos, O.~C.~M. Duarte, and G.~Pujolle, ``Profiling software defined
  networks for dynamic distributed-controller provisioning,'' in \emph{Network
  of the Future (NOF), 2016 7th International Conference on the}.\hskip 1em
  plus 0.5em minus 0.4em\relax IEEE, 2016, pp. 1--5.

\bibitem{wang2017efficient}
T.~Wang, F.~Liu, and H.~Xu, ``An efficient online algorithm for dynamic sdn
  controller assignment in data center networks,'' \emph{IEEE/ACM Transactions
  on Networking}, vol.~25, no.~5, pp. 2788--2801, 2017.

\bibitem{zhang2013moving}
L.~Zhang, C.~Wu, Z.~Li, C.~Guo, M.~Chen, and F.~C. Lau, ``Moving big data to
  the cloud: An online cost-minimizing approach,'' \emph{IEEE Journal on
  Selected Areas in Communications}, vol.~31, no.~12, pp. 2710--2721, 2013.

\bibitem{kumari2018optimizing}
A.~Kumari, J.~Chandra, and A.~S. Sairam, ``Optimizing flow setup time in
  software defined network,'' in \emph{Communication Systems \& Networks
  (COMSNETS), 2018 10th International Conference on}.\hskip 1em plus 0.5em
  minus 0.4em\relax IEEE, 2018, pp. 543--545.

\bibitem{ul2017large}
M.~T.~I. ul~Huque, W.~Si, G.~Jourjon, and V.~Gramoli, ``Large-scale dynamic
  controller placement,'' \emph{IEEE Transactions on Network and Service
  Management}, vol.~14, no.~1, pp. 63--76, 2017.

\bibitem{chaudet2013wireless}
C.~Chaudet and Y.~Haddad, ``Wireless software defined networks: Challenges and
  opportunities,'' in \emph{Microwaves, Communications, Antennas and
  Electronics Systems (COMCAS), 2013 IEEE International Conference on}.\hskip
  1em plus 0.5em minus 0.4em\relax IEEE, 2013, pp. 1--5.

\bibitem{detti2013wireless}
A.~Detti, C.~Pisa, S.~Salsano, and N.~Blefari-Melazzi, ``Wireless mesh software
  defined networks (wmsdn),'' in \emph{Wireless and Mobile Computing,
  Networking and Communications (WiMob), 2013 IEEE 9th International Conference
  on}.\hskip 1em plus 0.5em minus 0.4em\relax IEEE, 2013, pp. 89--95.

\bibitem{jin2013softcell}
X.~Jin, L.~E. Li, L.~Vanbever, and J.~Rexford, ``Softcell: Scalable and
  flexible cellular core network architecture,'' in \emph{Proceedings of the
  ninth ACM conference on Emerging networking experiments and
  technologies}.\hskip 1em plus 0.5em minus 0.4em\relax ACM, 2013, pp.
  163--174.

\bibitem{crowd}
H.~Ali-Ahmad, C.~Cicconetti, A.~d.~l. Oliva, M.~Dräxler, R.~Gupta, V.~Mancuso,
  L.~Roullet, and V.~Sciancalepore, ``Crowd: An sdn approach for densenets,''
  in \emph{2013 Second European Workshop on Software Defined Networks}, Oct
  2013, pp. 25--31.

\bibitem{auroux2014flow}
S.~Auroux and H.~Karl, ``Flow processing-aware controller placement in wireless
  densenets,'' in \emph{Personal, Indoor, and Mobile Radio Communication
  (PIMRC), 2014 IEEE 25th Annual International Symposium on}.\hskip 1em plus
  0.5em minus 0.4em\relax IEEE, 2014, pp. 1294--1299.

\bibitem{auroux2015efficient}
------, ``Efficient flow processing-aware controller placement in future
  wireless networks,'' in \emph{Wireless Communications and Networking
  Conference (WCNC), 2015 IEEE}.\hskip 1em plus 0.5em minus 0.4em\relax IEEE,
  2015, pp. 1787--1792.

\bibitem{johnston2015controller}
M.~Johnston and E.~Modiano, ``Controller placement for maximum throughput under
  delayed csi,'' in \emph{Modeling and Optimization in Mobile, Ad Hoc, and
  Wireless Networks (WiOpt), 2015 13th International Symposium on}.\hskip 1em
  plus 0.5em minus 0.4em\relax IEEE, 2015, pp. 521--528.

\bibitem{akyildiz2015softair}
I.~F. Akyildiz, P.~Wang, and S.-C. Lin, ``Softair: A software defined
  networking architecture for 5g wireless systems,'' \emph{Computer Networks},
  vol.~85, pp. 1--18, 2015.

\bibitem{ksentini2016using}
A.~Ksentini, M.~Bagaa, and T.~Taleb, ``On using sdn in 5g: the controller
  placement problem,'' in \emph{2016 IEEE Global Communications Conference
  (GLOBECOM)}.\hskip 1em plus 0.5em minus 0.4em\relax IEEE, 2016, pp. 1--6.

\bibitem{abdel2017stochastic}
M.~J. Abdel-Rahman, E.~A. Mazied, A.~MacKenzie, S.~Midkiff, M.~R. Rizk, and
  M.~El-Nainay, ``On stochastic controller placement in software-defined
  wireless networks,'' in \emph{Wireless Communications and Networking
  Conference (WCNC), 2017 IEEE}.\hskip 1em plus 0.5em minus 0.4em\relax IEEE,
  2017, pp. 1--6.

\bibitem{abdel2017robust}
M.~J. Abdel-Rahman, E.~A. Mazied, K.~Teague, A.~B. MacKenzie, and S.~F.
  Midkiff, ``Robust controller placement and assignment in software-defined
  cellular networks,'' in \emph{Computer Communication and Networks (ICCCN),
  2017 26th International Conference on}.\hskip 1em plus 0.5em minus
  0.4em\relax IEEE, 2017, pp. 1--9.

\bibitem{dvir2018wireless}
A.~Dvir, Y.~Haddad, and A.~Zilberman, ``Wireless controller placement
  problem,'' in \emph{Consumer Communications \& Networking Conference (CCNC),
  2018 15th IEEE Annual}.\hskip 1em plus 0.5em minus 0.4em\relax IEEE, 2018,
  pp. 1--4.

\bibitem{liu2018joint}
J.~Liu, Y.~Shi, L.~Zhao, Y.~Cao, W.~Sun, and N.~Kato, ``Joint placement of
  controllers and gateways in sdn-enabled 5g-satellite integrated network,''
  \emph{IEEE Journal on Selected Areas in Communications}, vol.~36, no.~2, pp.
  221--232, 2018.

\bibitem{ashrafplacing}
U.~Ashraf, ``Placing controllers in software-defined wireless mesh networks.''

\end{thebibliography}
\balance





\end{document}